%% file: draft.tex
\documentclass[11pt,reqno]{amsart}
\usepackage{graphics,amsmath,amssymb,epsfig,stmaryrd,color,amsaddr}
\usepackage{subfigure,amsfonts,geometry,array}
\usepackage{graphicx}
\graphicspath{ {./images/} }
\usepackage{amsthm}
\usepackage{mathabx} 
\usepackage{flafter}
\usepackage[svgnames]{xcolor}
\usepackage{bbm}
\usepackage{booktabs}
\usepackage{hyperref}
\newcommand{\indep}{\rotatebox[origin=c]{90}{$\models$}}
\newcommand{\nindep}{  \mathrel{\ooalign{    \hidewidth$/$\hidewidth\cr
    $\indep$\cr
  }}}
\usepackage{natbib}
\usepackage{ulem}
\usepackage{float}
\usepackage{threeparttable}
\usepackage{caption}
\usepackage{pdflscape}
\usepackage{enumerate}
\usepackage{multirow}

\usepackage[ruled,vlined,linesnumbered]{algorithm2e}

\makeatletter
\renewcommand\@endtheorem{\vvv@endmarker\endtrivlist\@endpefalse}
\newcommand\vvv@endmarker{  {\nobreak\hfil\penalty50
  \hskip2em\vadjust{}\nobreak\hfil\openbox
  \parfillskip=0pt \finalhyphendemerits=0 \par
  \penalty 10000 \parskip=0pt\noindent}\ignorespaces}
\makeatother

\theoremstyle{plain}

\newtheorem*{assumption*}{Assumption}
\newtheorem{assumption}{Assumption}

\newtheorem{corollary}{Corollary}

\newtheorem{definition}{Definition}
\newtheorem{example}{Example}

\newtheorem{lemma}{Lemma}

\newtheorem{proposition}{Proposition}
\newtheorem{remark}{Remark}

\numberwithin{equation}{section}

\newcommand{\figurenote}[1]{  \par\vspace{0.35em}  \begin{minipage}{0.92\linewidth}
  \footnotesize\textit{Notes:} #1
  \end{minipage}}

\newcommand{\MTE}{\operatorname{MTE}}

\begin{document}
\nonstopmode

\title[Treatment Effects with a Misclassified Treatment]{Local Average and Marginal Treatment Effects with a Misclassified Treatment}
\author{Kyunghoon Ban, D\'esir\'e K\'edagni, and Santiago Acerenza}
\address{Rochester Institute of Technology, UNC-Chapel Hill, and ORT Uruguay} 
\noindent \date{\scriptsize{The first draft was of May 1, 2021. The present version is as of \today. We are grateful to the Editor Xiaoxia Shi and three anonymous referees for their constructive feedback. We thank Isaiah Andrews, Alberto Abadie, Ot\'avio Bartalotti, Chris Bollinger, Helle Bunzel, Augustine Denteh, Sukjin Han, Guido Imbens, Shakeeb Khan, Brent Kreider, Matt Masten, Arnaud Maurel, Anna Mikusheva, Francesca Molinari, Ismael Mourifi\'e, Pierre Nguimkeu, Vitor Possebom, Thomas Richardson, Adam Rosen, Eric Tchetgen Tchetgen, Denni Tommasi, Takuya Ura, Lina Zhang, seminar participants at Cornell, Duke, Iowa State, UMass Amherst (Statistics), Syracuse, Tilburg, U. Iowa, Harvard-MIT, Rochester, U. Georgia, Wharton (Statistics), Simon Institute, AfES 2021, NASMES 2021, IAAE 2021, LACEA LAMES 2021, SEA 2021 for helpful discussions and comments. All errors are ours.
Corresponding address: Max Lowenthal Hall 3338, 107 Lomb Memorial Drive, Rochester, NY 14623, USA. Email address: kban@saunders.rit.edu}}

\begin{abstract}
This paper studies identification of the local average and marginal treatment effects (LATE and MTE) with a misclassified binary treatment variable.
We derive bounds on the (generalized) LATE and exploit its relationship with the MTE to further bound the MTE.
Indeed, under some standard assumptions, the MTE is a limit of the ratio of the variation in the conditional expectation of the observed outcome given the instrument to the variation in the true propensity score, which is partially identified. 
We characterize the identified set for the propensity score, and then for the MTE.
We show that our LATE bounds are tighter than the existing bounds and that the sign of the MTE is \textit{locally identified} under some mild regularity conditions. 
We use our MTE bounds to derive bounds on other commonly used parameters in the literature such as the policy relevant treatment parameter (PRTE) and illustrate the practical relevance of our derived bounds through numerical and empirical results.
\end{abstract}
\maketitle

{\footnotesize \textbf{Keywords}: Heterogeneous treatment effects, misclassification, instrumental variable, set identification.

}
\clearpage

\section{Introduction}
The existence of measurement error in a treatment variable makes the identification of many parameters used in the causal inference literature challenging. 
When the treatment variable is binary, it is well-understood in the literature that the measurement error is nonclassical, that is, it depends on the true treatment. 
Even in the homogeneous treatment effect framework, measurement errors in a binary regressor can result in severe identification deterioration of regression coefficients \citep{Kreider2010}. 
\citet{Ura2018} appears to be the first to investigate the identifying power of an instrumental variable (IV) in the (unobserved) heterogeneous treatment effect model when the treatment is endogenous and mismeasured. 
He derives bounds on the local average treatment effect (LATE) when a binary instrument is available.\footnote{\citet{Mahajan2006} and \citet{Lewbel2007} allow for heterogeneous treatment effect through observed covariates.} 
\citet{Calvi_al2018} propose a new estimand for the LATE, called the measurement robust LATE (MR-LATE), and obtain point-identification in an alternative framework. 
\citet{Yanagi2019} shows that with the help of exogenous covariates, point-identification of the LATE can be obtained under some conditions when there is misclassification in the treatment. 
\citet{tommasi2024bounding,tommasi2024identifying} extend the results in \citet{Ura2018}, and \citet{Calvi_al2018} to the case with multivalued discrete instruments. 
Building upon insights from the current paper and \citet{Mogstadal2018}, \citet{acerenza2024partial} derives bounds on the MTE in the presence of a misclassified treatment when a valid discrete instrument is available. 
More precisely, \citet{acerenza2024partial} uses results in Proposition \ref{prop1} of the current paper and some smoothness assumption on the marginal response function to derive bounds on the MTE. 
The framework developed in the current paper helps him provide nontrivial bounds on the propensity score and the variation in the propensity score, which are key ingredients to his bounding approach. 

This paper investigates the identification of the MTE (and PRTE) in settings where a binary treatment variable is misclassified, and a valid instrument is available. 
We start by deriving bounds on the (generalized) LATE. 
We then exploit the relationship between the MTE and the generalized LATE to bound the MTE. 
Indeed, under some standard assumptions, the MTE is a limit of the ratio of the variation in the conditional expectation of the observed outcome given the instrument to the variation in the true propensity score, which is partially identified. 
We provide a tractable characterization (which is an outer set) of the identified set for this propensity score, and then for the MTE. 
We show under some mild regularity conditions that the sign of the MTE is \textit{locally identified}.
We also derive functional sharp bounds for this propensity score when the misclassification is independent of the potential outcomes conditional on the actual treatment unobservable. 
We use our MTE bounds to derive bounds on other commonly used parameters in the literature such as the PRTE. 
In particular, we show that our bounds for the LATE are tighter than the existing \citeauthor{Ura2018}'s (\citeyear{Ura2018}) and \citeauthor{tommasi2024bounding}'s (\citeyear{tommasi2024bounding}) bounds when the instrument is discrete. 
The improvement over \citeauthor{Ura2018}'s (\citeyear{Ura2018}) bounds stems from the misclassification model we consider, which is more restrictive than his. 
The improvement over \citeauthor{tommasi2024bounding}'s (\citeyear{tommasi2024bounding}) bounds is a combination of the latter and some additional restrictions implied by the IV model in the discrete instrument setting that were missing from the paper.
We illustrate the practical relevance of our derived bounds through some numerical and empirical results. 
More precisely, we apply our methodology on data from the third wave of the Indonesia Family Life Survey to measure marginal returns to upper secondary or higher schooling while allowing for the possibility that education be mismeasured. 
We find that the return is heterogeneous and weakly decreases with the unobserved schooling cost. 

Several papers have extensively studied issues related to misclassification in treatment variables. See, for example, \citet{Aigner1973}, \citet{Bollinger1996}, \citet{Hausman_al1998}, \citet{Molinari2008}, \citet{Hu2008}, \citet{HuSchennach2008}, etc.
\citet{Mahajan2006} uses an additional instrument, which he called ``instrument-like variable,'' to nonparametrically identify the regression function in models with a misclassified binary regressor. 
Considering the same model as \citet{Mahajan2006} under different assumptions, \citet{Lewbel2007} also uses a ``second'' instrument to nonparametrically identify the average treatment effect (ATE) when the treatment is misclassified. 
However, their results hold only when the true treatment is exogenous.
\citet{DiTraglia2019} show that the identification result in \citet{Mahajan2006} does not extend to the case of an endogenous treatment. 
In that context, they derive bounds on the average treatment effect under standard assumptions. 
\citet{Nguimkeu_al2019} study a homogenous treatment effect linear regression model in which a  binary  regressor is potentially  misclassified and endogenous. 
They use exclusion restrictions for both the participation equation and measurement error equation to identify the regression coefficient with endogenous participation and one-sided endogenous misreporting. 
\citet{Millimet2011} studies the performance of several commonly used estimators in the causal inference literature when there is measurement error in a binary treatment, and warns researchers about the consequences of ignoring the presence of measurement errors.
\citet{Kreideral2012} partially identify the average effects of food stamps on health outcomes of children when participation is endogenous and misreported by using relatively weak nonparametric assumptions and information from auxiliary data.
Our paper studies potential heterogeneity in the treatment effect through the marginal treatment effect when the treatment is endogenous and mismeasured. 
Another related work to ours is \citet{Ura2020}, who investigates heterogeneous treatment effects in the presence of a misclassified endogenous binary treatment variable through the instrumental variable quantile regression model. 

Our paper also complements the work of \citet{Battistinal2011} and \citet{Battistinal2014}, who investigate the identification of average returns to education in the United Kingdom when attainment is potentially measured with error. 
While these authors focus on average returns, we investigate marginal returns, which may reveal (unobserved) heterogeneity in the treatment effects that would otherwise be hidden when looking at only average effects. 
Our work is also related to \citeauthor{Chalak2017}'s (\citeyear{Chalak2017}) who discusses the interpretation of various estimands (Wald, local IV) when the instrument is mismeasured, but the treatment variable is correctly measured. Differently from his framework, the instrument is observed with no error, while the treatment variable is potentially misreported. Recently, \citet{Jiangal2020} study identification in the binary IV model when allowing for simultaneous measurement errors in the instrument, treatment and/or outcome. They derive sharp bounds on the LATE assuming non-differential measurement errors and a valid IV. Our framework encompasses multivalued discrete/continuous outcomes and instruments, while the treatment variable is maintained binary. But, we only allow for misclassification in the treatment variable. Using a framework similar to ours, \citet{possebom2026crime} derives complementary identification results for the MTE, allowing for dependence between the instrument and the misclassification variable. While we focus on the case where the instrument is completely randomly assigned, we show in Appendix \ref{apx:zeps} how our approach can be extended to the situation where there is dependence between the instrument and the misclassification variable. This extension is similar to \citeauthor{Ura2018}'s (\citeyear{Ura2018}) in the LATE framework. Note that \citet{kasahara2022identification} study identification in regression models when an endogenous binary regressor is misclassified, allowing for correlation between the instrument and the misclassification error. They show identification of the regression coefficient when a ``special'' covariate in the outcome equation is excluded from the misclassification probability. While they allow for heterogeneity in the average effect through observed covariates, we focus on unobserved heterogeneity using marginal treatment effects.

The remainder of the paper is organized as follows. Section \ref{anaF} introduces the model and discusses the assumptions. Section \ref{Ident} presents the main nonparametric identification results, Section \ref{falserates} discusses some parametric identification approaches. Section \ref{ext:discrete} presents some extensions, Section \ref{App} provides a real world empirical example, and Section \ref{conclusion} concludes. Proofs of the main results are relegated to the Appendix.

\section{Analytical Framework}\label{anaF}
Consider the following model \begin{eqnarray}\label{seq1}
\left\{ \begin{array}{lcl}
     Y&=&Y_1D^*+Y_0(1-D^*),\\ \\
     D^*&=&\mathbbm{1}\left\{V\leq P(Z)\right\},\\ \\
     D&=&D^* (1-\varepsilon) +(1-D^*)\varepsilon,
     \end{array} \right.
\end{eqnarray}
where the vector $(Y,D,Z)$ represents the observed data, while the vector $(Y_1, Y_0, V, D^*,\varepsilon)$ is latent. 
In this model, $Y \in \mathcal Y$ is the observed outcome, $D^*\in \left\{0,1\right\}$ is the unobserved true treatment variable, while $D\in \left\{0,1\right\}$ is the observed mismeasured treatment, $\varepsilon \in \left\{0,1\right\}$ is an indicator for misreporting, $Y_0$ and $Y_1$ are the potential outcomes that would have been observed if the true treatment $D^*$ had been externally set to 0 and 1, respectively. 
The variable $Z \in \mathcal Z$ is an instrument, and $P(.)$ is a nontrivial/nonconstant function. 
In this paper, we are primarily interested in identifying the marginal treatment effect defined as 
$$MTE(p)\equiv \mathbb E[Y_1-Y_0\vert V=p].$$
However, when the instrument $Z$ is binary/discrete, we will focus on the (generalized) LATE parameter(s). 

\begin{example}[Marginal returns to schooling (leading example)]\label{ex1}
In this example, we assume that the researcher is interested in measuring marginal returns to college education. 
It is well-documented that education is usually mismeasured. 
For example, \citet{Black_etal2003} find that more than a third of respondents to the U.S. Census claiming to hold a professional degree have no such  degree. 
In this case, the variable $Y$ is earnings/wage, and $D$ is the indicator for college degree.
The variable $Z$ could be distance to college. 
The latent variable $V$ could be interpreted as an index for the cost of going to college (which includes the financial cost, the opportunity cost, the psychological cost, etc.), while $Y_1$ is the potential earnings for someone with a college degree, and $Y_0$ is the potential earnings for someone without a college degree. 
The variable $D^*$ is the individual's true indicator for college degree. 
\end{example}

\begin{example}[Marginal effects of masks]\label{ex2}
The variable $Y$ could be the indicator that an individual tests positive to Covid-19, $D^*$ the indicator that the individual actually wears masks, $D$ the indicator that the individual reports wearing masks, $V$ the disutility/discomfort (cost) of wearing masks, and $Z$ could be a shifter for the benefit of wearing masks (number of children).\footnote{Children are less likely to get the virus, and therefore less likely to contaminate others. Hence, the variable number of children in the households is likely to satisfy the exclusion restriction assumption. We may add the number of adults in the households as a control variable to the model. 
However, the validity of number of children as instrument remains questionable, as is the distance to college instrument in Example~\ref{ex1}.} 
People could report wearing masks while they actually do not (for example, because of social pressure). 
They could also pretend to wear mask while they do not wear it properly. 
On the other hand, someone could report not wearing regularly a mask (because of political reasons for example), while she actually does (because of her underlying health conditions).
These facts could lead to misclassification in the report of mask wearing, and therefore induce some bias in the measurement of the marginal effects of mask wearing on the positivity rate.
$Y_1$ could be the indicator that the individual tests positive to Covid-19 while she is wearing masks, and $Y_0$ could be the indicator that the individual tests positive to Covid-19 while she is not wearing masks. 
The effect of wearing masks on the positivity rate could be heterogeneous. 
Healthy individuals tend to think that they are immune or they will survive if they are infected.
For them, the disutility of wearing masks may be higher, and they may be less likely to wear masks.
We are interested in measuring the effect of wearing masks on the positivity rate for different levels of the disutility, i.e., $\mathbb E\left[Y_1-Y_0 \vert V=p\right]$. 
\end{example}

We will use the following assumptions for identification:
\begin{assumption}[Random assignment]\label{RA}
The instrument $Z$ is independent of $\left(Y_d,V,\varepsilon\right)$, i.e., $Z\ \indep\ \left(Y_d,V,\varepsilon\right)$, for each $d\in \left\{0,1\right\}$.
\end{assumption}
Assumption \ref{RA} requires that $Z$ be a valid instrument, in the sense that it is statistically independent of all the unobservables in the model. This is a commonly used assumption in the literature. 
Note that the model (\ref{seq1}) implicitly assumes that exclusion restriction holds, i.e., $Y_{dz}=Y_d$ for all $d$ and $z$. 
For this reason, we do not state this assumption explicitly. 
Assumption \ref{RA} requires more than the standard random assignment assumption when there is no misclassification (i.e., $\varepsilon=0$ a.s.), which states that $Z$ is statistically independent of $(Y_d,V)$ for each treatment arm $d$. 
Assumption \ref{RA} extends the standard random assignment assumption to include the misclassification variable $\varepsilon$.
Strictly speaking, we should have called this assumption ``\textit{extended random assignment}.''
We are abusing terminology by simply calling it \textit{random assignment}. 
As discussed in \citet{possebom2026crime}, the independence between the instrument $Z$ and the misclassification variable $\varepsilon$ could be too restrictive in practice.
We discuss how this assumption can be relaxed in Appendix \ref{apx:zeps}. 

Our model is based on the intuition that the misclassified treatment $D$ is a result of actual treatment choice $D^*$ and misreporting decision $\varepsilon$.
So, we can write $D=h(D^*,\varepsilon),$ where $h$ is an unknown function. 
Because misreporting is a binary decision, we have $D=h(D^*,1)\varepsilon+h(D^*,0)(1-\varepsilon)$.
This specification holds whether $D$ and $D^*$ are continuous, discrete, or mix. 
However, the fact that $D$ and $D^*$ are binary in our current setting implies $h(D^*,1)=1-D^*$ and $h(D^*,0)=D^*.$ 
Lemma \ref{misc:specific} in the Appendix proves this result: $D=(1-D^*)\varepsilon+D^*(1-\varepsilon)$. 
On the other hand, \citet{Ura2018} considers a potential outcome model for misreported treatment $D$: $D=D_1 D^* + D_0(1-D^*),$ where $D_1$ represents the potential misreported treatment when the actual treatment is $1$ while $D_0$ represents the potential misreported treatment when the actual treatment $0$. 
So conceptually, the two specifications are two different ways of tackling the same question. 
As pointed out by \citet{tommasi2024identifying}, without any assumption, the two specifications are equivalent. 
However, our model specification combined with the assumption $Z\ \indep\ (Y_d,V,\varepsilon)$ is more restrictive than \citeauthor{Ura2018}'s combined with the assumption $Z\ \indep\ (Y_d,D_0,D_1,\{D^*_z\equiv \mathbbm{1}\{V\leq P(z)\}: z\in \mathcal Z\}).$ 
In the Appendix, we provide Example \ref{ex:comparison} where our model assumptions fail while \citeauthor{Ura2018}'s model assumptions hold.

In our framework, the measurement error is nonclassical by definition of the model. 
Indeed, we can rewrite $D=D^*+(1-2D^*)\varepsilon$. 
So, the measurement error $(1-2D^*)\varepsilon$ depends on the true unobserved treatment $D^*$. 
This fact is well-documented and understood in the literature.
See \citet{Aigner1973}, \citet{Mahajan2006}, \citet{Lewbel2007}, \citet{Kreideral2012}, \citet{Ura2018}, \citet{Yanagi2019}, etc. 

\begin{assumption}[Absolute continuity of $V$]\label{Cont}
The latent variable $V$ is absolutely continuous. 
Without loss of generality, the unconditional distribution of $V$ is uniform over $[0,1]$, and the range of the function $P(z)$ is included in $[0,1]$.
\end{assumption}
This assumption is standard in the literature and has been considered in \citet{heckman1999,heckman2001,heckman2005structural}, \citet{carneirolee2009}, \citet{checkman2010,heckman2011}, etc. 
It does not require that the conditional density of $V$ given $\varepsilon$ exists, as will be apparent in the different specifications we consider in the Appendix. 
Using the law of total probability and Bayes' rule, this assumption implies 
\begin{eqnarray}\label{eq:mix}
\mathbb P(\varepsilon=1)F_{V\vert \varepsilon=1}(p)+\mathbb P(\varepsilon=0)F_{V\vert \varepsilon=0}(p)=p\ \text{ for all } p\in[0,1],
\end{eqnarray}
where $F_{V\vert \varepsilon}$ denotes the conditional distribution of $V$ given $\varepsilon$. 
Knowing this conditional distribution will help identify the threshold function $P(z)$, which is central to our analysis. 

\begin{assumption}[Continuous instrument]\label{Cont_inst}
The instrument $Z$ is continuous such that the support of the random variable $P(Z)$ is an interval.\end{assumption}
Assumption \ref{Cont_inst} is also standard in the literature and is crucial for our identification methodology for the MTE.  
In Section \ref{ext:discrete}, we show how our methodology can be used to identify multiple LATEs when the instrument is discrete.
Identification results for the MTE with discrete instruments have been developed in \citet{acerenza2024partial}, which built on insights from the current paper and \citet{Mogstadal2018}.

\begin{assumption}[Upper bound on misclassification rate]\label{Bound:mis}
The (unconditional) misclassification rate $\alpha \equiv \mathbb P(\varepsilon=1)$ has a known upper bound $\bar{\alpha}$, that is, $\alpha \in [0,\bar{\alpha}]$. 
\end{assumption}
A similar assumption to Assumption \ref{Bound:mis} has been considered in \citet{Horowitz1995}, \citet{Kreideral2007}, \citet{Molinari2008}, \citet{Kreideral2012}, etc. 
In this assumption, we only impose an upper bound on the extent of the misclassification, as we allow for the treatment to be correctly classified. 
One can alternatively place a lower bound on the misclassification probability too. 
For example, one can combine information from different sources (e.g., government, universities, etc.) to bound the extent of the misclassification from below as well as from above. 
For instance, universities can provide information on the number of individuals who actually have a college degree. 
This information can help identify $\mathbb P(D^*=1)$.
One can then use the absolute difference between $\mathbb P(D^*=1)$ and $\mathbb P(D=1)$ as $\underline{\alpha}$, a lower bound for the misclassification rate.
\footnote{Indeed, we have $D=D^*+(1-2D^*)\varepsilon$, which implies $\mathbb E[D-D^*]=\mathbb E[(1-2D^*)\varepsilon]=\mathbb E[(1-2D^*)\vert \varepsilon=1]\mathbb P(\varepsilon=1)$.
Therefore, $\vert \mathbb E[D-D^*] \vert= \vert \mathbb E[(1-2D^*) \vert \varepsilon=1]\vert \mathbb P(\varepsilon=1)\leq \mathbb P(\varepsilon=1)$, since $1-2D^* \in \{-1,1\}$.
Hence, we can set $\underline{\alpha}=\vert \mathbb E[D-D^*] \vert=\vert \mathbb P(D=1)-\mathbb P(D^*=1)\vert$.} 
We can also use this information to provide a value for $\bar{\alpha}$.
\footnote{We have $\mathbb P(\varepsilon=1)=\mathbb P(\varepsilon=1,D=1)+\mathbb P(\varepsilon=1,D=0)=\mathbb P(\varepsilon=1\vert D=1)\mathbb P(D=1)+\mathbb P(\varepsilon=1,D^*=1)=\mathbb P(\varepsilon=1\vert D=1)\mathbb P(D=1)+\mathbb P(\varepsilon=1\vert D^*=1)\mathbb P(D^*=1)$.
If $\mathbb P(\varepsilon=1\vert D=1) \leq \bar{\alpha}_0$, and $\mathbb P(\varepsilon=1\vert D^*=1) \leq \bar{\alpha}_1$, then we can set $\bar{\alpha}=\bar{\alpha}_0\mathbb P(D=1)+\bar{\alpha}_1\mathbb P(D^*=1)$.
In the context of example \ref{ex1}, we can set $\bar{\alpha}_0=1/3$ (following \citet{Black_etal2003}), and $\bar{\alpha}_1=0$ (assuming someone with a degree is unlikely to misreport).}
The case $\bar{\alpha}=1$ corresponds to the scenario where the researcher is agnostic about the range of the misclassification rate. All our derived results still hold in this case. 

\section{Nonparametric Identification Results}\label{Ident}

\subsection{Binary Instrument: Identification of the LATE}\label{sec:binary}
In this subsection, we consider the local average treatment effect (LATE) framework with a binary instrument $Z \in \mathcal Z= \{0, 1\}$. 
We can rewrite the model for the true treatment choice $D^*$ in the potential treatment framework:
$D^*=D^*_1 Z + D^*_0 (1-Z)$, where $D^*_z=\mathbbm{1}\{V \leq P(z)\}$ for $z \in \{0, 1\}$.

Suppose, without loss of generality, $P(0) < P(1)$.
Then $D^*_0 \leq D^*_1$, i.e., there are no defiers in the population. 
To compare our approach to existing results, we assume in this subsection that the researcher has no information about the misclassification rate $\alpha$. 

Using the law of total probability, the model specification \eqref{seq1}, and Assumption \ref{RA}, we show that the following equalities hold for any Borel set $A$:
\begin{eqnarray}
\mathbb{P}(Y \in A, D=1 \mid Z=1) &=& \mathbb{P}\left(Y_0 \in A, D_1^*=0, D_0^*=0, \varepsilon=1\right)+\mathbb{P}\left(Y_1 \in A, D_1^*=1, D_0^*=1, \varepsilon=0\right), \nonumber\\
&& \qquad \qquad \qquad + \mathbb{P}\left(Y_1 \in A, D_1^*=1, D_0^*=0, \varepsilon=0\right),\label{eq:late1main}\\
\mathbb{P}(Y \in A, D=1 \mid Z=0) &=& \mathbb{P}\left(Y_0 \in A, D_1^*=1, D_0^*=0, \varepsilon=1\right)+\mathbb{P}\left(Y_0 \in A, D_1^*=0, D_0^*=0, \varepsilon=1\right),\nonumber \\
&& \qquad \qquad \qquad +\mathbb{P}\left(Y_1 \in A, D_1^*=1, D_0^*=1, \varepsilon=0\right).\label{eq:late0main}
\end{eqnarray}
Denote $T \equiv\left(D_0^*, D_1^*\right) \in \{(0,1),(1,1),(0,0)\} \equiv \{c, a, n\} $, and take the summation of Equations \eqref{eq:late1main} and \eqref{eq:late0main} for $A=\mathcal Y$. The following implication holds,
\begin{eqnarray*}
    \mathbb P(D=1\vert Z=1)+\mathbb P(D=1\vert Z=0) &=& \mathbb P(T=c)+2[\mathbb P(T=a,\varepsilon=0)+\mathbb P(T=n,\varepsilon=1)],\\
    &\geq& \mathbb P(T=c).
\end{eqnarray*}
A similar result holds for the reported control group $D=0$. Hence, we obtain a nontrivial upper bound on the compliers share. 
\begin{eqnarray*}
    \mathbb P(T=c) \leq \min_{d\in\{0,1\}}\left\{\mathbb P(D=d\vert Z=1)+\mathbb P(D=d\vert Z=0)\right\}.
\end{eqnarray*}
To the best of our knowledge, this upper bound is new in the misclassification literature. It is a result of our model for misreported treatment and our model assumptions. 

Let $f_{X\vert W}\left(x \vert w\right)$ denote the conditional density of a generic random vector $X$ given $\left\{W=w\right\}$ that is absolutely continuous with respect to a known dominating measure $\mu_{X}$.
The density version of the above equalities \eqref{eq:late1main} and \eqref{eq:late0main} holds. We work with the density version because it provides tighter bounds than the probability version. We explain this in the next subsection.
We difference $f_{Y,D|Z}(y,d|z)$ across values of $z$ and obtain:
\begin{eqnarray*}
f_{Y, D\mid Z}(y, 1 \mid 1)-f_{Y, D\mid Z}(y, 1 \mid 0) =f_{Y_1, T, \varepsilon}(y, c, 0)-f_{Y_0, T, \varepsilon}(y, c, 1)
\end{eqnarray*}
Then, by the triangle inequality we have
$$
\left|f_{Y, D\mid Z}(y, 1 \mid 1)-f_{Y, D\mid Z}(y, 1 \mid 0)\right| \leq f_{Y_1, T, \varepsilon}(y, c, 0) + f_{Y_0, T, \varepsilon}(y, c, 1).
$$
By taking the integral with respect to $y$, we have
\begin{align*}
\int_{\mathcal Y} \left| f_{Y, D\mid Z}(y,1 \mid 1)-f_{Y, D\mid Z}(y, 1 \mid 0) \right| d \mu_Y(y)  &  \leq \int_{\mathcal Y} \left[f_{Y_0, T, \varepsilon}(y, c, 1)+f_{Y_1, T, \varepsilon}(y, c, 0)\right] d \mu_Y(y), \\
& =\mathbb{P}(T=c, \varepsilon=1)+\mathbb{P}(T=c, \varepsilon=0)=\mathbb{P}(T=c).
\end{align*}
A similar result holds for the observed control group $D=0$.
\begin{align*}
\int_{\mathcal Y} \left|f_{Y, D\mid Z}(y, 0 \mid 1)-f_{Y, D\mid Z}(y, 0 \mid 0)\right| d \mu_Y(y)  & \leq \mathbb{P}(T=c).
\end{align*}
Therefore, we have\footnote{Note that the upper bound is always less than or equal to one. By contradiction, suppose $\min_{d\in\{0,1\}}\left\{\mathbb P(D=d\vert Z=1)+\mathbb P(D=d\vert Z=0)\right\} >1$. Then for each $d$, $\mathbb P(D=d\vert Z=1)+\mathbb P(D=d\vert Z=0)>1$. Summing up this over $d\in\{0,1\}$ implies $2>2$ (contradiction).}
\begin{eqnarray}\label{eq:pcbounds}
\max_{d\in\{0,1\}} \left\{TV_{(Y,D=d)}(0,1) \right\} \leq \mathbb{P}(T=c) \leq \min_{d\in\{0,1\}}\left\{\mathbb P(D=d\vert Z=1)+\mathbb P(D=d\vert Z=0)\right\},
\end{eqnarray}
where
$
TV_{(Y,D=d)}(z',z) \equiv \int_{\mathcal Y}\left \lvert f_{Y,D\vert Z}\left(y, d \vert z\right) - f_{Y,D\vert Z}\left(y, d \vert z'\right) \right \rvert d \mu_{Y}(y).
$
Hence, we provide a nontrivial lower bound on the compliers probability.
This bound is a function of the joint distribution of the observed data $(Y,D,Z).$
Note that
\begin{align*}
TV_{(Y,D=d)}(0, 1) &\geq \left \lvert \int_{\mathcal Y} f_{Y,D\vert Z}\left(y, d \vert 1\right) - f_{Y,D\vert Z}\left(y, d \vert 0\right) d \mu_{Y}(y) \right \rvert, \\
&= \left \lvert \mathbb{P}(D=d \vert Z = 1) -  \mathbb{P}(D=d \vert Z = 0) \right \rvert=\lvert\Delta_{DZ}(0,1)\rvert,
\end{align*}
which implies that the first stage regression estimand of the observed mismeasured treatment on the instrument could be biased for the actual compliers proportion. The lower bound $p_c^{LB}$ is strictly positive if the observed treatment is correlated with the instrument. 

To see the intuition behind the lower bound, integrate the density difference equation before taking absolute values. 
For the observed first stage,
\[
\Delta_{DZ}(0,1)
\equiv \mathbb P(D=1\mid Z=1)-\mathbb P(D=1\mid Z=0)
=\mathbb P(T=c,\varepsilon=0)-\mathbb P(T=c,\varepsilon=1).
\]
Thus, in the absence of misclassification, $\varepsilon=0$ a.s., the observed
first stage is exactly the complier probability, $\Delta_{DZ}(0,1)=\mathbb P(T=c)$.
With misclassification, however, compliers with $\varepsilon=1$ are observed as
moving in the opposite direction; they are observed as treated when $Z=0$ and untreated when $Z=1$. Hence they subtract from, rather than add to, the observed first stage. 
The above equality implies
\[
|\Delta_{DZ}(0,1)|
=
\mathbb P(T=c)\left|1-2\mathbb P(\varepsilon=1\mid T=c)\right|
\leq \mathbb P(T=c).
\]
When truth-telling dominates among compliers, the first stage estimand $\Delta_{DZ}(0,1)$ suffers from attenuation bias of the compliers probability. The lower bound in \eqref{eq:pcbounds}
strengthens this first stage lower bound by exploiting changes in the joint
distribution of $(Y,D)$, rather than only the marginal distribution of $D$. 
If the density difference for each observed treatment status does not change sign over $\mathcal Y$, the lower bound collapses to the absolute observed first stage.

Notice that our model has a testable implication that the lower bound for the compliers share must be less than or equal to its upper bound. 
\begin{eqnarray}\label{eq:testimp}
p_c^{LB}\equiv \max_{d\in\{0,1\}} \left\{TV_{(Y,D=d)}(0,1) \right\} \leq \min_{d\in\{0,1\}}\left\{\mathbb P(D=d\vert Z=1)+\mathbb P(D=d\vert Z=0)\right\}\equiv p_c^{UB}.
\end{eqnarray}
Whenever this testable restriction fails to hold, our model specification together with the assumptions is rejected by the data.

We summarize the results in Proposition \ref{prop:LATE} below.
\begin{proposition}\label{prop:LATE}
Consider the LATE version of the model specification \eqref{seq1},
\begin{eqnarray*}
\left\{ \begin{array}{lcl}
     Y&=&Y_1D^*+Y_0(1-D^*),\\ \\
     D^*&=&D^*_1 Z + D^*_0 (1-Z),\\ \\
     D&=&D^* (1-\varepsilon) +(1-D^*)\varepsilon.
     \end{array} \right.
\end{eqnarray*}
Let $T\equiv(D_0^*,D_1^*)$ and $ c\equiv(0,1)$. 
Suppose $D^*_1 \geq D^*_0$ (no defiers), $\mathbb{P}(T=c)>0$, and random assignment, $Z\ \indep\ (Y_d,T,\varepsilon),$ holds. 

If \(0<p_c^{LB}\leq p_c^{UB}\), then the (sharp) identified set of $LATE\equiv \mathbb{E}\left[Y_1-Y_0 \mid T=c\right]$ is given as
\[
\left[
\min_{\ell \in \{LB,UB\}}\left\{\frac{\mathbb E[Y\mid Z=1]-\mathbb E[Y\mid Z=0]}{p_c^{\ell}}\right\},
\max_{\ell \in \{LB,UB\}}\left\{\frac{\mathbb E[Y\mid Z=1]-\mathbb E[Y\mid Z=0]}{p_c^{\ell}}\right\}
\right].
\] 
If \(p_c^{LB}>p_c^{UB}\), the maintained assumptions are
falsified. Inequality \eqref{eq:testimp} is a sharp testable implication of the model.

\end{proposition}
The proof of this proposition is in Appendix \ref{proofUra}.

\begin{remark}
The misclassification model we consider in this paper is a nonparametric version of the model in \cite[Equations (4)-(6)]{hausman1994misclassification} and is different from that in \citet{Ura2018}.
\citet{Ura2018} uses the potential outcome model for reporting $D=D^*D_1+(1-D^*)D_0$, which is more general than the misclassification model we consider.
As a result, we obtain tighter bounds for the proportion of compliers, and hence for the LATE.
Indeed, \citeauthor{Ura2018}'s (\citeyear{Ura2018}) lower bound on the proportion of compliers is $\frac{1}{2} TV_{(Y,D=1)}(0,1) + \frac{1}{2}TV_{(Y,D=0)}(0,1)$, which is less than or equal to our lower bound $p_c^{LB}$ defined above. His upper bound is the trivial bound $1,$ while ours is $\min_{d\in\{0,1\}}\left\{\mathbb P(D=d\vert Z=1)+\mathbb P(D=d\vert Z=0)\right\} \leq 1.$
It is worth pointing out that the data generating process (DGP) considered in Appendix E (page 6) in \citeauthor{Ura2018}'s (\citeyear{Ura2018}) supplementary material satisfies our model assumptions:
		$$
		\begin{aligned}
		Z & \sim \operatorname{Bernoulli}(0.5), \\
		D^* & =1\left\{-3 / 4+1 / 2 Z+U_1 \geq 0\right\}, \\
		Y & =2 D^*+\Phi\left(U_2\right), \\
		D & =D^*+\left(1-2 D^*\right) 1\left\{U_3 \leq \gamma\right\},
		\end{aligned}
		$$
		where $\Phi$ is the standard normal cdf, and conditional on $Z,\left(U_1, U_2, U_3\right)$ is drawn from the Gaussian copula with the correlation matrix
		$$
		\left(\begin{array}{ccc}
		1 & 0.25 & 0.25 \\
		0.25 & 1 & 0.25 \\
		0.25 & 0.25 & 1
		\end{array}\right)
		$$
As we can see, this DGP satisfies the model specification \eqref{seq1} and the assumptions we consider in this section of the paper, where $\varepsilon=1\left\{U_3 \leq \gamma\right\}$.		
\end{remark}

\begin{example}
   We provide a simple DGP where our bounds show a significant improvement over \citeauthor{Ura2018}'s. The variables $Y,D,Z$ are all binary. For $d,y,z\in \{0,1\},$
   \begin{eqnarray*}
       \mathbb P(Y_1=1,T=c,\varepsilon=0\vert Z=z)=0.3, && \mathbb P(Y_1=0,T=c,\varepsilon=0\vert Z=z)=0.2,\\
       \mathbb P(Y_0=y,T=c,\varepsilon=1\vert Z=z)=0, && \mathbb P(Y_1=y,T=c,\varepsilon=1\vert Z=z)=0,\\
       \mathbb P(Y_0=0,T=c,\varepsilon=0\vert Z=z)=0.5, && \mathbb P(Y_0=1,T=c,\varepsilon=0\vert Z=z)=0,\\
       \mathbb P(Y_d=y,T=n,\varepsilon=1\vert Z=z)=0 && \mathbb P(Y_d=y,T=a,\varepsilon=0\vert Z=z)=0,\\
       \mathbb P(Y_d=0,T=n,\varepsilon=0\vert Z=z)=0 && \mathbb P(Y_d=0,T=a,\varepsilon=1\vert Z=z)=0,\\
       \mathbb P(Y_d=1,T=n,\varepsilon=0\vert Z=z)=0.25 && \mathbb P(Y_d=1,T=a,\varepsilon=1\vert Z=z)=0.25.
   \end{eqnarray*}
   This DGP satisfies our model assumptions: $Z\ \indep\ (Y_d,T,\varepsilon),$ and there are no defiers. The observed data $(Y,D,Z)$ has the following distribution. 
   \begin{eqnarray*}
       \mathbb P(Y=1,D=1\vert Z=1)=0.3, && \mathbb P(Y=1,D=1\vert Z=0)=0,\\
       \mathbb P(Y=0,D=1\vert Z=1)=0.2, && \mathbb P(Y=0,D=1\vert Z=0)=0,\\
       \mathbb P(Y=1,D=0\vert Z=1)=0.5, && \mathbb P(Y=1,D=0\vert Z=0)=0.5,\\
       \mathbb P(Y=0,D=0\vert Z=1)=0, && \mathbb P(Y=0,D=0\vert Z=0)=0.5.
   \end{eqnarray*}
   Note that the distribution $\mathbb P(Y_d=,T=t,\varepsilon=e\vert Z=z)$ is consistent with the observed data distribution $\mathbb P(Y=y,D=d\vert Z=z)$. In this DGP, $TV_{(Y,D=1)}(0,1)=TV_{(Y,D=0)}(0,1)=0.5,$ $p^{LB}=0.5$, $p^{UB}=\min\{0.5,1.5\}=0.5,$ and $\mathbb E[Y\vert Z=1]-\mathbb E[Y\vert Z=0]=0.8-0.5=0.3.$ The actual compliers proportion is $\mathbb P(T=c)=0.5,$ and the actual LATE is $0.6$. Our identified set is a singleton $\{0.6\},$ while \citeauthor{Ura2018}'s bounds are $[0.3,0.6].$
\end{example}

\begin{remark}
    As we discuss above, our model assumptions can be rejected by the data when inequality \eqref{eq:testimp} fails to hold. As an example, consider the DGP below where $Y$ is binary, and 
  \begin{eqnarray*}
      \mathbb P(Y=0,D=1\mid Z=0)=0.1,&& \mathbb P(Y=0,D=0\mid Z=0)=0.9\\
      \mathbb P(Y=0,D=1\mid Z=1)=0.2,&&\mathbb P(Y=0,D=0\mid Z=1)=0.5,\\
\mathbb P(Y=1,D=0\mid Z=1)=0.3. &&
\end{eqnarray*}
In this DGP, we have $p_c^{LB}=0.7>0.3=p_c^{UB},$ implying that our model assumptions are rejected by the observed data $(Y,D,Z)$. 

On the other hand, \citeauthor{Ura2018}'s bounds are never empty. Indeed, from the triangle inequality, $TV_{(Y,D=d)}(0,1) \leq \mathbb P(D=d\vert Z=1)+\mathbb P(D=d\vert Z=0)$ for all $d\in\{0,1\}.$ Therefore,
$\frac{1}{2}TV_{(Y,D=1)}(0,1)+\frac{1}{2}TV_{(Y,D=0)}(0,1)\leq \frac{1}{2}[\mathbb P(D=1\vert Z=1)+\mathbb P(D=1\vert Z=0)]+\frac{1}{2}[\mathbb P(D=0\vert Z=1)+\mathbb P(D=0\vert Z=0)]=1$.
\end{remark}

\begin{remark}
To derive the bounds in Proposition \ref{prop:LATE}, we assume that the researcher observes a potentially misreported treatment $D$.
The approach can be extended to the extreme case where the researcher has no information about $D$, that is, the treatment variable is completely missing for all individuals.
Using a similar derivation as above, we provide the following bounds on $\mathbb P(T=c)$: 
\begin{align*}
\frac{1}{2}\int_{\mathcal Y} \left|f_{Y|Z}(y \mid 1)-f_{Y|Z}(y \mid 0)\right| d \mu_Y(y)  & \leq \mathbb{P}(T=c) \leq 1.
\end{align*}
Note that this result holds whether there is misclassification or not.
See proof in Appendix~\ref{apx:zeps}.
\end{remark}

\subsection{Continuous Instrument: Identification of the MTE}

In this subsection, we assume that the instrument $Z$ is continuous.\footnote{The analysis in this section still holds if $Z$ is a vector that contains at least one continuous instrument.} 
We extend the identification strategy in the previous subsection to the continuous instrument case. 
Define $\Delta_{XZ}(z',z)\equiv \mathbb E[X\vert Z=z]-\mathbb E[X\vert Z=z']$ for $X \in \{Y, D, D^*\}$, and consider the following Assumption \ref{morett}.
\begin{assumption}\label{morett} For all $v \in [0, 1]$,
$
\mathbb{P}(\varepsilon = 0 \vert V = v) > \mathbb{P}(\varepsilon = 1 \vert V = v).
$
\end{assumption}
Assumption \ref{morett} states that for any level of the unobserved heterogeneity $V$ that drives the selection into treatment, there are more truth-tellers than misreporters.
Under this assumption, the following Lemma \ref{morett_samedirection} establishes a crucial relationship between the way the instrument influences the true unobserved treatment and the observed treatment.

\begin{lemma}\label{morett_samedirection}
Under the model specification \eqref{seq1} and Assumption \ref{RA}, Assumption \ref{morett} implies 
$$
\Delta_{D^*Z}(z',z) \cdot \Delta_{DZ}(z',z) > 0,
$$
for any $z, z' \in \mathcal{Z}$ such that $P(z) \neq P(z')$.
\end{lemma}
The proof is relegated to Appendix \ref{proof_morett_samedirection}. 
An implication of Lemma \ref{morett_samedirection} is that the sign of $[P(z)-P(z')]$ is identified as the sign of $\Delta_{DZ}(z',z)$. 
Note that this implication can be seen as a generalization of Assumption 5 from \citet{possebom2026crime} as $z$ is not restricted to be a scalar, it could be a vector.
It is also related to the implication in Corollary 2.2(i) in \citet{tommasi2024bounding} derived in a discrete instrument setting where Assumption \ref{morett} is formulated in terms of relevant complier groups.  

Consider $z, z' \in \mathcal{Z}$ such that $P(z') < P(z)$.\footnote{It is always possible to find such $z$ and $z'$,
otherwise $P(z)$ would be constant across $z$, which contradicts our assumption that $P(.)$ is a nontrivial function. Using the result in Lemma \ref{morett_samedirection}, $z$ and $z'$ can be chosen from the observed propensity using $\Delta_{DZ}(z',z)>0$ since $P(z) - P(z') = \Delta_{D^*Z}(z',z)$. 
}
Define the generalized LATE, 
$$
LATE(z', z) \equiv \mathbbm{E}[Y_1 - Y_0 \vert P(z') < V \leq P(z)].
$$
From Proposition \ref{prop:LATE}, we can write 
\begin{align*}
LATE(z', z) &= \frac{\mathbb{E}[Y \mid Z=z]-\mathbb{E}[Y \mid Z=z']}{\mathbb{E}\left[D^* \mid Z=z\right]-\mathbb{E}\left[D^* \mid Z=z'\right]}, \\
&= \frac{\Delta_{YZ}(z',z) }{P(z)-P(z')},
\end{align*}
where $\mathbb{E}\left[D^* \mid Z=z\right] = \mathbbm{P}(V \leq P(Z) \vert Z = z) = P(z)$ under Assumptions \ref{RA} and \ref{Cont}.

In order to identify the MTE, we take the limit of $LATE(z', z)$ when $z'$ goes to $z$.
More specifically,
\begin{align*}
\lim_{z' \rightarrow z} LATE(z', z) &= \lim_{z' \rightarrow z} \mathbbm{E}[Y_1 - Y_0 \vert P(z') < V \leq P(z)],\\
&= \mathbbm{E}[Y_1 - Y_0 \vert V = P(z)], \\
&= MTE\big(P(z)\big).
\end{align*}
Hence,
$$
MTE\big(P(z)\big) = \lim_{z' \rightarrow z} \frac{\Delta_{YZ}(z',z)}{P(z)-P(z')},
$$
where we assume that the limit exists.

If we knew the true propensity score, $P(z)$, then the MTE would be identified.
Because of the presence of the misclassification in the treatment, the function $P(z)$ is only partially identified, and so is the MTE.
To proceed, we first discuss pointwise and functional identification of the function $P(z)$. 
Second, we provide analytical bounds for $MTE\big(P(z)\big)$.

\subsubsection{Identification of $P(z)$}
For any Borel set $A$, we have 
\begin{eqnarray}
\mathbb P(Y\in A, D=1 \vert Z=z)&=&\mathbb P(Y\in A, D=1, D^*=1 \vert Z=z) + \mathbb P(Y\in A, D=1, D^*=0 \vert Z=z),\nonumber\\
&=& \mathbb P(Y_1\in A, \varepsilon=0, V \leq P(z)) + \mathbb P(Y_0\in A, \varepsilon=1, V > P(z)), \label{eq1}
\end{eqnarray}
where the first equality holds from the law of total probability, and the second equality follows from the definition of the model and Assumption \ref{RA}.
In the special case where~$A=\mathcal Y$, we have
\begin{eqnarray}\label{eq:mis0}
\mathbb P(D=1 \vert Z=z)&=& \mathbb P(\varepsilon=0, V \leq P(z)) + \mathbb P(\varepsilon=1, V > P(z)).
\end{eqnarray}
When there is no misclassification in the treatment, i.e., $\varepsilon=0$ a.s., then $P(z)$ is identified as the propensity score $\mathbb P(D=1 \vert Z=z)$, since the distribution of $V$ is normalized to be uniform over $[0,1]$.
When the treatment is \textit{completely} misclassified, i.e., $\varepsilon=1$ a.s., $P(z)$ is identified under the previous normalization as $\mathbb P(D=0 \vert Z=z)$. We can rewrite the last equality as follows: 
\begin{eqnarray}
\mathbb P(D=1 \vert Z=z)&=& (1-\alpha) F_{V\vert \varepsilon=0}(P(z)) + \alpha (1-F_{V\vert \varepsilon=1}(P(z))). \label{eq2}
\end{eqnarray}

We have $\mathbb P(D^*=1\vert Z=z)=\mathbb P(V \leq P(z))=P(z)$. 
Thus, $P(z)$ is the true (unidentified) propensity score.
We show that the propensity score $P(z)$ is partially identified using Equations (\ref{eq:mix}) and (\ref{eq2}). 
Equation (\ref{eq2}) implies
\begin{eqnarray*}
\mathbb P(D=1 \vert P(Z)=p)&=& (1-\alpha) F_{V\vert \varepsilon=0}(p) + \alpha (1-F_{V\vert \varepsilon=1}(p)).
\end{eqnarray*}
For now, we assume $\alpha \in (0,1)$, since the cases where $\alpha\in \{0,1\}$ can be dealt with separately. Combining this with Equation (\ref{eq:mix}), and solving for $F_{V\vert \varepsilon=0}(p)$ and $F_{V\vert \varepsilon=1}(p)$ in the system of equations, we obtain:
\begin{eqnarray*}
F_{V\vert \varepsilon=1}(p)&=& \frac{p+\alpha-\mathbb P(D=1\vert P(Z)=p)}{2 \alpha},\\
F_{V\vert \varepsilon=0}(p)&=& \frac{p-\alpha+\mathbb P(D=1\vert P(Z)=p)}{2(1-\alpha)}.
\end{eqnarray*}
Therefore, the above functions need to satisfy all required conditions for a cumulative distribution on $[0,1]$: monotonicity, right-continuity, $F_{V\vert \varepsilon=1}(0)=F_{V\vert \varepsilon=0}(0)=0$, and $F_{V\vert \varepsilon=1}(1)=F_{V\vert \varepsilon=0}(1)=1$.
In general, it will be difficult to nonparametrically characterize the sharp identification region for the propensity score function $P(z)$ using those conditions. 
We therefore begin by exploiting only the range restrictions that the probabilities $F_{V\vert \varepsilon=1}(P(z))$ and $F_{V\vert \varepsilon=0}(P(z))$ lie between 0 and 1, which yield pointwise bounds on the propensity score $P(z)$. 
In the next subsection, we further use the montonicity condition to bound the variation in the propensity score $P(z)-P(z').$

The condition $F_{V\vert \varepsilon=1}(P(z)) \in [0, 1]$ implies
$$
\mathbb{P}(D=1 \vert Z = z) - \alpha \leq P(z) \leq \alpha + \mathbb{P}(D=1 \vert Z = z),
$$
because $\mathbb{P}(D=1 \vert P(Z) = P(z)) = \mathbb{P}(D=1 \vert Z = z)$ from the index sufficiency implied by the model.
Similarly, from the condition $F_{V\vert \varepsilon=0}(P(z)) \in [0, 1]$, we have
$$
\alpha - \mathbb{P}(D=1 \vert Z = z)  \leq P(z) \leq 1-\alpha + \mathbb{P}(D=0 \vert Z = z).
$$
Hence, the following proposition holds. 
\begin{proposition}\label{prop1}
Suppose that model (\ref{seq1}) along with Assumptions \ref{RA}, \ref{Cont}, and \ref{Bound:mis} hold. We have the following bounds on $P(z)$: $LB(z) \leq P(z) \leq UB(z)$, where 
\begin{eqnarray*}
\begin{array}{lcccl}
LB(z)\equiv \inf_{\alpha \in [0, \bar{\alpha}]}\max\left\{\mathbb P(D=1\vert Z=z)-\alpha, \alpha-\mathbb P(D=1\vert Z=z)\right\},\\ \\
UB(z)\equiv \sup_{\alpha \in [0, \bar{\alpha}]} \min\left\{\mathbb P(D=1\vert Z=z)+\alpha, (1-\alpha)+\mathbb P(D=0\vert Z=z)\right\}.
\end{array}
\end{eqnarray*}
These bounds are pointwise sharp. 
\end{proposition}
In Appendix \ref{proofprop1}, we provide two different specifications for the relationship between the decision to misreport $\varepsilon$ and the unobserved heterogeneity $V$ that achieve the above bounds on $P(z)$. However, these bounds are not necessarily \textit{functionally sharp} in the language of \citet{Mourifie2020}, as taking the difference of the bounds for $P(z)$ and $P(z')$ will not necessarily yield the tightest bounds for the difference $P(z)-P(z')$. We show this in Subsection \ref{anabounds} below. Intuitively, note that the pointwise bounds on $P(z)$ in Proposition \ref{prop1} are derived using only information from the first stage equation, namely the fact that the cdfs $F_{V\vert \varepsilon=1}(P(z))$ and $F_{V\vert \varepsilon=0}(P(z))$ lie between 0 and 1. We are going to use the monotonicity of the cdfs $F_{V\vert \varepsilon=0}$ and $F_{V\vert \varepsilon=1}$ and information from the second stage equation to tighten the bounds on $P(z)-P(z')$.  

\begin{remark}
 The bounds in Proposition \ref{prop1} are non-informative if $\bar{\alpha}=1$. Indeed, if $\alpha=\mathbb P(D=1\vert Z=z)$, then the lower bound on $P(z)$ is 0, and if $\alpha=\mathbb P(D=0\vert Z=z)$, then the upper bound on $P(z)$ is 1. Also, as expected, bigger values of $\bar{\alpha}$ yield weakly wider bounds. That is,  for two misclassification upper bounds $\bar{\alpha}_0 < \bar{\alpha}_1,$ we have $\Theta_I^{\bar{\alpha}_0}(P(z))\subseteq \Theta_I^{\bar{\alpha}_1}(P(z)),$ where $\Theta_I^{\bar{\alpha}}(P(z))$ denotes the identified set for $P(z)$ when the upper bound on the misclassification is set to $\bar{\alpha}.$
\end{remark}

For the rest of the paper, we derive our results for each value of $\alpha \in [0,\bar{\alpha}]$. As in Proposition \ref{prop1}, one can take the infimum of the lower bound on the parameter of interest over the range $[0,\bar{\alpha}]$, and similarly the supremum of the upper bound over $[0,\bar{\alpha}]$.  

\subsubsection{Analytical bounds for the MTE}\label{anabounds}
In this subsection, we provide analytical bounds on the MTE. 

For any $z$ and $z'$ such that $P(z') <P(z)$, the monotonicity of the cdfs $F_{V\vert \varepsilon=0}$ and $F_{V\vert \varepsilon=1}$ imply these inequalities below,
\begin{eqnarray*}
0 &\leq& F_{V\vert \varepsilon=0}(P(z))- F_{V\vert \varepsilon=0}(P(z')) \leq 1,\\
0 &\leq& F_{V\vert \varepsilon=1}(P(z))- F_{V\vert \varepsilon=1}(P(z')) \leq 1.
\end{eqnarray*}
These latter inequalities respectively imply
\begin{align*}
-\Delta_{DZ}(z',z) &\leq P(z)-P(z')  \leq 2(1-\alpha)-\Delta_{DZ}(z',z),\\
\Delta_{DZ}(z',z) &\leq P(z)-P(z') \leq 2 \alpha + \Delta_{DZ}(z',z). 
\end{align*}
Thus, we have the following bounds on the $P(z)-P(z')$:\begin{eqnarray*}
\left \lvert \Delta_{DZ}(z',z)\right \rvert \leq P(z)-P(z') \leq \min\left\{1,2\alpha+\Delta_{DZ}(z',z), 2(1-\alpha)-\Delta_{DZ}(z',z)\right\}. 
\end{eqnarray*}
These above bounds on the difference $P(z)-P(z')$ can be tightened using Equations \eqref{eq:fsharp2} and \eqref{eq:fsharp3} below. 
Notice that the model implies the following index sufficiency result:
\begin{eqnarray}
\mathbb P\left(Y\in A, D=d\vert P(Z)=P(z)\right)=\mathbb P\left(Y\in A, D=d\vert Z=z\right) \text{ for all  $z$ and $d$.} \label{eq:fsharp1}
\end{eqnarray} 
Indeed, similar to Equation (\ref{eq1}), the following holds under Assumption \ref{RA}:
\begin{eqnarray*}
\mathbb P(Y\in A, D=1 \vert P(Z)=P(z)) &=& \mathbb P(Y_1\in A, \varepsilon=0, V \leq P(z)) + \mathbb P(Y_0\in A, \varepsilon=1, V > P(z)).
\end{eqnarray*}
From Equation (\ref{eq1}), we can show under Assumptions \ref{RA} and \ref{Cont} that 
\begin{eqnarray}
\mathbb P(Y\in A, D=1 \vert Z=z) 
&=& \int^{P(z)}_0\mathbb P\left(Y_1\in A, \varepsilon=0 \vert V=v\right)dv \nonumber \\
&& \qquad \qquad + \int_{P(z)}^1\mathbb P\left(Y_0\in A, \varepsilon=1 \vert V=v\right)dv. \label{eq:fsharp2}
\end{eqnarray}
The outcome distribution in the observed treated group is decomposed as an integral of the treated potential outcome distribution for the truth-tellers plus an integral of the untreated potential outcome distribution for the misreporters.  
A similar result holds for the observed control group:
\begin{eqnarray}
\mathbb P(Y\in A, D=0 \vert Z=z) &=& \int^{P(z)}_0\mathbb P\left(Y_1\in A, \varepsilon=1 \vert V=v\right)dv \nonumber \\
&& \qquad \qquad + \int_{P(z)}^1\mathbb P\left(Y_0\in A, \varepsilon=0 \vert V=v\right)dv. \label{eq:fsharp3}
\end{eqnarray}
The above derived equalities allow us to characterize the functional identified set for $P(z)$.
\begin{definition}
The identified set for the function $P: \mathcal Z \rightarrow [0,1]$ is the collection $$\left\{P(z): \ 0\leq P(z) \leq 1,\ z \in \mathcal Z\right\}$$ such that there exists a joint distribution on $(Y_0, Y_1, \varepsilon, V, Z)$ that satisfies model (\ref{seq1}),  Assumptions \ref{RA}, \ref{Cont}, \ref{Bound:mis}, and Equations (\ref{eq:fsharp1})--(\ref{eq:fsharp3}).
\end{definition}
This characterization of the identified set for $P(z)$ is broad, but less tractable. We are going to derive analytical expressions for the MTE bounds based on \eqref{eq:fsharp2} and \eqref{eq:fsharp3}.

Consider Equation \eqref{eq:fsharp2} first. Take the difference of $\mathbb P(Y\in A,D=1\vert Z=z)$ between two values $z$ and $z'$ of the instrument $Z$ such that $P(z') < P(z).$
\begin{eqnarray*}
\mathbb P(Y\in A, D=1 \vert Z=z)-\mathbb P(Y\in A, D=1 \vert Z=z')
&=& \int^{P(z)}_{P(z')}\mathbb P\left(Y_1\in A, \varepsilon=0 \vert V=v\right)dv \nonumber \\
&& \qquad - \int^{P(z)}_{P(z')}\mathbb P\left(Y_0\in A, \varepsilon=1 \vert V=v\right)dv. 
\end{eqnarray*}
By taking the absolute value of each side, applying the triangle inequality, and using the monotonicity of a probability measure, we have
\begin{eqnarray*}
\lvert \Delta_{Y,D=1,Z}(z',z) \rvert \equiv \sup_{A}\lvert \mathbb P(Y\in A, D=1 \vert Z=z)-\mathbb P(Y\in A, D=1 \vert Z=z')\rvert
&\leq& P(z)-P(z').
\end{eqnarray*}
A similar result holds from Equation \eqref{eq:fsharp3} for $D=0$. These bounds can be further tightened using the density version of \eqref{eq:fsharp2}. We have
\begin{eqnarray*}
f_{Y,D\vert Z}\left(y, 1 \vert z\right) - f_{Y,D\vert Z}\left(y, 1 \vert z'\right)&=& \int^{P(z)}_{P(z')}f_{Y_1,\varepsilon\vert V}(y,0\vert v)dv - \int^{P(z)}_{P(z')}f_{Y_0,\varepsilon\vert V}(y,1\vert v)dv,
\end{eqnarray*}
where $f_{X\vert W}\left(x \vert w\right)$ is the conditional density of $X$ given $\left\{W=w\right\}$ that is absolutely continuous with respect to a known dominating measure $\mu_{X}$.
Therefore, by using the triangle inequality, integrating each side of the inequality over the support $\mathcal Y$, and using the Fubini-Tonelli theorem, we have
\begin{eqnarray*}
\int_{\mathcal Y}\left \lvert f_{Y,D\vert Z}\left(y, 1 \vert z\right) - f_{Y,D\vert Z}\left(y, 1 \vert z'\right) \right \rvert d \mu_{Y}(y) \leq \int^{P(z)}_{P(z')}\mathbb P(\varepsilon=0\vert V=v)+\mathbb P(\varepsilon=1\vert V=v)dv.
\end{eqnarray*}
Therefore, we have $TV_{(Y,D=1)}(z',z) \leq P(z)-P(z')$, where 
\begin{eqnarray*}
TV_{(Y,D=d)}(z',z) \equiv \int_{\mathcal Y}\left \lvert f_{Y,D\vert Z}\left(y, d \vert z\right) - f_{Y,D\vert Z}\left(y, d \vert z'\right) \right \rvert d \mu_{Y}(y).
\end{eqnarray*}
 
Using a similar argument on Equation \eqref{eq:fsharp3}, we have $TV_{(Y,D=0)}(z',z) \leq P(z)-P(z')$. Equation \eqref{eq:mis0} helps derive another upper bound for $P(z)-P(z').$ The argument is similar to the one derived for the compliers share in the previous subsection:
$P(z)-P(z') \leq \min_{d\in\{0,1\}}\left\{\mathbb P(D=d\vert Z=z)+\mathbb P(D=d\vert Z=z')\right\}.$
Hence, combining all the above bounds, we obtain the following bounds on the difference $P(z)-P(z')$:
\begin{eqnarray*}
&&\max_{d\in\{0,1\}}\max\left\{\left \lvert \Delta_{Y,D=d,Z}(z',z)\right \rvert, TV_{(Y,D=d)}(z',z)\right\}
 \leq P(z)-P(z') \leq \\
 && \min\bigg\{
\min_{d\in\{0,1\}}
\left[
\mathbb P(D=d\mid Z=z)+\mathbb P(D=d\mid Z=z')
\right], 
2\alpha+\Delta_{DZ}(z',z),\
2(1-\alpha)-\Delta_{DZ}(z',z)
\bigg\}.
\end{eqnarray*}
We can show that $TV_{(Y,D=d)}(z',z)\geq \lvert \Delta_{Y,D=d,Z}(z',z) \rvert \geq \left \lvert \Delta_{DZ}(z',z)\right \rvert$. See a formal proof in Appendix \ref{proof:mtebounds}. Consequently, we have
\begin{align}
LB_p(z',z)
&\leq P(z)-P(z') \leq UB_p(z',z),
\label{eq:pzbounds}
\end{align}
where 
\begin{align*}
    LB_p(z',z) &\equiv \max_{d\in\{0,1\}}\left\{TV_{(Y,D=d)}(z',z)\right\},\nonumber \\
UB_p(z',z)
&\equiv
\min\bigg\{
\min_{d\in\{0,1\}}
\left[
\mathbb P(D=d\mid Z=z)+\mathbb P(D=d\mid Z=z')
\right], \nonumber\\
&\qquad\qquad
2\alpha+\Delta_{DZ}(z',z),\
2(1-\alpha)-\Delta_{DZ}(z',z)
\bigg\}.
\nonumber
\end{align*}
These above bounds on $P(z)-P(z')$ are tighter than the ones we would get by taking the difference of the pointwise bounds derived previously in Proposition \ref{prop1}. 

Suppose $LB_p(z',z) \neq 0$ and $UB_p(z',z) \neq 0$. Then, the following holds. 
\begin{eqnarray*}
&& \min\left\{\frac{\Delta_{YZ}(z',z)}{UB_p(z',z)},  \frac{\Delta_{YZ}(z',z)}{LB_p(z',z)}\right\}\\
&& \qquad \qquad \leq \frac{\mathbb E[Y\vert P(Z)=P(z)]-\mathbb E[Y\vert P(Z)=P(z')]}{P(z)-P(z')} \leq \\
&& \qquad \qquad \qquad \qquad \qquad \qquad \qquad \qquad \qquad \max\left\{\frac{\Delta_{YZ}(z',z)}{UB_p(z',z)},  \frac{\Delta_{YZ}(z',z)}{LB_p(z',z)}\right\}.
\end{eqnarray*}
Therefore, we can take the limit of each side when $z'$ goes to $z$. Suppose that $\lim_{z' \rightarrow z} \frac{\Delta_{YZ}(z',z)}{UB_p(z',z)}$, $\lim_{z' \rightarrow z} \frac{\Delta_{YZ}(z',z)}{LB_p(z',z)}$, and $\lim_{z' \rightarrow z} \frac{\mathbb E[Y\vert P(Z)=P(z)]-\mathbb E[Y\vert P(Z)=P(z')]}{P(z)-P(z')}$ exist.\footnote{Then, we have $\lim_{z' \rightarrow z} \frac{\mathbb E[Y\vert P(Z)=P(z)]-\mathbb E[Y\vert P(Z)=P(z')]}{P(z)-P(z')}=\frac{\partial \mathbb E[Y \vert P(Z)=p] }{\partial p}\vert_{p=P(z)}=MTE(P(z))$. When the first two limits do not exist, we replace them by $\lim\inf$ and $\lim\sup$ in the lower and upper bounds of (\ref{eq:anabounds}).} Then, using the fact that the functions $\min$ and $\max$ are continuous, and assuming that $P(z)$ is continuous in $z$, we obtain
\begin{eqnarray}
&& \min\left\{\lim_{z' \rightarrow z} \frac{\Delta_{YZ}(z',z)}{UB_p(z',z)},  \lim_{z' \rightarrow z} \frac{\Delta_{YZ}(z',z)}{LB_p(z',z)}\right\} \nonumber\\
&& \qquad \qquad \qquad \qquad \qquad \leq MTE(P(z))\leq \label{eq:anabounds}\\
&& \qquad \qquad \qquad \qquad \qquad \qquad \max\left\{\lim_{z' \rightarrow z} \frac{\Delta_{YZ}(z',z)}{UB_p(z',z)},  \lim_{z' \rightarrow z} \frac{\Delta_{YZ}(z',z)}{LB_p(z',z)}\right\}. \nonumber
\end{eqnarray}
These bounds may not be sharp, but they provide a \textit{tractable outer set} of the identified set for $MTE(P(z))$. We summarize the results in the following proposition.

\begin{proposition}\label{prop:mtebounds}
Suppose that model (\ref{seq1}) along with Assumptions \ref{RA}, \ref{Cont}, \ref{Cont_inst}, \ref{Bound:mis}, and \ref{morett} hold. Also, suppose that $P(z)$ is continuous in $z$ and $\lim_{z' \rightarrow z} \frac{\Delta_{YZ}(z',z)}{UB_p(z',z)}$, $\lim_{z' \rightarrow z} \frac{\Delta_{YZ}(z',z)}{LB_p(z',z)}$, and $\lim_{z' \rightarrow z} \frac{\mathbb E[Y\vert P(Z)=P(z)]-\mathbb E[Y\vert P(Z)=P(z')]}{P(z)-P(z')}$ exist. The following statements hold:
\begin{enumerate}[(i)]
\item If $\bar{\alpha}=0$, then $MTE(P(z))$ is point-identified as
\begin{eqnarray}\label{eq:late}
\lim_{z' \rightarrow z} \frac{\mathbb E[Y\vert Z=z]-\mathbb E[Y\vert Z=z']}{\mathbb E[D\vert Z=z]-\mathbb E[D\vert Z=z']}.
\end{eqnarray}
\item If $\bar{\alpha} > 0$, then $MTE(P(z))$ is partially-identified:
\footnotesize{\begin{eqnarray}\label{eq:mtebounds}
 LB_{MTE}(z)\equiv \min\left\{0,  \lim_{z' \rightarrow z} \frac{\Delta_{YZ}(z',z)}{LB_p(z',z)}\right\} \leq MTE(P(z))\leq \max\left\{0,  \lim_{z' \rightarrow z} \frac{\Delta_{YZ}(z',z)}{LB_p(z',z)}\right\}\equiv UB_{MTE}(z).
\end{eqnarray}}
\end{enumerate}
\end{proposition}
The proof on Proposition \ref{prop:mtebounds} is shown in Appendix \ref{proof:mtebounds}. This proposition shows that when there is no misclassification ($\bar{\alpha}=0$), our bounds collapse to a point, which is the standard MTE estimand as a limit of a generalized LATE. In such a case, the point identification of $MTE(P(z))$ is achieved for $P(z)=\mathbb E[D\vert Z=z]$ where $z$ belongs to the support of $Z$. Outside the support of $\mathbb E[D\vert Z],$ we cannot say anything about $MTE(P(z))$ without further assumptions. \cite{Mogstadal2018} study policy-relevant treatment parameters using the MTE framework and combine the identified part of the MTE with additional restrictions to learn about policy parameters whose weights may place mass outside the support reached by the instrument. Thus, their framework is broader for global PRTE analysis in the correctly measured treatment case.  When $\bar{\alpha} >0$, the sign of the $MTE$ is \textit{locally identified} for each value $z$. Equation \eqref{eq:mtebounds} shows that our derived bounds are not changing with $\bar{\alpha} >0$. In particular, the bounds in Equation \eqref{eq:mtebounds} remain valid even when the researcher is agnostic about the misclassification probability ($\bar{\alpha}=1$).  The numerical example below illustrates how informative the bounds can be in practice, depending on the underlying structure in the data generating process. 

\begin{remark}
   If one considers the potential outcome model for reporting $D=D^*D_1+(1-D^*)D_0$ as in \citet{Ura2018}, s/he will obtain the following bounds on $P(z)-P(z')$: 
   $$\frac{1}{2} TV_{(Y,D=1)}(z',z) + \frac{1}{2}TV_{(Y,D=0)}(z',z)\leq P(z)-P(z') \leq 1.$$
   S/he can use these bounds instead of those in Equation \eqref{eq:pzbounds} to bound $MTE(P(z)).$ This approach will yield wider bounds than those of Proposition \ref{prop:mtebounds}, which is expected since our model is more restrictive. 
\end{remark}

\subsection{Continuous Instrument: Identification of the PRTE}\label{sec:prteid}
In this section, we illustrate how our methodology can be used to derive bounds on some policy relevant treatment effect (PRTE) parameters. Consider a new policy that induces $100 a\%$ of the untreated population under the current policy to treatment, so that we can write the new propensity score $P^a(z)$ as $P^a(z)=P(z)+a(1-P(z))$, where $a\in(0,1)$. 

Following \citet{heckman2005structural}, we maintain the policy invariance assumption ($(V^a, Y_1^a, Y_0^a)=(V,Y_1,Y_0)$) and write $Y^a=Y_1D^{*a}+Y_0(1-D^{*a})$, $D^{*a}=\mathbbm{1}\{V \leq P^a(Z)\}$. 
Under this policy invariance assumption, the PRTE defined as $PRTE^a\equiv \frac{\mathbb E[Y^a]-\mathbb E[Y]}{E[D^{*a}]-\mathbb E[D^*]}$ can be written as
\begin{align}
PRTE^a
&= \int_{0}^{1} MTE(v)\,
\left[\frac{F_{P(Z)}(v) - F_{P^a(Z)}(v)}{a\left(1-\mathbb E[P(Z)]\right)}\right]\, dv.
\label{eq:prte_v_def}
\end{align}
We can show that $F_{P^a(Z)}(p)=F_{P(Z)}(\frac{p-a}{1-a})=F_{P(Z)}(t_a(p))$ where $t_a(p)\equiv \max\left\{0,\frac{p-a}{1-a}\right\}$, since $P(Z) \in [0,1]$. To bound the PRTE, we use the pointwise bounds on $P(z)$ from Proposition \ref{prop1}, the functional bounds on $P(z)-P(z')$ from Equation \eqref{eq:pzbounds}, and the bounds on $MTE(P(z))$ from Proposition~\ref{prop:mtebounds}. 

Define the weight 
\begin{equation}\label{eq:wP_def_prop}
w_P(v)\equiv \frac{F_{P(Z)}(v)-F_{P^a(Z)}(v)}{a(1-\mathbb E[P(Z)])}
=\frac{F_{P(Z)}(v)-F_{P(Z)}(t_a(v))}{a(1-\mathbb E[P(Z)])},
\qquad v\in[0,1],
\end{equation}
and the MTE function $m(v)\equiv\mathbb E[Y_1-Y_0 \vert V=v]$ for $v\in[0,1]$.  

\begin{definition}
The identified set $\mathcal I(PRTE^a)$ for $PRTE^a$ is the collection $\left\{\int_0^1 m(v) w_P^a(v) dv\right\}$ such that
$m(v)=\mathbb E[Y_1-Y_0 \vert V=v]$ for $v\in[0,1]$, and the function $P: \mathcal Z \rightarrow [0,1]$ and the conditional joint distribution of $(Y_d,\varepsilon)$ given $V$ $(d=0,1)$ satisfy Equations (\ref{eq:fsharp1})--(\ref{eq:fsharp3}).
\end{definition}

For the sake of tractability, we now provide an outer set for the identified set $\mathcal{I}(PRTE^a)$.  Let $\mathcal Z_M\equiv \mathcal Z\cap[-M,M]$ and let
$\Pi:\ z_0<z_1<\cdots<z_k$ be a partition of $\mathcal Z_M$ with
norm $\|\Pi\|\equiv \max_i (z_i-z_{i-1})$. Define the Riemann-Stieltjes sum:
\begin{equation}\label{eq:RS_sum_prop}
S(\Pi;P,m)\equiv \sum_{i=1}^{k}
m(P(z_i))\,
\frac{F_{P(Z)}(P(z_i)) - F_{P^a(Z)}(P(z_i))}
{a\left(1-\mathbb E[P(Z)]\right)}\,
\big[P(z_i)-P(z_{i-1})\big].
\end{equation}

Define
\begin{align*}
    \mathcal P&\equiv \big\{P:\mathcal Z\to[0,1]: LB(z)\le P(z)\le UB(z), \\
& \qquad \qquad LB_p(z',z)\le P(z)-P(z')\le UB_p(z',z) \text{ for all $z'<z$}\big\},\\
    \mathcal M(P)&\equiv \{m:[0,1]\to\mathbb R: LB_{MTE}(z)\le m(P(z))\le UB_{MTE}(z)\},
\end{align*}

Proposition~\ref{prop:identified_set_RS} below provides outer bounds for the PRTE.
\begin{proposition}
\label{prop:identified_set_RS}
Suppose that the conditions in Proposition \ref{prop:mtebounds} hold. Suppose also that the propensity score function $P:\mathcal Z\to[0,1]$ belongs to $\mathcal P$, is continuous and has bounded variation on $\mathcal Z$ with endpoint normalization $\lim_{z\downarrow \underline z}P(z)=0$ and
$\lim_{z\uparrow \overline z}P(z)=1$ where $\underline z$ and $\overline z$ denote the lower and upper endpoints,
possibly in the extended reals, of $\mathcal Z$, and is such that the cdf $F_{P(Z)}(p)\equiv \mathbb P(P(Z) \leq p)$ is continuous on $[0,1]$, and $0< \mathbb P(D^*=1)=\mathbb E[P(Z)]<1$. Moreover, suppose that the MTE function $m: [0,1] \to \mathbb R$ belongs to $\mathcal M(P)$, satisfies $m(p)=\mathbb E[Y_1-Y_0 \vert V=p]$, and is bounded and continuous on $[0,1]$. For $a \in (0,1)$, define the new policy $P^a(Z) \equiv P(Z) + a (1-P(Z))$. 
Then, every feasible pair $(P,m)$ satisfying the above conditions admits the representation
\begin{equation}\label{eq:RS_limit_prop}
PRTE(P,m;a)
=
\lim_{M\to\infty}\ \lim_{\|\Pi\|\to 0,\ \Pi\subset\mathcal Z_M} S(\Pi;P,m),
\end{equation}
where the inner limit is taken over partitions of $\mathcal Z_M$ with norm $\|\Pi\|\to0$
and where the tags are fixed at right endpoints.

Moreover, define
\[
\underline{PRTE}^a\equiv \inf_{P\in\mathcal P,\ m\in\mathcal M(P)} PRTE(P,m;a),
\qquad
\overline{PRTE}^a\equiv \sup_{P\in\mathcal P,\ m\in\mathcal M(P)} PRTE(P,m;a).
\]
Then $\left[\underline{PRTE}^a, \overline{PRTE}^a \right]$ is a tractable outer set for $\mathcal I(PRTE^a)$.
\end{proposition}

The proof of Proposition \ref{prop:identified_set_RS} is shown in Appendix \ref{proof:identified_set_RS}.

For computational purposes, note that Assumption~\ref{morett} allows us to identify the sign of $\Delta_{D^*Z}(z',z)$ or $P(z)-P(z')$ for any given $(z', z)$ by Lemma~\ref{morett_samedirection}, and that Proposition~\ref{prop:mtebounds} locally identifies the sign of $MTE(P(z))$ for each value $z$.
Therefore, the above PRTE identification bounds can be expressed in a more elaborate explicit form as shown in Corollary~\ref{cor:sign_nodewise}.

\begin{corollary}
\label{cor:sign_nodewise}
Fix $a\in(0,1)$ and let $\Pi:\ z_0<z_1<\cdots<z_k$ be a partition of
$\mathcal Z_M\equiv \mathcal Z\cap[-M,M]$.
Suppose that the conditions in Proposition~\ref{prop:identified_set_RS} hold.
For a feasible pair $(P,m)$, define $p_i\equiv P(z_i)$, $\Delta p_i\equiv p_i-p_{i-1},$ and $s_i\equiv \operatorname{sgn}(\Delta p_i)$ for $i=0,\dots,k$.
Also define $B_i(P)\equiv
\frac{F_{P(Z)}(p_i)-F_{P^a(Z)}(p_i)}{a(1-\mathbb E[P(Z)])}\,\Delta p_i$,
and $T_i(P,m)\equiv m(p_i)\,B_i(P)$. 
Let
\begin{align*}
    \underline T_i(P)&\equiv
\Big[
\mathbbm{1}\{UB_{MTE}(z_i)=0,\ s_i=1\}\,LB_{MTE}(z_i)
-\mathbbm{1}\{LB_{MTE}(z_i)=0,\ s_i=-1\}\,UB_{MTE}(z_i)
\Big]|B_i(P)|, \\
    \overline T_i(P)&\equiv
\Big[
\mathbbm{1}\{LB_{MTE}(z_i)=0,\ s_i=1\}\,UB_{MTE}(z_i)
-\mathbbm{1}\{UB_{MTE}(z_i)=0,\ s_i=-1\}\,LB_{MTE}(z_i)
\Big]|B_i(P)|.
\end{align*}

Then, for each $i=1,\dots,k$, we have $\underline T_i(P)\le T_i(P,m)\le \overline T_i(P)$.

Consequently, defining the finite-partition RS lower and upper sums $\underline S(\Pi;P)\equiv \sum_{i=1}^k \underline T_i(P)$ and $\overline S(\Pi;P)\equiv \sum_{i=1}^k \overline T_i(P)$, we have $\underline S(\Pi;P)\le S(\Pi;P,m)\le \overline S(\Pi;P)$.

Therefore, the finite-partition PRTE bounds
\[
\underline{PRTE}^{a}_{\Pi,M}
\equiv
\inf_{P\in\mathcal P}\underline S(\Pi;P),
\qquad
\overline{PRTE}^{a}_{\Pi,M}
\equiv
\sup_{P\in\mathcal P}\overline S(\Pi;P)
\]
provide a tractable outer approximation to the RS approximation of the PRTE. In the RS limit, these bounds yield
\[
\inf_{P\in\mathcal P}
\liminf_{M\to\infty}\liminf_{\|\Pi\|\to0,\ \Pi\subset\mathcal Z_M}
\underline S(\Pi;P)
\le
PRTE^a
\le
\sup_{P\in\mathcal P}
\limsup_{M\to\infty}\limsup_{\|\Pi\|\to0,\ \Pi\subset\mathcal Z_M}
\overline S(\Pi;P).
\]

In particular, if $P$ is nondecreasing on $\mathcal Z$, so that $s_i=1$ for all $i$, then
\[
\mathbbm{1}\{UB_{MTE}(z_i)=0\}\,LB_{MTE}(z_i)\,|B_i(P)|
\le T_i(P,m)\le
\mathbbm{1}\{LB_{MTE}(z_i)=0\}\,UB_{MTE}(z_i)\,|B_i(P)|
\]
for all $i$.
\end{corollary}

The proof on Corollary~\ref{cor:sign_nodewise} is shown in Appendix \ref{proof:identified_set_RS}.

\subsubsection*{Algorithm to compute the PRTE bounds}
We provide an algorithm to numerically compute the PRTE bounds.

\begin{algorithm}[htbp]
\small
\SetAlgoNlRelativeSize{-1}
\caption{Numerical approximation of PRTE bounds}
\label{alg:prte_bounds}
\DontPrintSemicolon
\SetKwInOut{Input}{Input}
\SetKwInOut{Output}{Output}

\Input{Policy parameter $a\in(0,1)$; coarse partition $z_0<z_1<\cdots<z_K$ of $\mathcal Z_M$; local step $h_{\mathrm{MTE}}>0$; search parameters.}
\Output{Searched finite-grid bounds $\big(\widehat{\underline{PRTE}}^{\,a},\widehat{\overline{PRTE}}^{\,a}\big)$.}

Set $p_i\equiv P(z_i)$ for $i=0,\dots,K$ and $\Delta p_i\equiv p_i-p_{i-1}$ for $i=1,\dots,K$.\;

Compute pointwise score bounds $LB(z_i)\le p_i\le UB(z_i)$ for $i=0,\dots,K$, and adjacent increment bounds $LB_p(z_{i-1},z_i)\le \Delta p_i\le UB_p(z_{i-1},z_i)$ for $i=1,\dots,K$.\;

Compute nodewise MTE bounds $LB_{MTE}(z_i)\le m(p_i)\le UB_{MTE}(z_i)$ for the right-endpoint tags $i=1,\dots,K$ from Proposition~\ref{prop:mtebounds} using the local step $h_{\mathrm{MTE}}$.\;

If the support endpoints are normalized so that \(P(z_0)=0\) and \(P(z_K)=1\),
impose \(p_0=0\) and \(p_K=1\). Otherwise, impose the appropriate endpoint
restrictions \(p_0=P(z_0)\) and \(p_K=P(z_K)\), or their feasible bounds.\;

Define minimum and maximum remaining cumulative increases by $rem_{\min}(i)\equiv \sum_{j=i}^K LBd_j$ and $rem_{\max}(i)\equiv \sum_{j=i}^K UBd_j$.\;

For each interior node $i=1,\dots,K-1$, compute the admissible interval $[\ell_i,u_i]$ recursively from $p_{i-1}$, where $\ell_i\equiv \max\{LB(z_i),\,p_{i-1}+LBd_i,\,1-rem_{\max}(i+1)\}$ and $u_i\equiv \min\{UB(z_i),\,p_{i-1}+UBd_i,\,1-rem_{\min}(i+1)\}$; if some $[\ell_i,u_i]$ is empty, declare the current searched set infeasible on the grid.\;

Initialize $\mathcal P_{\mathrm{search}}\gets\emptyset$, then add: (i) the midpoint feasible path obtained by setting $p_i=(\ell_i+u_i)/2$ recursively; (ii) feasible random paths obtained from sequential draws $p_i\sim U[\ell_i,u_i]$ for $i=1,\dots,K-1$; (iii) feasible linear-programming-guided (LP-guided) boundary paths obtained by optimizing a collection of linear objectives over the pointwise, increment, and endpoint constraints; and (iv) feasible coordinate-wise refinements of existing candidate paths obtained by replacing each interior node $p_i$ with values in $\{p_i,\ell_i,u_i,(\ell_i+u_i)/2\}$ whenever feasible.\;

For each candidate path $p=(p_0,\dots,p_K)\in\mathcal P_{\mathrm{search}}$, compute $s_i\equiv\operatorname{sgn}(\Delta p_i)$ and the RS coefficients $B_i(p)\equiv \frac{F_{P(Z)}(p_i)-F_{P^a(Z)}(p_i)}{a(1-\mathbb E[P(Z)])}\Delta p_i$ for $i=1,\dots,K$, and use Corollary~\ref{cor:sign_nodewise} to obtain nodewise contribution bounds $\underline T_i(p)\le m(p_i)B_i(p)\le \overline T_i(p)$.\;

Form the pathwise RS lower and upper sums $\underline S(\Pi;p)\equiv \sum_{i=1}^K \underline T_i(p)$ and $\overline S(\Pi;p)\equiv \sum_{i=1}^K \overline T_i(p)$.\;

Return $\widehat{\underline{PRTE}}^{\,a}\equiv \min_{p\in\mathcal P_{\mathrm{search}}}\underline S(\Pi;p)$ and $\widehat{\overline{PRTE}}^{\,a}\equiv \max_{p\in\mathcal P_{\mathrm{search}}}\overline S(\Pi;p)$.\;

\end{algorithm}

\subsection{Numerical Illustration}\label{numeric1}
We consider the following data generating process (DGP):
\begin{eqnarray}\label{eq:ex}
\left\{
\begin{array}{lcl}
Y &=& \beta D^* + U, \\[2pt]
D^* &=& \mathbbm{1}\{V \leq P(Z)\}, \\[2pt]
D &=& D^*(1-\varepsilon) + (1-D^*)\varepsilon, \\[2pt]
\varepsilon &=& \mathbbm{1}\{\xi \leq \alpha\}, \\[2pt]
\beta &=& \beta_0 + \beta_1(1-\sqrt{V}) + \eta ,
\end{array}
\right.
\end{eqnarray}
where $Z\sim \mathrm{Unif}[0,1]$, $P(z)\equiv \Phi(\delta_0+\delta_1z)$, $V=\Phi(V^*)$, $\xi=\Phi(\xi^*)$,
$(V^*,\xi^*)$ follows the standard bivariate normal with correlation coefficient $\rho,$ $Z\ \indep\ (V,U,\eta,\xi),$ and $(U,\eta)\ \indep\ (V,\xi)$. Details on the DGP
and the parameter values used in the illustration are provided in
Subsection \ref{numeric:apx1} in the appendix. Under this specification,
\[
P(z)=\Pr(D^*=1\mid Z=z)=\Phi(\delta_0+\delta_1z),
\qquad
m(p)\equiv MTE(p)=\beta_0+\beta_1(1-\sqrt{p}),
\quad p\in[0,1],
\]
and
\[
MTE(P(z))=\beta_0+\beta_1(1-\sqrt{P(z)})
=\beta_0+\beta_1\left(1-\sqrt{\Phi(\delta_0+\delta_1z)}\right).
\]

\begin{figure}[h]
    \centering
\includegraphics[width=0.86\textwidth]{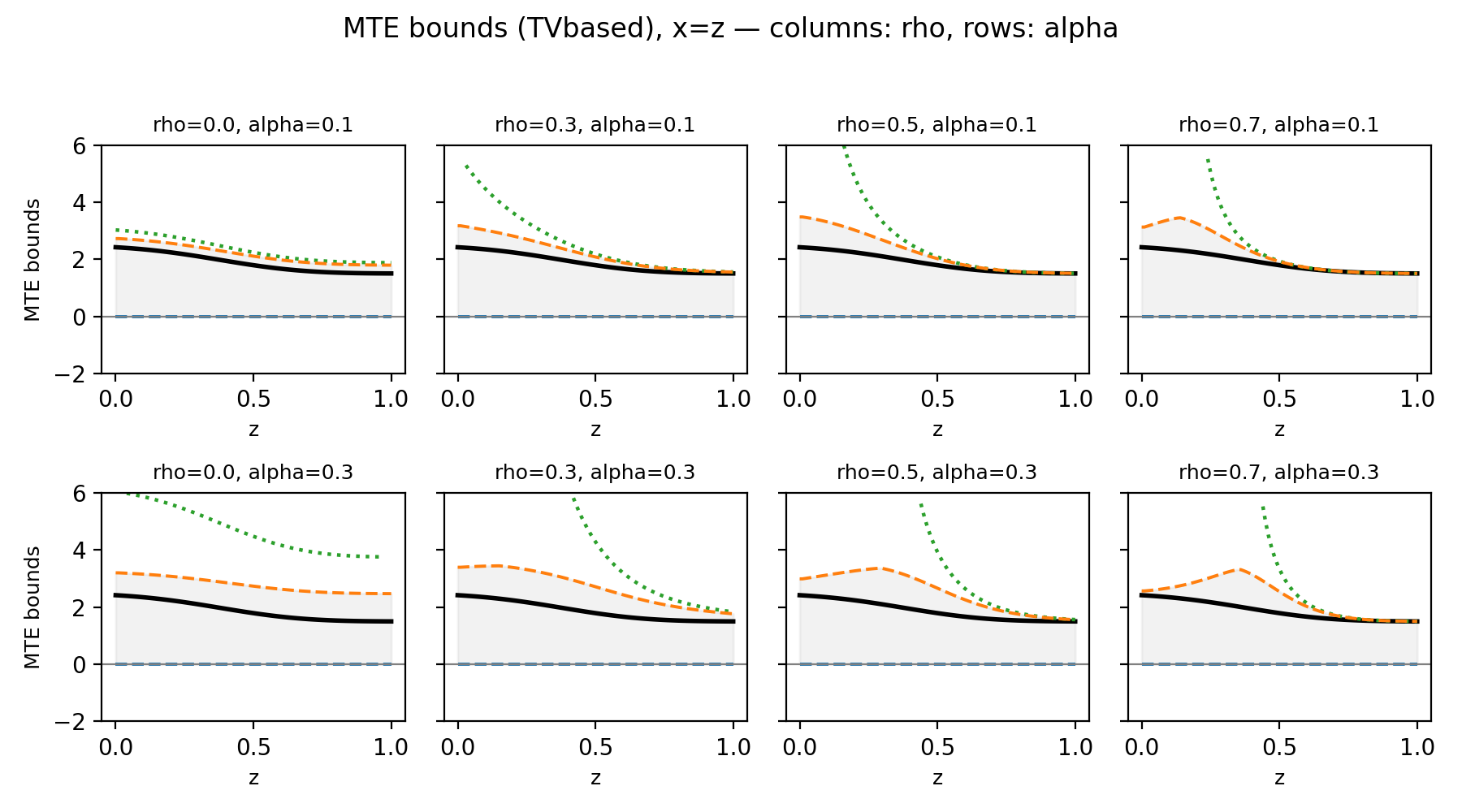}\\[0.75em]
\includegraphics[width=0.86\textwidth]{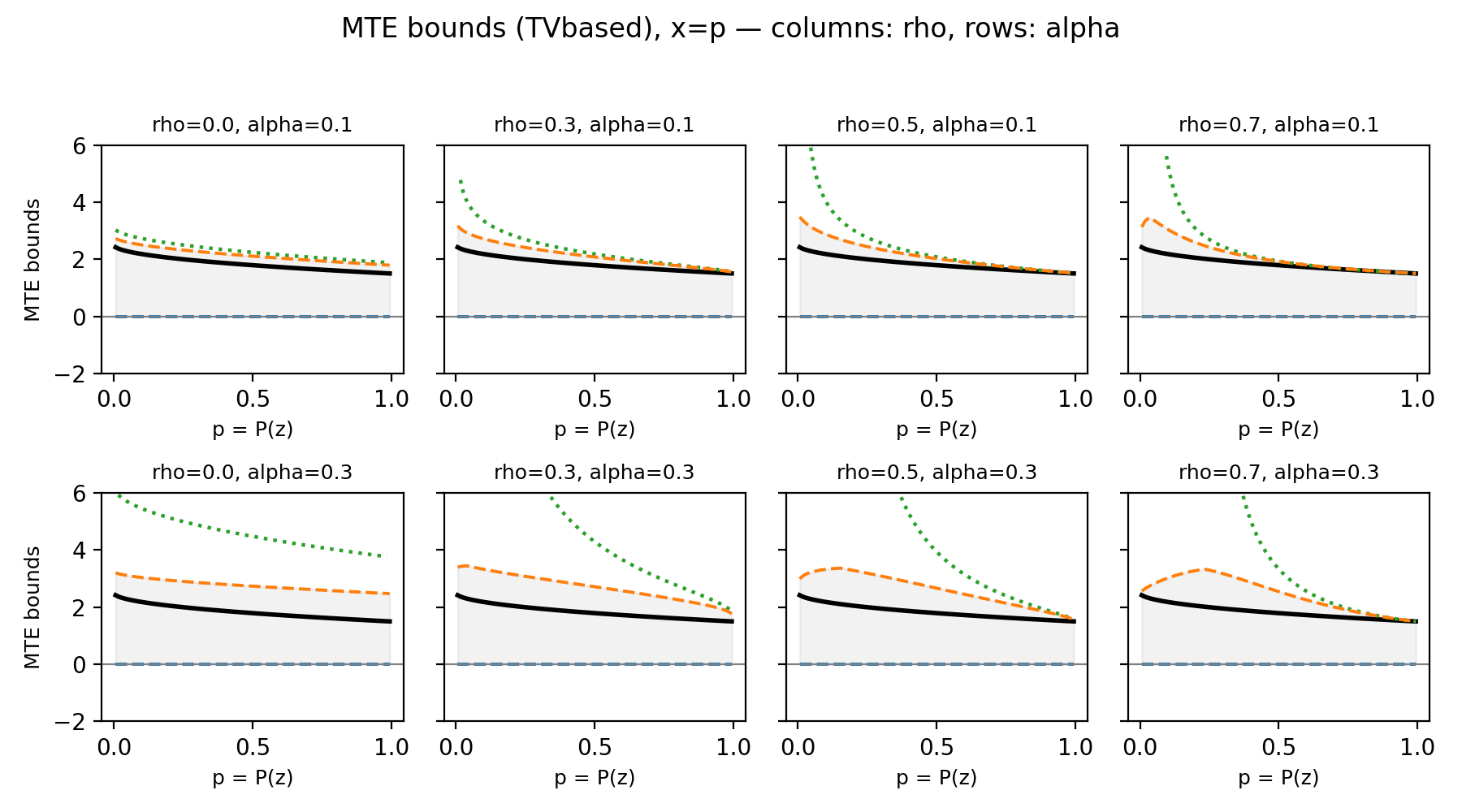}
\caption{TV-based bounds for $MTE(P(z))$ indexed by $z$ and by $p=P(z)$}
\label{fig.ey1.exam0921}
\figurenote{The black solid line represents the true MTE curve, $MTE(P(z))$. The blue and orange dashed lines represent the approximated lower and upper bounds for $MTE(P(z))$. The green dotted line represents the naive LIV estimand based on the observed treatment $D$; near-singular segments are omitted when the observed first stage is close to zero. The upper panel uses the raw instrument axis $z$, while the lower panel uses the latent propensity-score axis $p=P(z)$.}
\end{figure}

Figure \ref{fig.ey1.exam0921} reports the bounds in Proposition \ref{prop:mtebounds} for $MTE(P(z))$ using two equivalent horizontal axes. The upper panel indexes the bounds by the raw instrument value $z$, while the lower panel indexes the same object by the latent propensity score value $p=P(z)$.
The black solid line is the true curve $MTE(P(z))$, the blue and orange dashed lines represent the lower and upper bounds for $MTE(P(z))$, and the green dotted line is the naive LIV estimand constructed from the observed treatment $D$. 
In this design, the lower bound is equal to zero throughout, so the sign of the MTE is weakly identified. 
The upper envelope is generally lower when the misclassification rate is small, especially in the low-dependence panels, although this comparison is not uniform across all values of $\rho$.
By contrast, the effect of the dependence parameter $\rho$ on the MTE bounds is mixed rather than monotone; the outer bounds do not change monotonically with the dependence parameter $\rho$ between $V$ and the misreporting indicator $\varepsilon$. In general, our proposed outer bounds are informative for the MTE curve in the two panels, while the naive LIV that ignores misclassification is upward biased and lies outside the bounds.

\begin{figure}[!h]
    \centering
    \begin{minipage}{0.49\textwidth}
        \centering
        \includegraphics[width=\textwidth]{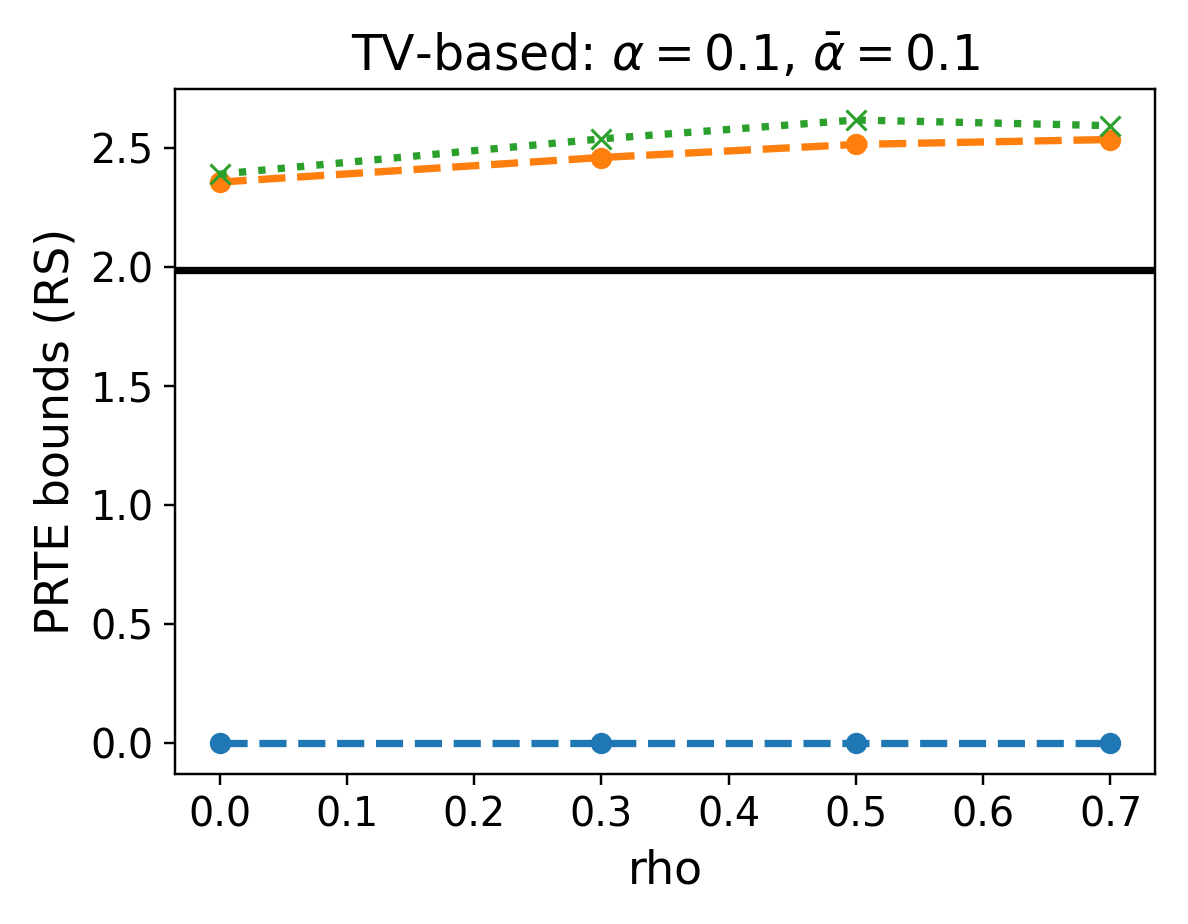}
    \end{minipage}\hfill
    \begin{minipage}{0.49\textwidth}
        \centering
        \includegraphics[width=\textwidth]{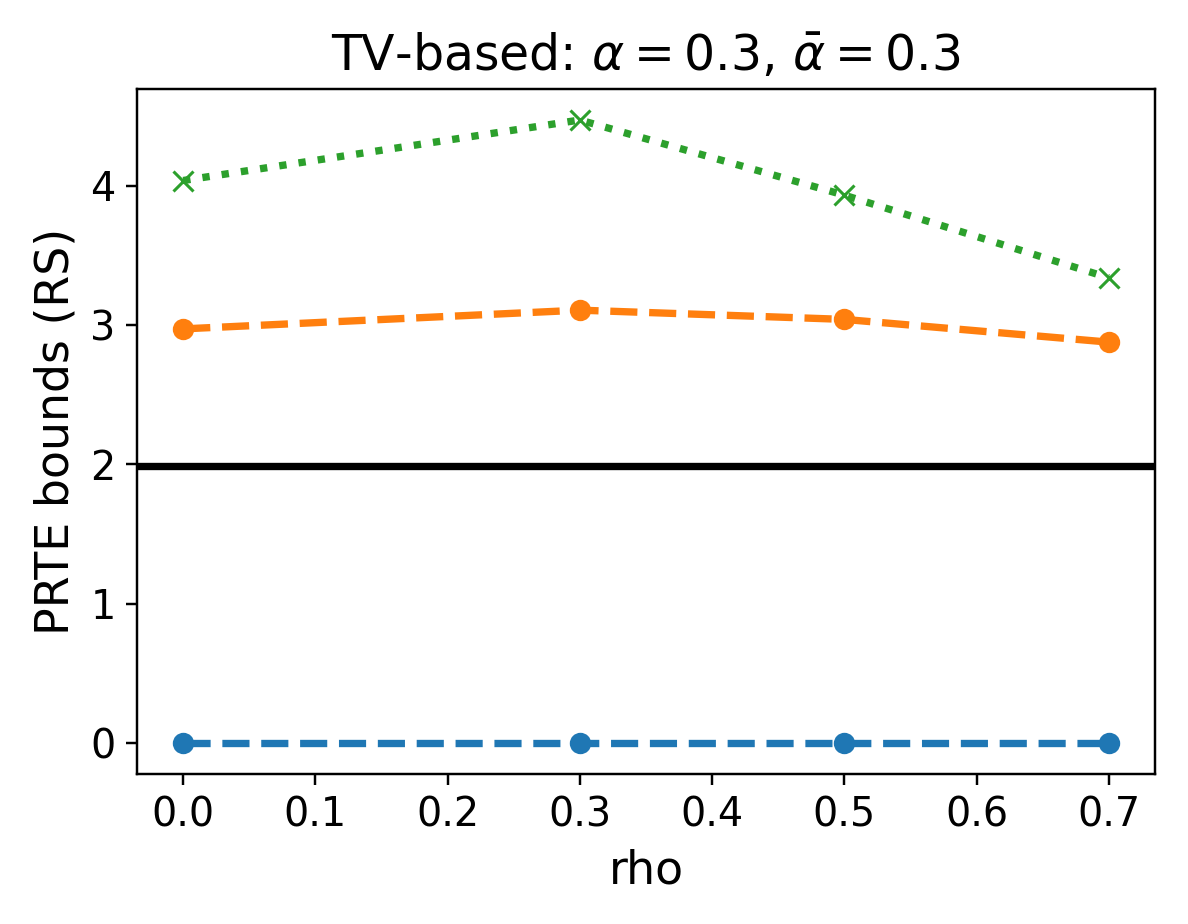}
            \end{minipage}
\caption{TV-based PRTE bounds}
\label{fig.prte.exam}
\figurenote{The black solid line represents the true PRTE. The blue and orange dashed lines represent the approximated lower and upper PRTE bounds. The green dotted line represents the naive PRTE estimand based on the observed treatment $D$.}
\end{figure}

Figure \ref{fig.prte.exam} reports the PRTE bounds for two benchmark configurations, $(\alpha,\bar\alpha)=(0.1,0.1)$ and $(0.3,0.3)$. 
In both cases, the lower bound is equal to zero. This is because under this DGP, the lower bound for $MTE(P(z))$ is always zero for $\alpha>0$, as shown in Figure~\ref{fig.ey1.exam0921}.   
Thus, in this design the PRTE outer bounds from Proposition \ref{prop:identified_set_RS} identify a weakly positive sign, but do not deliver a strictly positive lower bound.

The comparison between the two panels is informative about the role of misclassification. 
For $(\alpha,\bar\alpha)=(0.1,0.1)$, our proposed upper PRTE bound ranges from approximately $2.36$ to $2.54$ over the reported values of $\rho$, while the true PRTE is approximately $1.99$.
For $(\alpha,\bar\alpha)=(0.3,0.3)$, the corresponding upper bound ranges from approximately $2.83$ to $3.08$.
Hence, as expected, higher misclassification rates widen the PRTE outer bounds.
At the same time, the naive PRTE constructed from the observed treatment lies above the upper PRTE bound in the reported cases, which illustrates that ignoring misclassification can generate a substantial upward bias in the policy relevant treatment parameter estimand based on the LIV.
As for the identification of the MTE in Figure \ref{fig.ey1.exam0921}, the PRTE outer bounds do not change monotonically with $\rho$. Note that our PRTE outer bounds cover the true PRTE value as expected. 

\section{Identification under Parametric Misclassification Assumptions}\label{falserates}

In this section, we provide some identification results under flexible parametric assumptions. Before we do that, we discuss that some restrictions imposed in the literature could be too restrictive in our current framework.

Equation (\ref{eq:mis0}) implies
\begin{eqnarray*}
\mathbb P(D=1\vert Z=z) &=& \mathbb P(\varepsilon=0\vert V \leq P(z))\mathbb P(V\leq P(z)) + \mathbb P(\varepsilon=1\vert V > P(z))\mathbb P(V > P(z)),\\
&=& (1-\alpha_1(z)) P(z) +\alpha_0(z) (1-P(z)),\\
&=& (1-\alpha_0(z)-\alpha_1(z)) P(z) + \alpha_0(z),
\end{eqnarray*}
where the second equality holds from Assumption \ref{Cont}, $\alpha_1(z)\equiv \mathbb P(\varepsilon=1\vert V \leq P(z))$ is the false negative misclassification rate, and $\alpha_0(z)\equiv \mathbb P(\varepsilon=1\vert V > P(z))$ is the false positive misclassification rate. Suppose that $\alpha_0(z)+\alpha_1(z) < 1$.\footnote{This constraint is known as the \textit{monotonicity condition} in the misclassification literature and is distinct from the monotonicity restriction imposed on the treatment selection in model (\ref{seq1}).} Then,
\begin{eqnarray*}
P(z)= \frac{\mathbb P(D=1\vert Z=z)-\alpha_0(z)}{1-\alpha_0(z)-\alpha_1(z)}.
\end{eqnarray*}
As is customary in the literature \citep[e.g.,][]{Hausman_al1998}, suppose that the misclassification rates $\alpha_0(z)$ and $\alpha_1(z)$ are constant across $z$. Then the true propensity score $P(z)$ is identified up to the misclassification probabilities $\alpha_0$ and $\alpha_1$ as follows:
\begin{eqnarray*}
P(z)= \frac{\mathbb P(D=1\vert Z=z)-\alpha_0}{1-\alpha_0-\alpha_1}.
\end{eqnarray*}
Lemma \ref{symmetry} below shows that under Assumption \ref{RA} the false positive rate $\alpha_0(z)$ and false negative rate $\alpha_1(z)$ are constant across $z$ if and only if the misclassification is symmetric, i.e., $\alpha_0=\alpha_1=\alpha$. The misclassification is symmetric if $\varepsilon$ is independent of $V$.
\begin{lemma}\label{symmetry}
Suppose Assumption \ref{RA} holds under the model specification \eqref{seq1}. Then, false positive rate $\alpha_0(z)$ and false negative rate $\alpha_1(z)$ do not depend on $z$ if and only if the misclassification is symmetric, i.e., $\alpha_0(z)=\alpha_1(z)$. \end{lemma}
The result of this lemma illustrates that the standard practice which consists of assuming that the false positive rate $\alpha_0(z)$ and false negative rate $\alpha_1(z)$ do not depend on $z$ is restrictive in our setting. There is evidence in the literature that supports this claim. For instance, \citet{HaiderStephens2020} show that the assumption that $\alpha_0(z)$ and $\alpha_1(z)$ are constant in $z$ is invalid in routine empirical settings. For these reasons, we discuss how one can allow the false positive/negative rates to depend on $z$ in a flexible parametric model.

When the misclassification is asymmetric and false positive and false negative rates depend on the instrument, without further assumptions the researcher can use the bounds we derive in Section \ref{Ident}. However, there may exist some parametrization of the misclassification probabilities that can yield tighter bounds and potentially point identification. 

\subsection{Parametric Copula}\label{paracopula}
Consider the structural model for the treatment choice and the misreporting decision
\begin{eqnarray}\label{eq:structuralmis}
\left\{
\begin{array}{lcl}
D^* &=& \mathbbm{1}\{V \leq P(Z)\}, \\[2pt]
D &=& D^*(1-\varepsilon) + (1-D^*)\varepsilon, \\[2pt]
\varepsilon &=& \mathbbm{1}\{\xi \leq \alpha\},
\end{array}
\right.
\end{eqnarray}
where $\xi$ and $V$ are normalized to be uniformly distributed over $[0,1],$ with a \textit{known} parametric copula $C_{\xi,V}(u,v)\equiv C_{\rho}(u,v).$ In this specification, assuming $0<P(z)<1,$ the false negative and positive misclassification rates $\alpha_1(z)$ and $\alpha_0(z)$ are
\begin{eqnarray*}
    \alpha_1(z)=\frac{C_{\rho}(\alpha,P(z))}{P(z)}, \qquad \qquad \alpha_0(z)=\frac{\alpha-C_{\rho}(\alpha,P(z))}{1-P(z)}.
\end{eqnarray*}

From Equation (\ref{eq:mis0}), we have
\begin{eqnarray}
    \mathbb E[D\vert Z=z]&=&\mathbb P(\xi > \alpha, V\leq P(z))+\mathbb P(\xi \leq \alpha, V > P(z)), \nonumber\\
    &=& \mathbb P(V\leq P(z))-\mathbb P(\xi \leq \alpha, V\leq P(z))\nonumber\\
    && + \mathbb P(\xi \leq \alpha)-\mathbb P(\xi \leq \alpha, V\leq P(z)),\nonumber\\
    &=& P(z)+\alpha - 2 C_{\rho}(\alpha,P(z)). \label{eq:copula}
\end{eqnarray}
We consider a Gaussian copula where $C_{\rho}(u,v)=\Phi_2(\Phi^{-1}(u),\Phi^{-1}(v);\rho),$ $\Phi_2$ is the bivariate standard normal cdf, $\Phi$ is the standard normal cdf, and $\Phi^{-1}$ its inverse. The parameter $\rho\in [-1,1]$ is the correlation coefficient. We focus on three scenarios:
\begin{enumerate}[(i)]
\item $V\ \indep\ \xi$\ \  $(\rho=0)$ 

In this case, $C_{0}(u,v)=uv,$ and Equation \eqref{eq:copula} becomes $$\mathbb E[D\vert Z=z]=P(z)+\alpha-2\alpha P(z).$$
Assuming $\alpha \leq \overline{\alpha} < 1/2,$ we have $P(z)\in \left\{\frac{\mathbb E[D\vert Z=z]-\alpha}{1-2\alpha}: 0\leq \frac{\mathbb E[D\vert Z=z]-\alpha}{1-2\alpha} \leq 1\right\}.$

\item $\xi=V$\ \  $(\rho=1)$

This scenario corresponds to the Fr\'echet upper bound copula, $C_{1}(u,v)=\min\{u,v\}.$ Equation \eqref{eq:copula} becomes $$\mathbb E[D\vert Z=z]=P(z)+\alpha-2\min\{\alpha, P(z)\}=\max\{P(z)-\alpha,\alpha-P(z)\},$$
which implies $$
P(z) \in \{\alpha - \mathbb E[D\vert Z=z]: 0 \leq \alpha - \mathbb E[D\vert Z=z] \leq 1 \} \cup \{\alpha + \mathbb E[D\vert Z=z]: 0 \leq \alpha + \mathbb E[D\vert Z=z] \leq 1 \}.
$$

\item $\xi=1-V$\ \ $(\rho=-1)$

This case corresponds to the Fr\'echet lower bound $C_{-1}(u,v)=\max\{u+v-1,0\},$ and Equation \eqref{eq:copula} becomes $$\mathbb E[D\vert Z=z]=P(z)+\alpha-2\max\{\alpha+P(z)-1,0\}=\min\{(1-\alpha)+1-P(z),P(z)+\alpha\},$$
which implies $$
P(z) \in \{E[D\vert Z=z] - \alpha : 0 \leq E[D\vert Z=z] - \alpha  \leq 1 \} \cup \{2- \alpha - \mathbb E[D\vert Z=z]: 0 \leq 2- \alpha - \mathbb E[D\vert Z=z] \leq 1 \}.
$$

\end{enumerate}

\subsection{Parametric Copula and Propensity Score}

In addition to parameterizing the copula between $V$ and $\xi,$ suppose $P(z)=F(\delta' z),$ where $F$ is \textit{known} (e.g., logit, probit).

From Equation (\ref{eq:mis0}), we have
\begin{eqnarray*}
    \mathbb E[D\vert Z=z]           &=& F(\delta' z)+\alpha - 2 C_{\rho}(\alpha,F(\delta' z))\equiv \mu(z,\alpha,\delta,\theta).
\end{eqnarray*}
Let $D=F(\delta' Z)+\alpha - 2 C_{\rho}(\alpha,F(\delta' Z))+e$. Under our model assumptions, $\mathbb E[e\vert Z=z]=0.$ Then $$(\alpha, \delta,\theta)\in \arg\min_{(\tilde{\alpha},\tilde{\delta},\tilde{\theta})} \mathbb E[(D-\mu(Z,\tilde{\alpha},\tilde{\delta},\tilde{\theta}))^2].$$ 
To see this, note that we can write 
\begin{eqnarray*}
    \mathbb E[(D-\mu(Z,\tilde{\alpha},\tilde{\delta},\tilde{\theta}))^2]=\mathbb E[e^2]+\mathbb E[(\mu(Z,\alpha,\delta,\theta)-\mu(Z,\tilde{\alpha},\tilde{\delta},\tilde{\theta}))^2],
\end{eqnarray*}
which is minimized if $\mu(Z,\alpha,\delta,\theta)=\mu(Z,\tilde{\alpha},\tilde{\delta},\tilde{\theta})$ a.s., i.e., $\mathbb E[D\vert Z]=\mu(Z,\tilde{\alpha},\tilde{\delta},\tilde{\theta})$ a.s. We denote $\Theta_I(\alpha,\delta,\theta)$ the set of parameter values $(\tilde{\alpha},\tilde{\delta},\tilde{\theta})$ that minimize $\mathbb E[(D-\mu(Z,\tilde{\alpha},\tilde{\delta},\tilde{\theta}))^2].$ When $\Theta_I(\alpha,\delta,\theta)$ is a singleton, we achieve point identification. The idea we propose here is similar to the point identification setting in \citet{hausman1994misclassification} where false positive and false negative rates do not vary with $z$. These rates are allowed to vary with $z$ in our setting.

\textbf{Additional evidence.}
To keep the exposition focused on the identification argument, we relegate the Monte Carlo exercise and the empirical illustration to Appendix~\ref{sec:mc_copula_validation} and Appendix~\ref{sec:empirical_copula_illustration}, respectively. The simulations assess the finite-sample performance of the proposed sequential estimator, while the empirical illustration applies the copula-based first stage to the earnings application and highlights both its empirical implications and its sensitivity to the imposed parametric restrictions.

\section{Extension to Multivalued Discrete Instruments}\label{ext:discrete}
\begin{assumption}[Discrete instrument]\label{Discrete}
The instrument $Z$ is discrete with support $\{z_1, z_2, \ldots, z_K\}$ and the propensity score $p_\ell \equiv \mathbb P\left[D^*=1 \vert Z=z_\ell\right]$ satisfies $0 \leq p_{1} < p_{2} < \ldots < p_{K} \leq1.$
\end{assumption}
This assumption states that the ordering of the true propensity score is known, but the support $\{z_1, z_2, \ldots, z_K\}$ of the instrument does not necessarily have the same ranking. This assumption does not require monotonicity in the propensity score. The result in Lemma \ref{morett_samedirection} implies that under Assumptions \ref{RA} and \ref{morett}, $p_{\ell}-p_k$ has the same sign as $\mathbb P(D=1 \vert Z=z_{\ell})-\mathbb P(D=1 \vert Z=z_k).$ For example, when the false positive and negative rates do not depend on $z$, we have shown is the previous section that the propensity score can be written as:
\begin{eqnarray*}
P(z)= \frac{\mathbb P(D=1\vert Z=z)-\alpha}{1-2\alpha}.
\end{eqnarray*}
As we can see, in such a case, the ordering of the true propensity score $p_\ell$ is the same as that of the reported propensity score $\mathbb P(D=1\vert Z=z_\ell)$. 
	
We sum up Equations \eqref{eq:fsharp2} and \eqref{eq:fsharp3}, and take the difference for $z_{\ell}$ and $z_{\ell-1}$, respectively. Combining this with the index sufficiency result \eqref{eq:fsharp1}, we have
\begin{eqnarray*}
&&\mathbb P(Y\in A\vert P(Z)=p_{\ell}) - \mathbb P(Y\in A \vert P(Z)=p_{\ell-1})\\
&&\qquad \qquad= \int^{p_{\ell}}_{p_{\ell-1}}\mathbb P\left(Y_1\in A \vert V=v\right)dv - \int_{p_{\ell-1}}^{p_{\ell}}\mathbb P\left(Y_0\in A \vert V=v\right)dv,\\
&& \qquad \qquad= (p_{\ell}-p_{\ell-1})\mathbb P\left(Y_1\in A \vert p_{\ell-1}<V\leq p_{\ell}\right) - (p_{\ell}-p_{\ell-1})\mathbb P\left(Y_0\in A \vert p_{\ell-1}<V\leq p_{\ell}\right)\nonumber.
\end{eqnarray*}
Therefore,
\begin{eqnarray*}
\mathbb P\left(Y_1\in A \vert p_{\ell-1}<V\leq p_{\ell}\right) - \mathbb P\left(Y_0\in A \vert p_{\ell-1}<V\leq p_{\ell}\right)= \frac{\mathbb P(Y\in A\vert P(Z)=p_{\ell}) - \mathbb P(Y\in A \vert P(Z)=p_{\ell-1})}{p_{\ell}-p_{\ell-1}}.
\end{eqnarray*}
The analog of the result holds with expectations. Hence, we identify the MTE up to the function $P(z)$ as follows:
\begin{eqnarray*}
\mathbb E\left[Y_1-Y_0 \vert p_{\ell-1}<V\leq p_{\ell}\right]&=& \frac{\mathbb E[Y\vert P(Z)=p_{\ell}] - \mathbb E[Y \vert P(Z)=p_{\ell-1}]}{p_{\ell}-p_{\ell-1}},\\
&=& \frac{\mathbb E[Y\vert Z=z_{\ell}] - \mathbb E[Y \vert Z=z_{\ell-1}]}{p_{\ell}-p_{\ell-1}}.
\end{eqnarray*}

We have
\begin{eqnarray}\label{eq:newpzbounds}
p_\ell-p_{\ell-1}=\left(p_L-p_1\right)-\sum_{k\neq \ell}\left(p_k-p_{k-1}\right).
\end{eqnarray}
Equations (\ref{eq:pzbounds}) and (\ref{eq:newpzbounds}) imply the following additional bounds on $p_\ell-p_{\ell-1}$:
\begin{eqnarray*}
&&\max_{d\in\{0,1\}}\left\{TV_{(Y,D=d)}(z_1,z_L)\right\}\\
&& -\sum_{k\neq \ell}\min_{d\in\{0,1\}}\left\{\mathbb P(D=d\vert Z=z_{k})+\mathbb P(D=d\vert Z=z_{k-1}),2\alpha+\Delta_{DZ}(z_{k-1},z_k), 2(1-\alpha)-\Delta_{DZ}(z_{k-1},z_k)\right\}\nonumber\\
 && \qquad \leq p_\ell-p_{\ell-1} \leq\\ 
 && \qquad \qquad \min_{d\in\{0,1\}}\left\{\mathbb P(D=d\vert Z=z_L)+\mathbb P(D=d\vert Z=z_1),2\alpha+\Delta_{DZ}(z_1,z_L), 2(1-\alpha)-\Delta_{DZ}(z_1,z_L)\right\} \\
 && \qquad \qquad \qquad \qquad  - \sum_{k\neq \ell}\max_{d\in\{0,1\}}\left\{TV_{(Y,D=d)}(z_{k-1},z_k)\right\}. 
\end{eqnarray*}
This latter implication is missing from \citet{tommasi2024bounding} as they do not utilize the restriction imposed by Equation \eqref{eq:newpzbounds}. Combining it with the bounds in Equation (\ref{eq:pzbounds}) yields the following bounds for $p_\ell-p_{\ell-1}$: 
\begin{eqnarray*}
LB_p(z_{\ell-1},z_\ell) \leq p_\ell-p_{\ell-1}  \leq UB_p(z_{\ell-1},z_\ell),
\end{eqnarray*}
 where the bounds are defined in Appendix \ref{anaboundsdiscrete}.
The proposition below holds.
\begin{proposition}\label{propdiscrete}
Suppose that model (\ref{seq1}) along with Assumptions \ref{RA}--\ref{Cont_inst}, and \ref{Discrete} hold. Then, we have the following bounds for $LATE(p_{\ell-1},p_{\ell})\equiv\mathbb E\left[Y_1-Y_0 \vert p_{\ell-1}<V\leq p_{\ell}\right]$:
\begin{eqnarray}
&& \min\left\{\frac{\Delta_{YZ}(z_\ell,z_{\ell-1})}{UB_p(z_\ell,z_{\ell-1})}, \frac{\Delta_{YZ}(z_\ell,z_{\ell-1})}{LB_p(z_\ell,z_{\ell-1})}\right\} \nonumber\\
&& \qquad \qquad \leq LATE(p_{\ell-1},p_{\ell})\leq \\
&& \qquad \qquad \qquad \qquad \max\left\{\frac{\Delta_{YZ}(z_\ell,z_{\ell-1})}{UB_p(z_\ell,z_{\ell-1})}, \frac{\Delta_{YZ}(z_\ell,z_{\ell-1})}{LB_p(z_\ell,z_{\ell-1})}\right\}. \nonumber
\end{eqnarray}
\end{proposition}
At this point, we do not have a result on the sharpness of the bounds in Proposition~\ref{propdiscrete}. This could be investigated in future work. However, when $\alpha$ is completely unknown (i.e., $\bar{\alpha}=1$), these bounds are tighter than the existing bounds in \citet{tommasi2024bounding}.

\section{Empirical Illustration: Returns to Upper Secondary Schooling in Indonesia}\label{App}
To illustrate our methodology, we use data from the third wave of the Indonesia Family Life Survey (IFLS) fielded from June through November 2000. We build upon \citet{Carneiroal2017} who estimate average and marginal returns to schooling in Indonesia using a semiparametric selection model. The authors use exogenous geographic variation in access to upper secondary schools to identify their model when ignoring the presence of measurement errors in the treatment variable. In their analysis, these researchers control for several family and village characteristics, namely father’s and mother’s education, an indicator of whether the community of residence was a village, religion, whether the location of residence is rural, province dummies, and distance from the village of residence to the nearest health post.

 The IFLS is a household and community level panel survey that was conducted in 1993, 1997 and 2000. The sample was drawn from 321 randomly selected villages, spread among 13 Indonesian provinces containing 83\%  of the population, and consists of males aged 25--60 employees in public and private sectors. Females are excluded from the sample because of low labor force participation, self-employed workers are also excluded because it is difficult to measure their earnings. The sample size is 2608.

Following \citet{Carneiroal2017}, we define the dependent variable in the  analysis as the log of the hourly wage $(Y)$, which is constructed from self-reported monthly wages and hours worked per week. The treatment variable $(D)$ is the indicator that the individual has an upper secondary or higher education (i.e., s/he completed at least 10 years of education). As we argue in Example~\ref{ex1}, people often misreport their education level. So, we observe their true education level with some measurement errors.
The control variables $(X)$ are indicator variables for age, indicators for the level of schooling completed by each of the parents (no education, elementary education, secondary education, and an indicator for unreported parental education), an indicator for whether the individual was living in a village at age 12, indicators for the province of residence, an indicator of rural residence, and distance (in kilometers) from the office of the head of the community of residence to the nearest community health post. 

The instrumental variable $(Z)$ for schooling is the distance (in kilometers) from the office of the community head to the nearest secondary school. The main assumption from \citet{Carneiroal2017} is that if we consider two individuals with equally educated parents, with the same religion, living in a village which is located in an area that is equally rural, in the same province, and at the same distance of a health post, then distance to the nearest secondary school is uncorrelated with direct determinants of wages other than schooling. 
The authors present evidence that this assumption is likely to hold, suggesting that the IV is valid. In particular, they show that, once the listed control variables are held fixed, there is no dependence between the distance to the nearest secondary school and whether the individual ever failed a grade in elementary school, how many times he repeated a grade in elementary school, and whether he had to work while attending elementary school. In addition, they show (using a different sample) that the distance variable is unrelated to test scores (Math, Bahasa, Science, and Social Studies) in elementary school.  The validity of the distance-to-secondary-school instrument remains a maintained and contestable assumption, so this exercise should be interpreted as illustrative.

\subsection*{Estimation Results}
The estimation procedure follows the parametric copula approach discussed in Subsection~\ref{paracopula}, with each first-stage and second-stage object estimated from the observed data. Let
$q_i \equiv \widehat{\Pr}(D_i=1\mid Z_i,X_i)$
denote the fitted observed-treatment propensity score from the logit first stage. The benchmark specification imposes symmetric misclassification with \(\varepsilon\ \indep\ V\) $(\rho=0)$. In this case, the estimated propensity score for each misclassification rate $\alpha$ is
$
P_i(\alpha)=\frac{q_i-\alpha}{1-2\alpha}.
$
Appendix~\ref{apx:sup} reports two additional specifications that allow \(\varepsilon\) to depend on the latent cost variable. These specifications include \(\varepsilon=\mathbbm{1}\{V\leq\alpha\}\) $(\rho=1)$, which concentrates misclassification at lower values of \(V\), and \(\varepsilon=\mathbbm{1}\{V>1-\alpha\}\) $(\rho=-1)$, which concentrates misclassification at higher values of \(V\). The implemented estimated propensity scores are\footnote{
For \(\rho=\pm1\), the map from \(P(z)\) to \(Q(z)\) is generally not one-to-one; i.e., for some $q_i$, there could be multiple compatible $P_i$.
The full inversion is therefore set-valued pointwise and, without additional functional restrictions, creates a branching problem over the support of \(Z\).
In this empirical illustration, we do not solve this full set-valued inversion.
Instead, we report two single-branch transformations as sensitivity specifications.
A pure full inversion is generally not tractable but could be implemented by imposing a parametric form for \(P(z)\) or by searching over monotone admissible paths.
}
\[
P_i^{\leq}(\alpha)=
\begin{cases}
\alpha-q_i, & q_i<\alpha,\\
q_i+\alpha, & q_i\ge \alpha,
\end{cases}
\qquad
P_i^{>} (\alpha)=
\begin{cases}
q_i-\alpha, & q_i<1-\alpha,\\
1-q_i+1-\alpha, & q_i\ge 1-\alpha.
\end{cases}
\]

For each specification, the transformed propensity score is winsorized to \([10^{-4},1-10^{-4}]\) before MTE and aggregate parameters are computed.

\subsubsection*{Choice of \(\bar{\alpha}\)} 
Following the discussion in Section~\ref{anaF} and the comments below Assumption~\ref{Bound:mis}, we assume that a person with at least upper secondary education in Indonesia truthfully reports this achievement. Hence, the false negative rate \(\Pr(\varepsilon=1\mid D^*=1)\) is set to zero. Following \citet{Black_etal2003}, we also assume that among those who report at least upper secondary schooling, the proportion misclassified is bounded by one third, \(\Pr(\varepsilon=1\mid D=1)\leq 1/3\). Therefore,
\[
\alpha=\Pr(\varepsilon=1)
=\Pr(\varepsilon=1\mid D=1)\Pr(D=1)
+\Pr(\varepsilon=1\mid D^*=1)\Pr(D^*=1)
\leq \frac{1}{3}\Pr(D=1).
\]
In the estimation sample, \(\Pr(D=1)=1085/2608=0.4160\), so the data-based upper bound is \(\Pr(D=1)/3=0.138676\), reported as \(\bar{\alpha}\approx0.139\). We also report sensitivity exercises for \(\bar{\alpha}\in\{0.100,0.050,0.025\}\).

\subsubsection*{Implementation}
For each \(\bar{\alpha}\), the code evaluates a grid of admissible \(\alpha\) values with step size 0.01 and includes the exact endpoint \(\bar{\alpha}\). The reported envelope for each \(\bar{\alpha}\) uses every computed grid point satisfying \(\alpha\leq\bar{\alpha}\). This convention preserves nesting of the identified sets and confidence regions across \(\bar{\alpha}\).

The observed-treatment propensity score \(q_i\) is estimated by a logit regression of \(D_i\) on distance to the nearest secondary school, predetermined covariates, and the distance--covariate interaction terms in the empirical specification of \citet{Carneiroal2017}. For each misclassification specification and each \(\alpha\), the fitted value \(q_i\) is mapped into a candidate true propensity score by the displayed formula and trimmed to \([10^{-4},1-10^{-4}]\), following the endpoint trimming convention in \citet{carneirolee2009}.

To control for exogenous covariates, the code applies Robinson residualization within each candidate \(P_i(\alpha)\), following \citet{robinson1988} and the empirical implementation in \citet{Carneiroal2017}. Specifically, write
\[
Y_1=\lambda_1+X'\beta_1+U_1,
\qquad
Y_0=\lambda_0+X'\beta_0+U_0,
\]
with \((U_0,U_1)\indep(Z,X)\). Since \(Y=Y_0+(Y_1-Y_0)D^*\),
\[
\begin{aligned}
\mathbb E[Y\mid X,P]
&=\lambda_0+X'\beta_0+P(\lambda_1-\lambda_0)
+PX'(\beta_1-\beta_0)+K(P),
\end{aligned}
\]
where \(K(P)\) collects the nonparametric selection component. Hence,
\[
MTE(x,p)=(\lambda_1-\lambda_0)+x'(\beta_1-\beta_0)+K'(p).
\]
Operationally, for each candidate \(P_i(\alpha)\), the code residualizes \(Y_i\), each component of \(X_i\), and each component of \(P_iX_i\) on \(P_i\), regresses residualized \(Y_i\) on residualized \(X_i\) and \(P_iX_i\), constructs
\[
\widehat R_i=Y_i-X_i'\widehat\beta_0-P_iX_i'(\widehat\beta_1-\widehat\beta_0),
\]
and estimates the derivative of \(\mathbb E[\widehat R_i\mid P_i=p]\). The reported average MTE curve adds \(\bar X'(\widehat\beta_1-\widehat\beta_0)\) and divides by the estimated education-year difference, 7.792, as in \citet{Carneiroal2017}.

The Gaussian branch uses a local quadratic Gaussian-kernel estimator with bandwidth 0.27, as in \citet{Carneiroal2017} and consistent with the local-polynomial recommendation in \citet{fan1996local} for derivative estimation. The spline branch uses a degree-2 penalized B-spline with five interior knots and a second-difference penalty \(\lambda=10\). For the spline branch, the knots are chosen from the estimation-sample distribution of the relevant \(P_i(\alpha)\) and held fixed across bootstrap draws for that misclassification-specification--method--alpha configuration. In both branches, the same smoothing rule is used in the Robinson residualization step and in the final derivative step.

The aggregate parameters are computed by the simulation-based integration approach used by \citet{Carneiroal2017}, where we evaluate the estimated individual-level MTE surface on a uniform \(V\)-grid of 1000 points. With \(\widehat m_i(v)\) denoting the estimated MTE for individual \(i\) at margin \(v\), we have
\[
\hat{ATE}=\mathbb E[\widehat m_i(V)],
\qquad
\hat{ATT}=\mathbb E[\widehat m_i(V)\mid P_i>V],
\qquad
\hat{ATU}=\mathbb E[\widehat m_i(V)\mid P_i<V].
\]
We also report \(\hat{AMTE}_\zeta=\mathbb E[\widehat m_i(V)\mid |P_i-V|<\zeta]\) for \(\zeta\in\{0.10,0.05,0.01\}\); the main table displays \(\zeta=0.05\). 
As in Subsection \ref{sec:prteid}, we consider a policy relevant treatment effect parameter for a counterfactual policy that induces $100a\%$ of the currently untreated population into treatment. More precisely, we consider a counterfactual policy that exposes fraction $a$ of people with less than upper secondary schooling to upper secondary or higher education: $
P^a(z)=P(z)+a(1-P(z))$.\footnote{The same policy has been considered in \citet{SasakiUra2021}. \citet{SasakiUra2021} estimate the PRTE on the same data using a double-debiased orthogonal-score estimator. Here the PRTE is recovered from the MTE surface with cluster-bootstrap inference}
We report the PRTE values for each $a\in\{0.1,0.2,0.3\}$.
The aggregate parameters are computed by the simulation-based integration approach used by \citet{Carneiroal2017}. See Appendix \ref{sec:implementation} for more details on the implementation. 

Inference uses 1000 cluster bootstrap replications at the community level. Each bootstrap draw re-estimates the logit first stage, the Robinson residualization, and the second-stage MTE. For MTE curves, the confidence region is the bootstrap envelope over the admissible alpha grid. For aggregate parameters, the confidence region is the outer union of alpha-wise bootstrap intervals over \(\alpha\in[0,\bar{\alpha}]\). This rule is conservative and is used in the table and aggregate-summary figures.

\begin{figure}[!htbp]
\centering
\includegraphics[width=\textwidth]{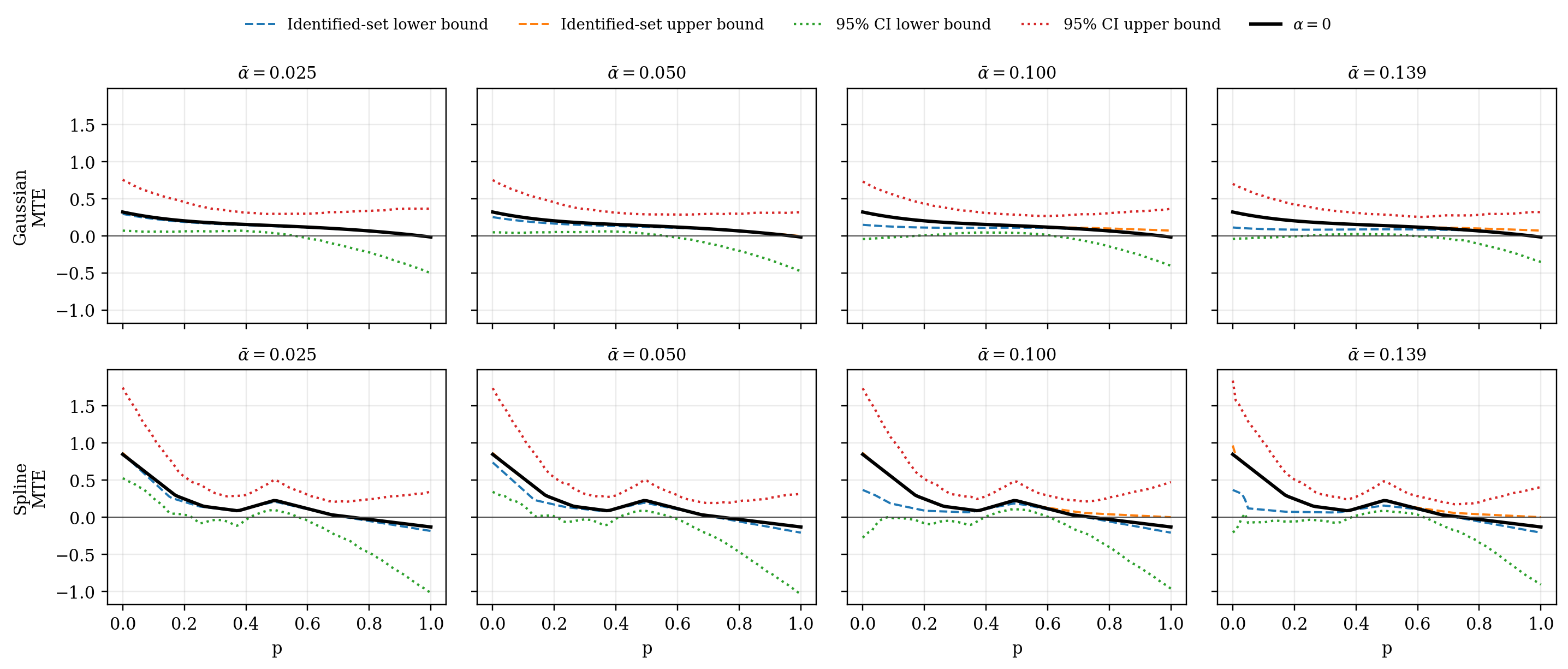}
\caption{MTE identified sets and conservative 95\% confidence regions under $\varepsilon\ \indep\ V$.}
\label{fig:main-mte-exogenous}
\figurenote{The upper row reports the Gaussian local-quadratic estimator and the lower row reports the degree-2 penalized B-spline estimator. The columns correspond to $\bar\alpha\in\{0.025,0.050,0.100,0.139\}$. Dashed curves are the lower and upper endpoints of the identified set, dotted curves are the lower and upper endpoints of the conservative 95\% confidence region, and the solid black curve is the estimate at $\alpha=0$. The confidence region uses 1000 community-level cluster-bootstrap replications and the alpha-grid envelope. The horizontal axis is the latent true propensity score $p$.}
\end{figure}

\input{tables/refined/table_main_aggregate_exogenous.tex}

\begin{figure}[!htbp]
\centering
\includegraphics[width=0.92\textwidth]{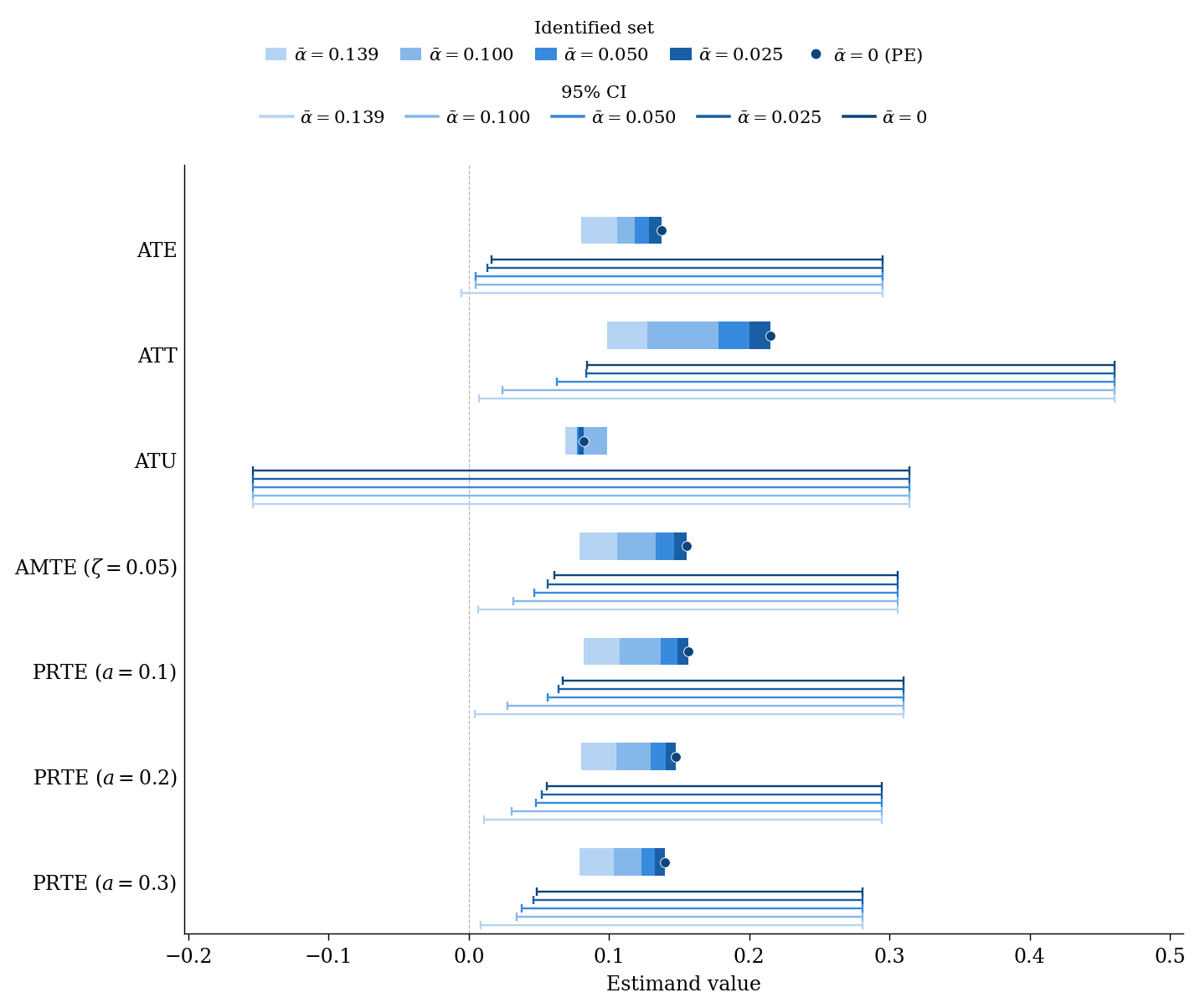}
\caption{Aggregate-return summaries under $\varepsilon\ \indep\  V$ using the Gaussian local-quadratic estimator.}
\label{fig:main-aggregate-exogenous-gaussian}
\figurenote{Each row is an estimand. Shaded horizontal bands report identified sets for $\bar\alpha\in\{0.139,0.100,0.050,0.025\}$, with lighter shades for larger upper bounds and darker shades for smaller upper bounds. The dot reports the point estimate at $\alpha=0$. Stacked horizontal intervals report conservative 95\% confidence regions for $\bar\alpha\in\{0.139,0.100,0.050,0.025\}$ and the 95\% confidence interval for $\alpha=0$, using 1000 community-level cluster-bootstrap replications.}
\end{figure}

\begin{figure}[!htbp]
\centering
\includegraphics[width=0.92\textwidth]{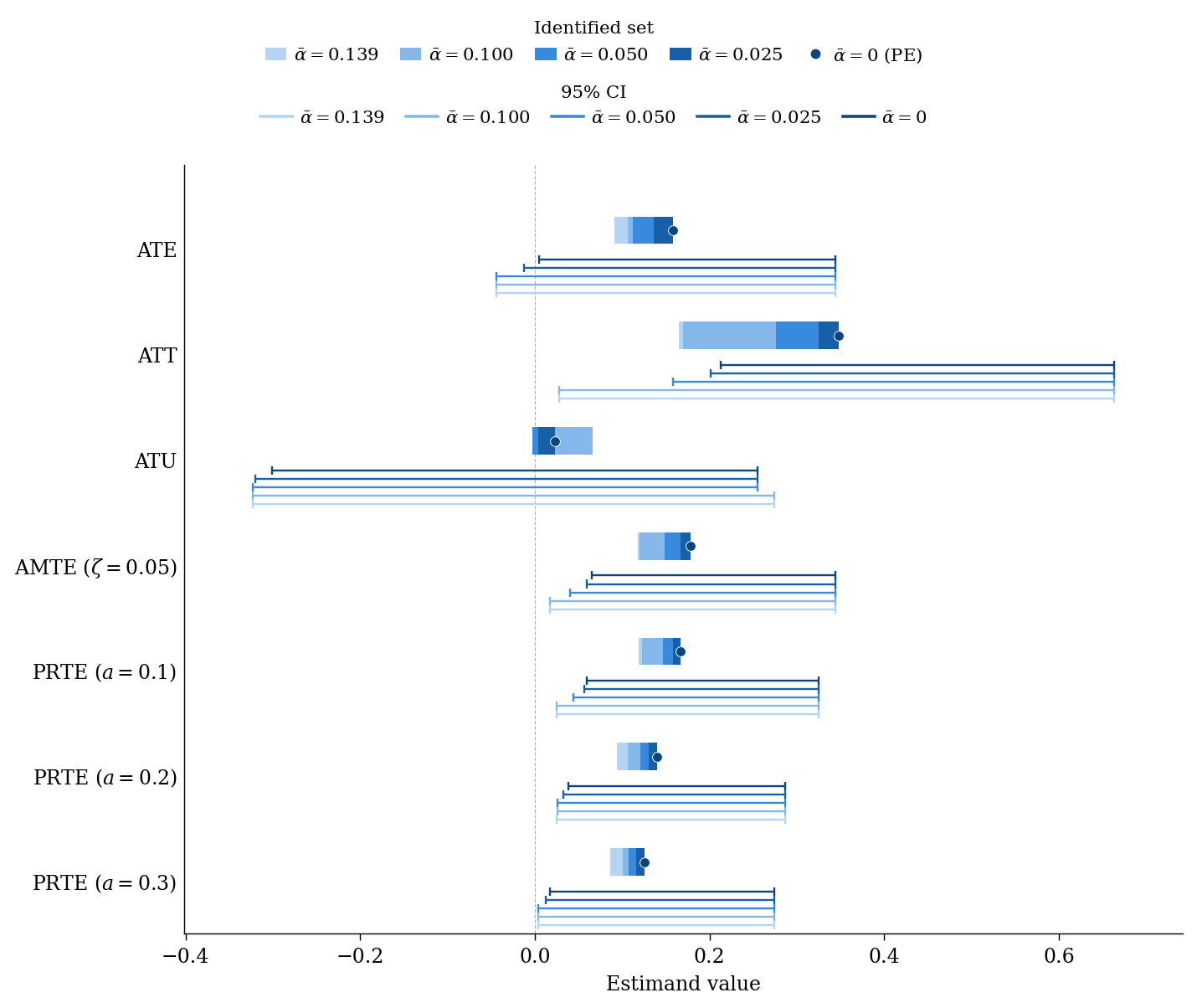}
\caption{Aggregate-return summaries under $\varepsilon\ \indep\ V$ using the degree-2 penalized B-spline estimator.}
\label{fig:main-aggregate-exogenous-spline}
\figurenote{Each row is an estimand. Shaded horizontal bands report identified sets for $\bar\alpha\in\{0.139,0.100,0.050,0.025\}$, with lighter shades for larger upper bounds and darker shades for smaller upper bounds. The dot reports the point estimate at $\alpha=0$. Stacked horizontal intervals report conservative 95\% confidence regions for $\bar\alpha\in\{0.139,0.100,0.050,0.025\}$ and the 95\% confidence interval for $\alpha=0$, using 1000 community-level cluster-bootstrap replications.}
\end{figure}

\subsection*{Discussion}
Figure~\ref{fig:main-mte-exogenous} shows that the estimated MTE is generally decreasing in the latent true propensity score \(p\). The returns are largest for individuals at low cost margins and decline as \(p\) increases. This pattern is visible under both the Gaussian local-quadratic and degree-2 penalized B-spline estimators. As expected, higher values of \(\bar{\alpha}\) lead to wider identified sets.

Table~\ref{tab:main-aggregate-exogenous} and Figures~\ref{fig:main-aggregate-exogenous-gaussian}--\ref{fig:main-aggregate-exogenous-spline} summarize the average and policy-relevant parameters under \(\varepsilon\ \indep\ V\). At \(\alpha=0\), the ATE, ATT, AMTE, and PRTE estimates are positive for both estimators. The ATT is statistically significant while the ATU is not. Given the upper bound on the misclassification rate \(\bar{\alpha}=0.139\), the Gaussian-kernel estimated identified sets are \([0.080,0.137]\) for ATE, \([0.099,0.215]\) for ATT, and \([0.069,0.099]\) for ATU. The corresponding spline estimated identified sets are \([0.091,0.158]\), \([0.165,0.348]\), and \([-0.003,0.067]\), respectively. The confidence regions are mostly significant for all parameters except the ATU, suggesting statistically significant returns to upper secondary schooling. The 95\% confidence region for the ATT under the spline specification suggests that the average return for individuals with upper secondary schooling varies between 2.8\% and 66.3\%. 

The 95\% confidence sets for PRTE under the spline specification when $a=0.1$ suggest that inducing 10\% of individuals who currently do not have an upper secondary education to reach this education level will increase the average return by 2.5\% to 32.5\%. These ranges decrease as the share $a$ increases but remain statistically significant.  
This is consistent with larger shifts drawing in higher-propensity individuals whose estimated marginal returns are smaller.\footnote{These estimates should be interpreted as illustrative because the validity of the distance instrument could be questionable.}

Appendix~\ref{apx:sup} shows the results for the two specifications in which measurement error depends on \(V\): \(\varepsilon=\mathbbm{1}\{V\leq\alpha\}\) ($\rho=1,$ perfect positive dependence), and \(\varepsilon=\mathbbm{1}\{V>1-\alpha\}\) ($\rho=-1,$ perfect negative dependence). Although the MTE follows a pattern similar to that in the independence case $(\varepsilon\ \indep\ V)$ and the estimated identified sets show positive values,  the 95\% confidence sets for most parameters including the PRTE are not statistically significant. 
This suggests that the conclusions about the PRTE and other causal parameters could sensitive to the dependence between the unobservables that drive selection into treatment and misreporting.

\section{Conclusion}\label{conclusion}
In this paper, we show that the MTE is generally partially identified in the presence of misclassification. We start by deriving bounds on the (generalized) LATE. We then exploit the relationship between the MTE and the generalized LATE to bound the MTE. Indeed, under some standard assumptions, the MTE is a limit of the ratio of the variation in the conditional expectation of the observed outcome given the instrument to the variation in the true propensity score, which is partially identified. We provide nonparametric characterization of the identified set for the propensity score and the MTE. We show under some mild regularity conditions that the sign of the MTE is \textit{locally identified}. We use our MTE bounds to derive bounds on other commonly used parameters in the literature, namely the PRTE. We show that  our bounds are tighter than the existing bounds for the local average treatment effect. We illustrate the methodology numerically and empirically. We investigate the measurement of the return to upper secondary schooling in Indonesia, and find that the return is heterogeneous for people at the cost margin. Overall, marginal returns seem weakly decreasing with the schooling cost.   

We have not developed a formal inference method for the analytical bounds for the MTE and PRTE in this work. We believe that constructing a confidence set for these bounds could be worth exploring in future research. Also, future research could explore identification of the MTE in the presence of misclassification when imperfect instruments are available. 

\bibliographystyle{jpe}
\bibliography{mybib}

\appendix

\section{Discussion about the model specification}\label{apx:spec}
One might think that the specification $D=D^* (1-\varepsilon) +(1-D^*)\varepsilon$ is too restrictive. But, without additional restrictions, this representation always holds. To show this, we prove the following lemma.
\begin{lemma}\label{misc:specific}
For any two binary random variables $D$ and $D^*$, there exists a binary random variable $\varepsilon$ such that 
\begin{eqnarray}
D=D^* (1-\varepsilon) +(1-D^*)\varepsilon.
\end{eqnarray}
\end{lemma}

\begin{proof}
Let $D$ and $D^*$ be two binary variables. We can write $D=D^*+(D-D^*)=D^*+\xi$, where $\xi=D-D^*$. Since $D$ and $D^*$ are binary, we have $$\left(D,D^*,\xi\right) \in \left\{(0,0,0),(1,1,0),(1,0,1),(0,1,-1)\right\}.$$ We can see that $D=1-D^*$ if $\xi\in\{-1,1\}$ and $D=D^*$  if $\xi\in\{0\}$. Hence, we can write $D=D^*\left(\mathbbm{1}\left\{\xi\in\{0\}\right\}\right)+(1-D^*) \mathbbm{1}\left\{\xi\in\{-1,1\}\right\}$. By setting $\varepsilon=\mathbbm{1}\left\{\xi\in\{-1,1\}\right\}$, we have $D=D^* (1-\varepsilon) +(1-D^*)\varepsilon$, because $\mathbbm{1}\left\{\xi\in\{-1,1\}\right\}+\mathbbm{1}\left\{\xi\in\{0\}\right\}=\mathbbm{1}\left\{\xi\in\{-1,0,1\}\right\}=1$. 
\end{proof}

\subsection{Comparison between \citeauthor{Ura2018}'s assumptions and ours}
Below is an example of DGP where our model assumptions fail while Ura's hold.
\begin{example}\label{ex:comparison}
  Let $V\sim \mathrm{Unif}(0,1)$ and  $Z\sim\mathrm{Bernoulli}(1/2)$, and $D^*=\mathbbm{1}\{V\le P(Z)\}$ with $P(0)=0.2,$ and $P(1)=0.8$. 
        Also, let $D = D_1D^* + D_0 (1-D^*)$ with $D_1 \sim {\rm Bernoulli}(0.95)$ and $D_0 \sim {\rm Bernoulli}(0.2)$, and $Y = Y_1D^* +Y_0(1-D^*)$ with $Y_1 = 1 + U$, $Y_0 = U$, $Z\ \indep\  (U,V,D_0,D_1),$ and $V\ \indep\ (D_0,D_1)$.

        Then, we have
        \begin{align*}
        \mathbb P(\varepsilon = 1 \mid Z=z ) &= \mathbb P(\varepsilon = 1, D^*=1 \mid Z=z )+\mathbb P(\varepsilon = 1, D^*=0 \mid Z=z ) \\
         &= \mathbb P(D = 0, D^*=1 \mid Z=z )+\mathbb P(D = 1, D^*=0 \mid Z=z ) \\
          &= \mathbb P(D_1 = 0, D^*=1 \mid Z=z )+\mathbb P(D_0 = 1, D^*=0 \mid Z=z ) \\
          &= \mathbb P(D^*=1 \mid D_1=0, Z=z )\mathbb P(D_1=0 \mid Z=z ) \\
            & \quad + \mathbb P(D^*=0 \mid D_0 = 1, Z=z )\mathbb P(D_0 = 1\vert Z=z)\\
         &= \mathbb P(D^*_z = 1)\mathbb P(D_1=0) \\
            & \quad + [1-\mathbb P(D^*_z = 1)]\mathbb P(D_0=1),\\
            &=0.2 - 0.15\cdot \mathbb P(D^* = 1 \mid Z=z),
        \end{align*}
where the first equality holds from the law of total probability, the second and third from the definition of the model, the fourth from Bayes' rule, the fifth from the two independence assumptions, and the last from the model parameters.
        
        Therefore, $Z\nindep \varepsilon$ and our model assumptions fail, while \citeauthor{Ura2018}'s hold.  
\end{example}

\section{Proof of Proposition \ref{prop:LATE}}\label{proofUra}

\subsection{Validity of the bounds}
First, note that 
\begin{align*}
& \mathbb{P}(Y \in A, D=1 \mid Z=1) \\
&= \mathbb{P}(Y \in A, D=1, \varepsilon=1 \mid Z=1)+\mathbb{P}(Y \in A, D=1, \varepsilon=0 \mid Z=1), \\
&= \mathbb{P}\left(Y_0 \in A, D^*=0, \varepsilon=1 \mid Z=1\right)+\mathbb{P}\left(Y_1 \in A, D^*=1, \varepsilon=0 \mid Z=1\right), \\
&= \mathbb{P}\left(Y_0 \in A, D_1^*=0, \varepsilon=1 \mid Z=1\right)+\mathbb{P}\left(Y_1 \in A, D_1^*=1, \varepsilon=0 \mid Z=1\right), \\
&= \mathbb{P}\left(Y_0 \in A, D_1^*=0, \varepsilon=1\right)+\mathbb{P}\left(Y_1 \in A, D_1^*=1, \varepsilon=0\right), \\
&= \mathbb{P}\left(Y_0 \in A, D_1^*=0, D_0^*=0, \varepsilon=1\right)+\mathbb{P}\left(Y_1 \in A, D_1^*=1, D_0^*=1, \varepsilon=0\right), \\
& \quad \quad \quad + \mathbb{P}\left(Y_1 \in A, D_1^*=1, D_0^*=0, \varepsilon=0\right).
\end{align*}

Moreover, we have
\begin{align*}
&\mathbb{P}(Y \in A, D=1 \mid Z=0) \\
& = \mathbb{P}\left(Y_0 \in A, D_0^*=0, \varepsilon=1\right)+\mathbb{P}\left(Y_1 \in A, D_0^*=1, \varepsilon=0\right), \\
& =\mathbb{P}\left(Y_0 \in A, D_1^*=1, D_0^*=0, \varepsilon=1\right)+\mathbb{P}\left(Y_0 \in A, D_1^*=0, D_0^*=0, \varepsilon=1\right), \\
& \quad \quad \quad +\mathbb{P}\left(Y_1 \in A, D_1^*=1, D_0^*=1, \varepsilon=0\right).
\end{align*}

Hence, we have 
\begin{align*}
&\mathbb{P}(Y \in A, D=1 \mid Z=1)-\mathbb{P}(Y \in A, D=1 \mid Z=0) \\
&= \mathbb{P}\left(Y_1 \in A, D_1^*=1, D_0^*=0, \varepsilon=0\right)-\mathbb{P}\left(Y_0 \in A, D_1^*=1, D_0^*=0, \varepsilon=1\right).
\end{align*}

The density version of this would be
\begin{equation}
f_{Y, D\vert Z}(y, 1 \mid 1)-f_{Y, D\vert Z}(y, 1 \mid 0) =f_{Y_1, T, \varepsilon}(y, c, 0)-f_{Y_0, T, \varepsilon}(y, c, 1), \label{eq:late.1}
\end{equation}
where $T \equiv\left(D_0^*, D_1^*\right) \in \{(0,1),(1,1),(0,0)\} \equiv \{c, a, n\} $.

In a similar manner, we have
\begin{align*}
&\mathbb{P}(Y \in A, D=0 \mid Z=1) \\
& =\mathbb{P}\left(Y_1 \in A, D_1^*=1, \varepsilon=1\right)+\mathbb{P}\left(Y_0 \in A, D_1^*=0, \varepsilon=0\right), \\
& =\mathbb{P}\left(Y_1 \in A, D_1^*=1, D_0^*=0, \varepsilon=1\right)+\mathbb{P}\left(Y_1 \in A, D_1^*=1,D_0^*=1, \varepsilon=1\right) \\
& \quad \quad \quad +\mathbb{P}\left(Y_0 \in A, D_1^*=0, D_0^*=0, \varepsilon=0\right),
\end{align*}
and
\begin{align*}
&\mathbb{P}(Y \in A, D=0 \mid Z=0) \\
& = \mathbb{P}\left(Y_1 \in A, D_0^*=1, \varepsilon=1\right)+\mathbb{P}\left(Y_0 \in A, D_0^*=0, \varepsilon=0\right), \\
& = \mathbb{P}\left(Y_1 \in A, D_1^*=1, D_0^*=1, \varepsilon=1\right)+\mathbb{P}\left(Y_0 \in A, D_1^*=0, D_0^*=0, \varepsilon=0\right) \\
& \quad \quad \quad +\mathbb{P}\left(Y_0 \in A, D_1^*=1, D_0^*=0, \varepsilon=0\right).
\end{align*}

Thus, we have
\begin{align*}
&\mathbb{P}(Y \in A, D=0 \mid Z=1)-\mathbb{P}(Y \in A, D=0 \mid Z=0) \\
& =\mathbb{P}\left(Y_1 \in A, D_1^*=1, D_0^*=0, \varepsilon=1\right)-\mathbb{P}\left(Y_0 \in A, D_1^*=1, D_0^*=0, \varepsilon=0\right),
\end{align*}
or
\begin{equation}
f_{Y, D\vert Z}(y, 0 \mid 1)-f_{Y, D \vert Z}(y, 0 \mid 0) =f_{Y_1, T, \varepsilon}(y, c, 1)-f_{Y_0, T, \varepsilon}(y, c, 0). \label{eq:late.2}
\end{equation}

Therefore, by triangle inequality we have
$$
\left|f_{Y, D\vert Z}(y, 1 \mid 1)-f_{Y, D\vert Z}(y, 1 \mid 0)\right| \leq f_{Y_1, T, \varepsilon}(y, c, 0) + f_{Y_0, T, \varepsilon}(y, c, 1),
$$
and
$$
\left|f_{Y, D\vert Z}(y, 0 \mid 1)-f_{Y, D\vert Z}(y, 0 \mid 0)\right| \leq f_{Y_1, T, \varepsilon}(y, c, 1) + f_{Y_0, T, \varepsilon}(y, c, 0).
$$

By taking the integral with respect to $y$ for each of them, we have
\begin{align*}
\int \left| f_{Y, D\vert Z}(y,1 \mid 1)-f_{Y, D\vert Z}(y, 1 \mid 0) \right| d \mu_Y(y)  &  \leq \int\left[f_{Y_0, T, \varepsilon}(y, c, 1)+f_{Y_1, T, \varepsilon}(y, c, 0)\right] d \mu_Y(y), \\
& =\mathbb{P}(T=c, \varepsilon=1)+\mathbb{P}(T=c, \varepsilon=0), \\
& =\mathbb{P}(T=c),
\end{align*}
and
\begin{align*}
\int\left|f_{Y, D\vert Z}(y, 0 \mid 1)-f_{Y, D\vert Z}(y, 0 \mid 0)\right| d \mu_Y(y)  & \leq \int\left[f_{Y_1, T, \varepsilon}(y, c, 1)+f_{Y_0, T, \varepsilon}(y, c, 0)\right] d \mu_Y(y), \\
& =\mathbb{P}(T=c).
\end{align*}

Therefore, we have
\begin{eqnarray*}
\max_{d\in\{0,1\}} \left\{\int \left| f_{Y, D\vert Z}(y,d \mid 1)-f_{Y, D\vert Z}(y, d \mid 0) \right| d \mu_Y(y) \right\}  \leq \mathbb{P}(T=c),
\end{eqnarray*}
that is, $\max_{d\in\{0,1\}}\{TV_{(Y,D=d)}(0, 1)\} \leq \mathbb P(T=c).$
In the main text, we prove the validity of the upper bound,
\begin{eqnarray*}
    \mathbb P(T=c) \leq \min_{d\in\{0,1\}}\left\{\mathbb P(D=d\vert Z=1)+\mathbb P(D=d\vert Z=0)\right\}.
\end{eqnarray*}

Under the standard assumptions of LATE framework, we have
\begin{align*}
\mathbb{E}\left[Y_1-Y_0 \mid T=c\right] & =\frac{\mathbb{E}[Y \mid Z=1]-\mathbb{E}[Y \mid Z=0]}{\mathbb{E}\left[D^* \mid Z=1\right]-\mathbb{E}\left[D^* \mid Z=0\right]}, \\ 
& =\frac{\mathbb{E}[Y \mid Z=1]-\mathbb{E}[Y \mid Z=0]}{\mathbb{P}(T=c)},
\end{align*}
and thus the LATE, $\mathbb{E}\left[Y_1-Y_0 \mid T=c\right]$, is partially identified because $\mathbb P(T=c)$ is.

\subsection{Sharpness of the bounds}\label{proofUraSharp}
We show that any given element in the identified set can be obtained as the LATE of a counterfactual distribution $(\tilde{Y}_d,\tilde{T}, \tilde{\varepsilon})$ that is consistent with the observed data $(Y,D,Z)$. Recall that our model has a testable implication: $\max_{d\in\{0,1\}}\{TV_{(Y,D=d)}(0, 1)\}\leq \min_{d\in\{0,1\}}\left\{\mathbb P(D=d\vert Z=1)+\mathbb P(D=d\vert Z=0)\right\}.$ Suppose that this testable implication holds. 

To prove the sharpness of the bounds, we need to provide a counterfactual distribution $(\tilde{Y}_d,\tilde{T}, \tilde{\varepsilon})$ such that the analogue of Assumption \ref{RA}, $Z\ \indep\ (\tilde{Y}_d,\tilde{T}, \tilde{\varepsilon}),$ holds, $\mathbb P(\tilde{T}=df)=0$ (no defiers), and any given element in the identified set is obtained as the LATE for this counterfactual. 

The counterfactual distribution $f_{\tilde{Y}_d,\tilde{T},\tilde{\varepsilon}\vert Z}(y,t,e\vert z)$ must be consistent with the observed data $(Y,D,Z)$, that is, it must satisfy the following equality constraints.

\begin{eqnarray}
    f_{Y,D\vert Z}(y,1\vert 1) &=& f_{\tilde{Y}_0,\tilde{T},\tilde{\varepsilon}}(y,n,1)+f_{\tilde{Y}_1,\tilde{T},\tilde{\varepsilon}}(y,a,0) + f_{\tilde{Y}_1,\tilde{T},\tilde{\varepsilon}}(y,c,0),\label{eq:sharp1}\\
    f_{Y,D\vert Z}(y,1\vert 0) &=& f_{\tilde{Y}_0,\tilde{T},\tilde{\varepsilon}}(y,n,1)+f_{\tilde{Y}_1,\tilde{T},\tilde{\varepsilon}}(y,a,0) + f_{\tilde{Y}_0,\tilde{T},\tilde{\varepsilon}}(y,c,1),\label{eq:sharp2}\\
    f_{Y,D\vert Z}(y,0\vert 1) &=& f_{\tilde{Y}_0,\tilde{T},\tilde{\varepsilon}}(y,n,0)+f_{\tilde{Y}_1,\tilde{T},\tilde{\varepsilon}}(y,a,1) + f_{\tilde{Y}_1,\tilde{T},\tilde{\varepsilon}}(y,c,1),\label{eq:sharp3}\\
    f_{Y,D\vert Z}(y,0\vert 0) &=& f_{\tilde{Y}_0,\tilde{T},\tilde{\varepsilon}}(y,n,0)+f_{\tilde{Y}_1,\tilde{T},\tilde{\varepsilon}}(y,a,1) + f_{\tilde{Y}_0,\tilde{T},\tilde{\varepsilon}}(y,c,0).\label{eq:sharp4}
\end{eqnarray}

Define $$\tilde{p}_c\equiv \lambda \max_{d\in\{0,1\}}\{TV_{(Y,D=d)}(0, 1)\}+(1-\lambda)\min_{d\in\{0,1\}}\left\{\mathbb P(D=d\vert Z=1)+\mathbb P(D=d\vert Z=0)\right\},$$ where $\lambda \in [0,1].$ Note that $\tilde{p}_c \in [0,1]$ since $$\max_{d\in\{0,1\}}\{TV_{(Y,D=d)}(0, 1)\} \leq \min_{d\in\{0,1\}}\left\{\mathbb P(D=d\vert Z=1)+\mathbb P(D=d\vert Z=0)\right\}.$$ Suppose $\max_{d\in\{0,1\}}\{TV_{(Y,D=d)}(0, 1)\} > 0$. Then $\tilde{p}_c>~0.$

We use a ``guess and verify'' approach to construct the counterfactual distribution. Define
\begin{eqnarray*}
    &&f_{\tilde{Y}_1,\tilde{T},\tilde{\varepsilon}\vert Z}(y,c,0\vert z)=f_{Y,D\vert Z}(y,1\vert 1)-\\
    &&\qquad \frac{(\mathbb P(D=1\vert Z=1)+\mathbb P(D=1\vert Z=0)-\tilde{p}_c)\min\left\{f_{Y,D \vert Z}(y,1 \vert 1),f_{Y,D \vert Z}(y,1 \vert 0)\right\}}{2\int \min\left\{f_{Y,D \vert Z}(y,1 \vert 1),f_{Y,D \vert Z}(y,1 \vert 0)\right\}d\mu_Y(y)+\mathbbm{1}\left\{\int \min\left\{f_{Y,D \vert Z}(y,1 \vert 1), f_{Y,D \vert Z}(y,1 \vert 0)\right\}d\mu_Y(y)=0\right\}},\\
    &&f_{\tilde{Y}_0,\tilde{T},\tilde{\varepsilon}\vert Z}(y,c,1\vert z)=f_{Y,D\vert Z}(y,1\vert 0)-\\
    &&\qquad\frac{(\mathbb P(D=1\vert Z=1)+\mathbb P(D=1\vert Z=0)-\tilde{p}_c)\min\left\{f_{Y,D \vert Z}(y,1 \vert 1),f_{Y,D \vert Z}(y,1 \vert 0)\right\}}{2\int \min\left\{f_{Y,D \vert Z}(y,1 \vert 1),f_{Y,D \vert Z}(y,1 \vert 0)\right\}d\mu_Y(y)+\mathbbm{1}\left\{\int \min\left\{f_{Y,D \vert Z}(y,1 \vert 1),f_{Y,D \vert Z}(y,1 \vert 0)\right\}d\mu_Y(y)=0\right\}},\\
   &&f_{\tilde{Y}_1,\tilde{T},\tilde{\varepsilon}\vert Z}(y,c,1\vert z) =  f_{Y,D \vert Z}(y,0 \vert 1)-\\
   &&\qquad \frac{(\mathbb P(D=0\vert Z=1)+\mathbb P(D=0\vert Z=0)-\tilde{p}_c)\min\left\{f_{Y,D \vert Z}(y,0 \vert 1),f_{Y,D \vert Z}(y,0 \vert 0)\right\}}{2\int \min\left\{f_{Y,D \vert Z}(y,0 \vert 1),f_{Y,D \vert Z}(y,0 \vert 0)\right\}d\mu_Y(y)+\mathbbm{1}\left\{\int \min\left\{f_{Y,D \vert Z}(y,0 \vert 1),f_{Y,D \vert Z}(y,0 \vert 0)\right\}d\mu_Y(y)=0\right\}},\\
   &&f_{\tilde{Y}_0,\tilde{T},\tilde{\varepsilon}\vert Z}(y,c,0\vert z) =  f_{Y,D \vert Z}(y,0 \vert 0)-\\
   &&\qquad \frac{(\mathbb P(D=0\vert Z=1)+\mathbb P(D=0\vert Z=0)-\tilde{p}_c)\min\left\{f_{Y,D \vert Z}(y,0 \vert 1),f_{Y,D \vert Z}(y,0 \vert 0)\right\}}{2\int \min\left\{f_{Y,D \vert Z}(y,0 \vert 1),f_{Y,D \vert Z}(y,0 \vert 0)\right\}d\mu_Y(y)+\mathbbm{1}\left\{\int \min\left\{f_{Y,D \vert Z}(y,0 \vert 1),f_{Y,D \vert Z}(y,0 \vert 0)\right\}d\mu_Y(y)=0\right\}},\\
 &&f_{\tilde{Y}_0,\tilde{T},\tilde{\varepsilon}\vert Z}(y,n,1\vert z)=f_{\tilde{Y}_1,\tilde{T},\tilde{\varepsilon}\vert Z}(y,a,0\vert z)=f_{\tilde{Y}_0,\tilde{T},\tilde{\varepsilon}\vert Z}(y,a,0\vert z)=f_{\tilde{Y}_1,\tilde{T},\tilde{\varepsilon}\vert Z}(y,n,1\vert z)\\
 &&\qquad =\frac{(\mathbb P(D=1\vert Z=1)+\mathbb P(D=1\vert Z=0)-\tilde{p}_c)\min\left\{f_{Y,D \vert Z}(y,1 \vert 1),f_{Y,D \vert Z}(y,1 \vert 0)\right\}}{4\int \min\left\{f_{Y,D \vert Z}(y,1 \vert 1),f_{Y,D \vert Z}(y,1 \vert 0)\right\}d\mu_Y(y)+\mathbbm{1}\left\{\int \min\left\{f_{Y,D \vert Z}(y,1 \vert 1),f_{Y,D \vert Z}(y,1 \vert 0)\right\}d\mu_Y(y)=0\right\}},\\
 &&f_{\tilde{Y}_0,\tilde{T},\tilde{\varepsilon}\vert Z}(y,n,0\vert z)=f_{\tilde{Y}_1,\tilde{T},\tilde{\varepsilon}\vert Z}(y,a,1\vert z)=f_{\tilde{Y}_0,\tilde{T},\tilde{\varepsilon}\vert Z}(y,a,1\vert z)=f_{\tilde{Y}_1,\tilde{T},\tilde{\varepsilon}\vert Z}(y,n,0\vert z)\\
 &&\qquad=\frac{(\mathbb P(D=0\vert Z=1)+\mathbb P(D=0\vert Z=0)-\tilde{p}_c)\min\left\{f_{Y,D \vert Z}(y,0 \vert 1),f_{Y,D \vert Z}(y,0 \vert 0)\right\}}{4\int \min\left\{f_{Y,D \vert Z}(y,0 \vert 1),f_{Y,D \vert Z}(y,0 \vert 0)\right\}d\mu_Y(y)+\mathbbm{1}\left\{\int \min\left\{f_{Y,D \vert Z}(y,0 \vert 1),f_{Y,D \vert Z}(y,0 \vert 0)\right\}d\mu_Y(y)=0\right\}}.
\end{eqnarray*}

By construction, $Z\ \indep\ (\tilde{Y}_d,\tilde{T}, \tilde{\varepsilon})$. To complete the proof, we need to show that the function $f_{\tilde{Y}_d,\tilde{T},\tilde{\varepsilon}\vert Z}(y,t,e\vert z)$ is non-negative, integrates to one over the support $\mathcal Y \times \{a,c,n\} \times \{0,1\},$ satisfies the restrictions \eqref{eq:sharp1}-\eqref{eq:sharp4}, and $\int \sum_{e \in \{0,1\}} f_{\tilde{Y}_d,\tilde{T},\tilde{\varepsilon}\vert Z}(y,c,e\vert z)d \mu_Y(y)=\tilde{p}_c.$

The nonegativity of $f_{\tilde{Y}_d,\tilde{T},\tilde{\varepsilon}\vert Z}(y,t,e\vert z)$ holds by definition for $t\in\{a,n\}$ because $\tilde{p}_c \leq \min_{d\in\{0,1\}}\left\{\mathbb P(D=d\vert Z=1)+\mathbb P(D=d\vert Z=0)\right\}.$  To show the nonnegativity of $f_{\tilde{Y}_d,\tilde{T},\tilde{\varepsilon}\vert Z}(y,c,e\vert z)$, we need to show that for each $d,z\in \{0,1\},$
\small{$$f_{Y,D\vert Z}(y,d\vert z)
    -\frac{(\mathbb P(D=d\vert Z=1)+\mathbb P(D=d\vert Z=0)-\tilde{p}_c)\min\left\{f_{Y,D \vert Z}(y,d \vert 1),f_{Y,D \vert Z}(y,d \vert 0)\right\}}{2\int \min\left\{f_{Y,D \vert Z}(y,d \vert 1),f_{Y,D \vert Z}(y,d \vert 0)\right\}d\mu_Y(y)+\mathbbm{1}\left\{\int \min\left\{f_{Y,D \vert Z}(y,d \vert 1),f_{Y,D \vert Z}(y,d \vert 0)\right\}d\mu_Y(y)=0\right\}} \geq 0.$$}
    \normalsize
If $\int \min\left\{f_{Y,D \vert Z}(y,d \vert 1),
f_{Y,D \vert Z}(y,d \vert 0)\right\}d\mu_Y(y)=0,$
the result holds trivially as $f_{Y,D\vert Z}(y,d\vert z)\geq~0$. If $\int \min\left\{f_{Y,D \vert Z}(y,d \vert 1),
f_{Y,D \vert Z}(y,d \vert 0)\right\}d\mu_Y(y)>0$, it suffices to show that

$0\leq \frac{\mathbb P(D=d\vert Z=1)+\mathbb P(D=d\vert Z=0)-\tilde{p}_c}{2\int \min\left\{f_{Y,D \vert Z}(y,d \vert 1),f_{Y,D \vert Z}(y,d \vert 0)\right\}d\mu_Y(y)} \leq 1,$ that is, 
$$0 \leq \mathbb P(D=d\vert Z=1)+\mathbb P(D=d\vert Z=0)-\tilde{p}_c \leq 2\int \min\left\{f_{Y,D \vert Z}(y,d \vert 1),f_{Y,D \vert Z}(y,d \vert 0)\right\}d\mu_Y(y).$$
We use this equality $\min\{a,b\}=\frac{a+b-|a-b|}{2}$. Hence, 
$$2\min\left\{f_{Y,D \vert Z}(y,d \vert 1),f_{Y,D \vert Z}(y,d \vert 0)\right\}=f_{Y,D \vert Z}(y,d \vert 1)+f_{Y,D \vert Z}(y,d \vert 0)-\lvert f_{Y,D \vert Z}(y,d \vert 1)-f_{Y,D \vert Z}(y,d \vert 0) \rvert.$$
Therefore, it suffices to show the following: $\mathbb P(D=d\vert Z=1)+\mathbb P(D=d\vert Z=0)-\tilde{p}_c \geq 0,$ and
$$\mathbb P(D=d\vert Z=1)+\mathbb P(D=d\vert Z=0)-\tilde{p}_c \leq \int f_{Y,D \vert Z}(y,d \vert 1)+f_{Y,D \vert Z}(y,d \vert 0)-\lvert f_{Y,D \vert Z}(y,d \vert 1)-f_{Y,D \vert Z}(y,d \vert 0) \rvert d\mu_Y(y),$$
which is equivalent to $\tilde{p}_c \leq \mathbb P(D=d\vert Z=1)+\mathbb P(D=d\vert Z=0),$ and
$$\tilde{p}_c \geq \int \lvert f_{Y,D \vert Z}(y,d \vert 1)-f_{Y,D \vert Z}(y,d \vert 0) \rvert d\mu_Y(y)=TV_{(Y,D=d)}(0,1),$$ because $\int f_{Y,D \vert Z}(y,d \vert z) d\mu_Y(y)=\mathbb P(D=d \vert Z=z).$
From our model testable implication, we know by the definition of $\tilde{p}_c$ that 
$TV_{(Y,D=d)}(0,1) \leq \tilde{p}_c \leq \mathbb P(D=d\vert Z=1)+\mathbb P(D=d\vert Z=0).$

Therefore,
\begin{eqnarray*}
    &&f_{Y,D\vert Z}(y,d\vert z)
    -\frac{\mathbb P(D=d\vert Z=1)+\mathbb P(D=d\vert Z=0)-\tilde{p}_c}{2\int \min\left\{f_{Y,D \vert Z}(y,d \vert 1),f_{Y,D \vert Z}(y,d \vert 0)\right\}d\mu_Y(y)} \min\left\{f_{Y,D \vert Z}(y,d \vert 1),f_{Y,D \vert Z}(y,d \vert 0)\right\}\\
    && \qquad \qquad \qquad \geq f_{Y,D\vert Z}(y,d\vert z)
    - \min\left\{f_{Y,D \vert Z}(y,d \vert 1),f_{Y,D \vert Z}(y,d \vert 0)\right\} \geq 0.
\end{eqnarray*}
As a result, $f_{\tilde{Y}_d,\tilde{T},\tilde{\varepsilon}\vert Z}(y,t,e\vert z) \geq 0$ for all $y\in \mathcal Y,$ $t\in \{a,c,n\},$ and $e\in \{0,1\}$.

By construction, we have 
\begin{eqnarray*}
    &&\mathbb P(\tilde{T}=c)=\int \sum_{e \in \{0,1\}} f_{\tilde{Y}_d,\tilde{T},\tilde{\varepsilon}\vert Z}(y,c,e\vert z)d \mu_Y(y)=\tilde{p}_c,\\
    &&\mathbb P(\tilde{T}=a)=\int \sum_{e \in \{0,1\}}  f_{\tilde{Y}_d,\tilde{T},\tilde{\varepsilon}\vert Z}(y,a,e\vert z)d \mu_Y(y)=\frac{1-\tilde{p}_c}{2},\\
    &&\mathbb P(\tilde{T}=n)=\int \sum_{e \in \{0,1\}} f_{\tilde{Y}_d,\tilde{T},\tilde{\varepsilon}\vert Z}(y,n,e\vert z)d \mu_Y(y)=\frac{1-\tilde{p}_c}{2}.
\end{eqnarray*}
Since $\mathbb P(\tilde{T}=c)+\mathbb P(\tilde{T}=a)+\mathbb P(\tilde{T}=n)=\tilde{p}_c+\frac{1-\tilde{p}_c}{2}+\frac{1-\tilde{p}_c}{2}=1,$ we conclude that $\mathbb P(\tilde{T}=df)=0,$ and 
$\sum_{t\in \{a,c,n\}}\int \sum_{e \in \{0,1\}} f_{\tilde{Y}_d,\tilde{T},\tilde{\varepsilon}\vert Z}(y,t,e\vert z)d \mu_Y(y)=1.$

Note that the distributions $f_{\tilde{Y}_0,\tilde{T},\tilde{\varepsilon}\vert Z}(y,t,e\vert z)$ and $f_{\tilde{Y}_1,\tilde{T},\tilde{\varepsilon}\vert Z}(y,t,e\vert z)$ are mutually compatible.
For every \(t\in\{a,c,n\}\), \(e\in\{0,1\}\), and \(z\in\{0,1\}\), we can verify that
\[
\int f_{\tilde{Y}_0,\tilde{T},\tilde{\varepsilon}\mid Z}
(y,t,e\mid z)\,d\mu_Y(y)
=
\int f_{\tilde{Y}_1,\tilde{T},\tilde{\varepsilon}\mid Z}
(y,t,e\mid z)\,d\mu_Y(y).
\]

The joint distribution $f_{\tilde{Y}_0,\tilde{Y}_1,\tilde{T},\tilde{\varepsilon}\vert Z}(y_0,y_1,t,e\vert z)$ can be defined as $$f_{\tilde{Y}_0\vert \tilde{T},\tilde{\varepsilon},Z}(y_0\vert t,e,z)f_{\tilde {Y}_1 \vert \tilde{T},\tilde{\varepsilon}, Z}(y_1\vert t,e,z)f_{\tilde{T},\tilde{\varepsilon}\vert Z}(t,e\vert z):$$
\[
f_{\widetilde Y_0,\widetilde Y_1,\widetilde T,\widetilde\varepsilon\mid Z}
(y_0,y_1,t,e\mid z)
=
\frac{
f_{\tilde{Y}_0,\tilde{T},\tilde{\varepsilon}\vert Z}(y_0,t,e\vert z)
f_{\tilde{Y}_1,\tilde{T},\tilde{\varepsilon}\vert Z}(y_1,t,e\vert z)
}{
\int f_{\tilde{Y}_0,\tilde{T},\tilde{\varepsilon}\vert Z}
(y,t,e\vert z)\,d\mu_Y(y)+\mathbbm{1}\left\{\int f_{\tilde{Y}_0,\tilde{T},\tilde{\varepsilon}\vert Z}
(y,t,e\vert z)\,d\mu_Y(y)=0\right\}}.\]

Finally, it is easy to check by direct substitution that
\(f_{\tilde{Y}_d,\tilde{T},\tilde{\varepsilon}\vert Z}(y,t,e\vert z)\)
satisfies the restrictions \eqref{eq:sharp1}--\eqref{eq:sharp4}.
Summing the two \(Z=1\) equations, \eqref{eq:sharp1} and \eqref{eq:sharp3},
and subtracting the two \(Z=0\) equations, \eqref{eq:sharp2} and
\eqref{eq:sharp4}, gives
\begin{align*}
    \mathbb E[\widetilde Y_1-\widetilde Y_0\mid \widetilde T=c]
&=
\frac{
\int y\sum_{e\in\{0,1\}}
\left[
f_{\widetilde Y_1,\widetilde T,\widetilde\varepsilon\mid Z}(y,c,e\mid z)
-
f_{\widetilde Y_0,\widetilde T,\widetilde\varepsilon\mid Z}(y,c,e\mid z)
\right]\,d\mu_Y(y)
}{\tilde p_c}\\
&=
\frac{\mathbb E[Y\mid Z=1]-\mathbb E[Y\mid Z=0]}{\tilde p_c}.
\end{align*}
Thus the constructed distribution attains the desired LATE value. This completes the proof.

\section{Proof of Lemma \ref{morett_samedirection}}
\label{proof_morett_samedirection}
\begin{proof}
Assumption \ref{morett} states
$$
\mathbb{P}(\varepsilon = 0 \vert V = v) > \mathbb{P}(\varepsilon = 1 \vert V = v)\ \text{ for all } v \in [0, 1]. 
$$
Then, for any non-empty Borel set $A \subseteq [0, 1]$, we have
$$
\int_A\mathbb{P}(\varepsilon = 0 \vert V = v) f_V(v)dv > \int_A\mathbb{P}(\varepsilon = 1 \vert V = v) f_V(v) dv,\ \ \text{ where } f_V(v)=1 \text{ under Assumption \ref{Cont}},
$$
or equivalently
$$
\mathbb{P}(\varepsilon = 0 \vert V \in A) > \mathbb{P}(\varepsilon = 1 \vert V \in A).
$$
By multiplying each side of the above equation by $\mathbb P(V \in A)>0$, this latter inequality is equivalent to $\mathbb{P}(\varepsilon = 0, V \in A) > \mathbb{P}(\varepsilon = 1, V \in A).$
For any $z, z' \in \mathcal{Z}$ such that $P(z) \neq P(z')$, we take 
\begin{align*}
A &= \big( \min\{P(z), P(z')\}, \max\{P(z), P(z')\} \big] \subseteq [0, 1],
\end{align*} 
and we have
\begin{align*}
& \mathbb{P}\big(\varepsilon = 0, \min\{P(z), P(z')\} < V \leq \max\{P(z), P(z')\}\big) \\
& \qquad > \mathbb{P}\big(\varepsilon = 1, \min\{P(z), P(z')\} < V \leq \max\{P(z), P(z')\}\big).
\end{align*}
We can easily check that for any $\ell \in \{0,1\},$ $$\Big\vert \mathbb{P}\big(\varepsilon = \ell, V \leq P(z) \big) - \mathbb{P}\big(\varepsilon = \ell, V \leq P(z') \big) \Big\vert =\mathbb{P}\big(\varepsilon = \ell, \min\{P(z), P(z')\} < V \leq \max\{P(z), P(z')\}\big).$$

Thus, Assumption \ref{morett} implies
\begin{align*}
& \Big\vert \mathbb{P}\big(\varepsilon = 0, V \leq P(z) \big) - \mathbb{P}\big(\varepsilon = 0, V \leq P(z') \big) \Big\vert \\
& \qquad > \Big\vert \mathbb{P}\big(\varepsilon = 1, V \leq P(z) \big) - \mathbb{P}\big(\varepsilon = 1, V \leq P(z') \big) \Big\vert,
\end{align*}
which, under the model specification \eqref{seq1} and Assumption \ref{RA}, implies
\begin{align*}
& \Big\vert \mathbb{P}\big(D^* = 1, \varepsilon = 0 \vert Z = z \big) - \mathbb{P}\big(D^* = 1, \varepsilon = 0 \vert Z = z' \big) \Big\vert \\
& \qquad > \Big\vert \mathbb{P}\big(D^* = 1, \varepsilon = 1 \vert Z = z  \big) - \mathbb{P}\big(D^* = 1, \varepsilon = 1 \vert Z = z' \big) \Big\vert.
\end{align*}
Note that the last inequality is equivalent to 
\begin{align}
& \Big[ \mathbb{P}\big(D^* = 1, \varepsilon = 0 \vert Z = z \big) - \mathbb{P}\big(D^* = 1, \varepsilon = 0 \vert Z = z' \big) \Big]^2 \nonumber \\
& \qquad - \Big[ \mathbb{P}\big(D^* = 1, \varepsilon = 1 \vert Z = z  \big) - \mathbb{P}\big(D^* = 1, \varepsilon = 1 \vert Z = z' \big) \Big]^2 > 0. \label{eq.lem1}
\end{align}

Lastly, we have
\begin{align*}
\Delta_{D^*Z}(z',z)& = \mathbb{P}(D^*=1 \vert Z = z) - \mathbb{P}(D^*=1 \vert Z = z'), \\
& = \big[\mathbb{P}(D^*=1, \varepsilon = 0 \vert Z = z) - \mathbb{P}(D^*=1, \varepsilon = 0 \vert Z = z') \big] \\
& \qquad + \big[\mathbb{P}(D^*=1, \varepsilon = 1 \vert Z = z) - \mathbb{P}(D^*=1, \varepsilon = 1 \vert Z = z') \big],
\end{align*}
and
\begin{align*}
\Delta_{DZ}(z',z) &= \mathbb{P}(D=1 \vert Z = z) - \mathbb{P}(D=1 \vert Z = z'), \\
 &= \mathbb{P}(D^*(1-\varepsilon)+(1-D^*)\varepsilon = 1 \vert Z = z) \\
 & \qquad - \mathbb{P}(D^*(1-\varepsilon)+(1-D^*)\varepsilon=1 \vert Z = z'), \\
 &= \big[ \mathbb{P}(D^*= 1, \varepsilon=0 \vert Z = z) + \mathbb{P}(D^*=0, \varepsilon=1 \vert Z = z) \big] \\
 & \qquad - \big[ \mathbb{P}(D^*= 1, \varepsilon=0 \vert Z = z') + \mathbb{P}(D^*=0, \varepsilon=1 \vert Z = z') \big], \\
 &= \big[ \mathbb{P}(D^*= 1, \varepsilon=0 \vert Z = z) - \mathbb{P}(D^*= 1, \varepsilon=0 \vert Z = z') \big] \\
 & \qquad - \big[\mathbb{P}(D^*=0, \varepsilon=1 \vert Z = z') - \mathbb{P}(D^*=0, \varepsilon=1 \vert Z = z)\big], \\
 &= \big[ \mathbb{P}(D^*= 1, \varepsilon=0 \vert Z = z) - \mathbb{P}(D^*= 1, \varepsilon=0 \vert Z = z') \big] \\
 & \qquad - \big[\mathbb{P}(D^*=1, \varepsilon=1 \vert Z = z) - \mathbb{P}(D^*=1, \varepsilon=1 \vert Z = z')\big], 
\end{align*}
where the second equality holds from the model \eqref{seq1}, and the last equality holds because
\begin{eqnarray*}
\mathbb{P}(D^*=0, \varepsilon=1 \vert Z = z) &=& \mathbb{P}(\varepsilon=1 \vert Z = z) - \mathbb{P}(D^*=1, \varepsilon=1 \vert Z = z),\\
&=& \mathbb{P}(\varepsilon=1) - \mathbb{P}(D^*=1, \varepsilon=1 \vert Z = z)\ \ \text{ under Assumption \ref{RA}},
\end{eqnarray*}
for any $z \in \mathcal{Z}$.

Therefore, 
\begin{align*}
\Delta_{D^*Z}(z',z)\Delta_{DZ}(z',z) &= \Big[ \mathbb{P}\big(D^* = 1, \varepsilon = 0 \vert Z = z \big) - \mathbb{P}\big(D^* = 1, \varepsilon = 0 \vert Z = z' \big) \Big]^2 \nonumber \\
& \qquad - \Big[ \mathbb{P}\big(D^* = 1, \varepsilon = 1 \vert Z = z  \big) - \mathbb{P}\big(D^* = 1, \varepsilon = 1 \vert Z = z' \big) \Big]^2 .
\end{align*}
Hence, \eqref{eq.lem1} is equivalent to $\Delta_{D^*Z}(z',z)\Delta_{DZ}(z',z)>0$. This completes the proof.
\end{proof}

\section{Proofs of Proposition \ref{prop1}}\label{proofprop1}
Since the function $\max$ and $\min$ are continuous, there exist $\alpha^*$ the $\inf$ in the lower bound on $P(z)$ is attained. Similar result holds for the upper. We propose two misclassification scenarios that yield the lower or upper bound for each value of $\alpha$.
\subsection{$\varepsilon=\mathbbm{1}\{V\leq \alpha\}$} Here, we assume that the misclassification occurs when the unobserved heterogeneity $V$ is less than or equal to $\alpha$. Then, we have $\mathbb P(\varepsilon=1)=\alpha$, and $F_{V\vert \varepsilon=1}(p)=\min\{p/\alpha,1\}$. Hence, the conditional distribution of $V$ given $\varepsilon=1$ is concave. Using Equation (\ref{eq:mix}), we obtain that $F_{V\vert \varepsilon=0}(p)=\frac{p-\min\{p,\alpha\}}{1-\alpha}$. 

If $\alpha \leq P(z)$, then Equation (\ref{eq2}) implies
\begin{eqnarray*}
\mathbb P(D=1\vert Z=z)=(1-\alpha)\left(\frac{P(z)-\alpha}{1-\alpha}\right)+\alpha(1-\min\{P(z)/\alpha,1\})=P(z)-\alpha,
\end{eqnarray*}
which leads to 
\begin{eqnarray}
P(z)=\mathbb P(D=1\vert Z=z)+\alpha.
\end{eqnarray}

If $\alpha \geq P(z)$, then Equation (\ref{eq2}) implies
\begin{eqnarray*}
\mathbb P(D=1\vert Z=z)=(1-\alpha)\left(\frac{P(z)-P(z)}{1-\alpha}\right)+\alpha(1-P(z)/\alpha)=\alpha-P(z),
\end{eqnarray*}
which in turn implies 
\begin{eqnarray}
P(z)=\alpha-\mathbb P(D=1\vert Z=z).
\end{eqnarray}

\subsection{$\varepsilon=\mathbbm{1}\{V> 1-\alpha\}$} Given this specification for the misclassification, we have $\mathbb P(\varepsilon=1)=\alpha$, and $F_{V\vert \varepsilon=0}(p)=\min\{\frac{p}{1-\alpha},1\}$. From there, we have $F_{V\vert \varepsilon=1}(p)=\max\{0,1-\frac{1-p}{\alpha}\}$. Hence, the conditional distribution of $V$ given $\varepsilon=1$ is convex.

From Equation (\ref{eq2}), we have:
\begin{eqnarray*}
(1-\alpha)\min\left\{\frac{P(z)}{1-\alpha},1\right\}+\alpha \left(1-\max\left\{0,1-\frac{1-P(z)}{\alpha}\right\}\right)=\mathbb P(D=1\vert Z=z).
\end{eqnarray*}

If $1-\alpha \leq P(z)$, the above equation becomes:
\begin{eqnarray*}
(1-\alpha)+\alpha \left(1-1+\frac{1-P(z)}{\alpha}\right)=\mathbb P(D=1\vert Z=z),
\end{eqnarray*}
which implies 
\begin{eqnarray}
P(z)=\mathbb P(D=0\vert Z=z)+1-\alpha.
\end{eqnarray}

If $1-\alpha \geq P(z)$, the equation becomes:
\begin{eqnarray*}
(1-\alpha)\frac{P(z)}{1-\alpha}+\alpha (1-0)=\mathbb P(D=1\vert Z=z),
\end{eqnarray*}
which implies 
\begin{eqnarray}
P(z)=\mathbb P(D=1\vert Z=z)-\alpha.
\end{eqnarray}

\section{Proof of Proposition \ref{prop:mtebounds}}\label{proof:mtebounds}

Consider Equation \eqref{eq:fsharp2}. Take the difference of $\mathbb P(Y\in A,D=1\vert Z=z)$ between two values $z$ and $z'$ of the instrument $Z$ such that $P(z') < P(z).$
\begin{eqnarray*}
\mathbb P(Y\in A, D=1 \vert Z=z)-\mathbb P(Y\in A, D=1 \vert Z=z')
&=& \int^{P(z)}_{P(z')}\mathbb P\left(Y_1\in A, \varepsilon=0 \vert V=v\right)dv \nonumber \\
&& \qquad - \int^{P(z)}_{P(z')}\mathbb P\left(Y_0\in A, \varepsilon=1 \vert V=v\right)dv. 
\end{eqnarray*}
By taking the absolute value and applying the triangle inequality, we obtain
\begin{eqnarray*}
\lvert \mathbb P(Y\in A, D=1 \vert Z=z)-\mathbb P(Y\in A, D=1 \vert Z=z')\rvert
&\leq& \int^{P(z)}_{P(z')}\mathbb P\left(Y_1\in A, \varepsilon=0 \vert V=v\right)dv \nonumber \\
&& \qquad + \int^{P(z)}_{P(z')}\mathbb P\left(Y_0\in A, \varepsilon=1 \vert V=v\right)dv, 
\end{eqnarray*}
Using the monotonicity of a probability measure, we know that $\mathbb P\left(Y_d\in A, \varepsilon=\ell \vert V=v\right) \leq \mathbb P\left(\varepsilon=\ell \vert V=v\right)$. Therefore,
\begin{eqnarray*}
\lvert \mathbb P(Y\in A, D=1 \vert Z=z)-\mathbb P(Y\in A, D=1 \vert Z=z')\rvert
&\leq& \int^{P(z)}_{P(z')}\mathbb P\left(\varepsilon=0 \vert V=v\right)+\mathbb P\left(\varepsilon=1 \vert V=v\right)dv, \\
&=& P(z)-P(z').
\end{eqnarray*}
We can take the supremum over $A$ and obtain the following.
\begin{eqnarray*}
\sup_{A}\lvert \mathbb P(Y\in A, D=1 \vert Z=z)-\mathbb P(Y\in A, D=1 \vert Z=z')\rvert
&\leq& P(z)-P(z').
\end{eqnarray*}
This bound can be further tightened using the density version of \eqref{eq:fsharp2}. The density versions of Equations \eqref{eq:fsharp2} and \eqref{eq:fsharp3} hold.

The density versions of Equations \eqref{eq:fsharp2} and \eqref{eq:fsharp3} hold. We first consider Equation \eqref{eq:fsharp2}. For two values $z$ and $z'$ of the instrument $Z$ such that $P(z') < P(z)$, we have

\begin{eqnarray*}
f_{Y,D\vert Z}\left(y, 1 \vert z\right) - f_{Y,D\vert Z}\left(y, 1 \vert z'\right)&=& \int^{P(z)}_{P(z')}f_{Y_1,\varepsilon\vert V}(y,0\vert v)dv - \int^{P(z)}_{P(z')}f_{Y_0,\varepsilon\vert V}(y,1\vert v)dv,
\end{eqnarray*}
where $f_{X\vert W}\left(x \vert w\right)$ is the conditional density of $X$ given $\left\{W=w\right\}$ that is absolutely continuous with respect to a known dominating measure $\mu_{X}$.
Using the triangle inequality, we have
\begin{eqnarray*}
\left \lvert f_{Y,D\vert Z}\left(y, 1 \vert z\right) - f_{Y,D\vert Z}\left(y, 1 \vert z'\right) \right \rvert  &\leq&  \int^{P(z)}_{P(z')}f_{Y_1,\varepsilon\vert V}(y,0\vert v)dv + \int^{P(z)}_{P(z')}f_{Y_0,\varepsilon\vert V}(y,1\vert v)dv.
\end{eqnarray*}
Therefore, by integrating each side of the last inequality over the support $\mathcal Y$, and using the Fubini-Tonelli theorem, we have
\begin{eqnarray*}
\int_{\mathcal Y}\left \lvert f_{Y,D\vert Z}\left(y, 1 \vert z\right) - f_{Y,D\vert Z}\left(y, 1 \vert z'\right) \right \rvert d \mu_{Y}(y) &\leq&  \int^{P(z)}_{P(z')}\mathbb P(\varepsilon=0\vert V=v)dv+\int^{P(z)}_{P(z')}\mathbb P(\varepsilon=1\vert V=v)dv,\\
&=& \int^{P(z)}_{P(z')}dv= P(z)-P(z').
\end{eqnarray*}
Hence, we have $TV_{(Y,D=1)}(z',z) \leq P(z)-P(z')$, where 
\begin{eqnarray*}
TV_{(Y,D=d)}(z',z) \equiv \int_{\mathcal Y}\left \lvert f_{Y,D\vert Z}\left(y, d \vert z\right) - f_{Y,D\vert Z}\left(y, d \vert z'\right) \right \rvert d \mu_{Y}(y).
\end{eqnarray*}
Using a similar argument on Equation \eqref{eq:fsharp3}, we have $TV_{(Y,D=0)}(z',z) \leq P(z)-P(z')$.
Therefore, we obtain the following bounds on the difference $P(z)-P(z')$:
\begin{eqnarray*}
&&\max\left\{\left \lvert \Delta_{DZ}(z',z)\right \rvert, TV_{(Y,D=1)}(z',z), TV_{(Y,D=0)}(z',z)\right\}\\
 && \qquad \qquad \qquad \leq P(z)-P(z') \leq \min\left\{1,2\alpha+\Delta_{DZ}(z',z), 2(1-\alpha)-\Delta_{DZ}(z',z)\right\}. 
\end{eqnarray*}
We can show that $\max\left\{TV_{(Y,D=1)}(z',z), TV_{(Y,D=0)}(z',z) \right\} \geq \left \lvert \Delta_{DZ}(z',z)\right \rvert$. 
In general, for any Borel set $A \subseteq \mathcal Y$ we have 
\begin{eqnarray*}
\int_{\mathcal Y}\left \lvert f_{Y,D\vert Z}\left(y, d \vert z\right) - f_{Y,D\vert Z}\left(y, d \vert z'\right) \right \rvert d \mu_{Y}(y) &\geq& \int_{A}\left \lvert f_{Y,D\vert Z}\left(y, d \vert z\right) - f_{Y,D\vert Z}\left(y, d \vert z'\right) \right \rvert d \mu_{Y}(y),\\
&\geq& \left \lvert \int_{A} (f_{Y,D\vert Z}\left(y, d \vert z\right) - f_{Y,D\vert Z}\left(y, d \vert z'\right))d \mu_{Y}(y) \right \rvert ,\\
&=& \lvert \mathbb P(Y\in A, D=d\vert Z=z)-\mathbb P(Y\in A, D=d\vert Z=z')\rvert,
\end{eqnarray*}
where the first inequality holds because $A \subseteq \mathcal Y,$ the second holds because for any $\mu$-integrable function $h$, $\left \lvert \int_{A} h(y) d \mu_{Y}(y) \right \rvert \leq \int_{A} \left \lvert h(y) \right \rvert d \mu_{Y}(y),$ and the last holds because $\int_{A} f_{Y,D\vert Z}\left(y, d \vert z\right) d \mu_{Y}(y) = \mathbb P(Y\in A, D=d\vert Z=z).$ Therefore, 
\begin{eqnarray*}
  \int_{\mathcal Y}\left \lvert f_{Y,D\vert Z}\left(y, d \vert z\right) - f_{Y,D\vert Z}\left(y, d \vert z'\right) \right \rvert d \mu_{Y}(y)
&\geq& \sup_A \lvert \mathbb P(Y\in A, D=d\vert Z=z)-\mathbb P(Y\in A, D=d\vert Z=z')\rvert,\\
&\geq& \lvert \mathbb P(Y\in \mathcal Y, D=d\vert Z=z)-\mathbb P(Y\in \mathcal Y, D=d\vert Z=z')\rvert,\\
&=& \lvert \mathbb P(D=d\vert Z=z)-\mathbb P(D=d\vert Z=z')\rvert=\lvert \Delta_{DZ}(z',z)\rvert.
\end{eqnarray*}

On the other hand, Equation \eqref{eq:mis0} implies
\begin{eqnarray*}
\mathbb P(D=1 \vert Z=z)&=& \mathbb P(\varepsilon=0, P(z') < V \leq P(z))+\mathbb P(\varepsilon=0, V \leq P(z')) + \mathbb P(\varepsilon=1, V > P(z)),\\
\mathbb P(D=1 \vert Z=z')&=& \mathbb P(\varepsilon=0, V \leq P(z')) + \mathbb P(\varepsilon=1, P(z) \geq V > P(z'))+\mathbb P(\varepsilon=1, V > P(z)).
\end{eqnarray*}
Summing up these two equations, we obtain
\begin{eqnarray*}
\mathbb P(D=1 \vert Z=z)+\mathbb P(D=1 \vert Z=z')&=& \mathbb P(P(z') < V \leq P(z))+2[\mathbb P(\varepsilon=0, V \leq P(z')) + \mathbb P(\varepsilon=1, V > P(z))],\\
&\geq& \mathbb P(P(z') < V \leq P(z))=P(z)-P(z').
\end{eqnarray*}
Similarly, for $D=0,$ we have
\begin{eqnarray*}
\mathbb P(D=0 \vert Z=z)+\mathbb P(D=0 \vert Z=z')&\geq& P(z)-P(z').
\end{eqnarray*}

Consequently, combining all the bounds, we have
\begin{eqnarray*}
&&\max_{d\in\{0,1\}}\left\{TV_{(Y,D=d)}(z',z)\right\}\leq P(z)-P(z') \leq  \min_{d\in \{0,1\}}\bigg\{\mathbb P(D=d \vert Z=z)+\mathbb P(D=d \vert Z=z'),\nonumber\\
 &&\qquad \qquad \qquad \qquad \qquad \qquad \qquad \qquad \qquad \qquad \qquad 2\alpha+\Delta_{DZ}(z',z), 2(1-\alpha)-\Delta_{DZ}(z',z)\bigg\}. 
\end{eqnarray*}
These above bounds on $P(z)-P(z')$ are tighter than the ones one would get by taking the difference of the pointwise bounds derived previously in Proposition \ref{prop1}. 

Define
\begin{eqnarray*}
LB_p(z',z) &\equiv& \max_{d\in\{0,1\}}\left\{TV_{(Y,D=d)}(z',z)\right\},\\
UB_p(z',z) &\equiv& \min_{d\in\{0,1\}}\left\{\mathbb P(D=d \vert Z=z)+\mathbb P(D=d \vert Z=z'),2\alpha+\Delta_{DZ}(z',z), 2(1-\alpha)-\Delta_{DZ}(z',z)\right\}.
\end{eqnarray*} 
Suppose $LB_p(z',z) \neq 0$ and $UB_p(z',z) \neq 0$. Then, the following holds. 
\begin{eqnarray*}
\left\{ \begin{array}{lcl}
     \frac{\Delta_{YZ}(z',z)}{UB_p(z',z)} \leq \frac{\Delta_{YZ}(z',z)}{P(z)-P(z')} \leq \frac{\Delta_{YZ}(z',z)}{LB_p(z',z)}\ \text{ if }\ \Delta_{YZ}(z',z) \geq 0,&&\\ \\
      \frac{\Delta_{YZ}(z',z)}{LB_p(z',z)} \leq \frac{\Delta_{YZ}(z',z)}{P(z)-P(z')}  \leq \frac{\Delta_{YZ}(z',z)}{UB_p(z',z)}\ \text{ if }\ \Delta_{YZ}(z',z) < 0.&&
     \end{array} \right.
\end{eqnarray*}
Hence, we have
\begin{eqnarray*}
&& \min\left\{\frac{\Delta_{YZ}(z',z)}{UB_p(z',z)},  \frac{\Delta_{YZ}(z',z)}{LB_p(z',z)}\right\}\\
&& \qquad \qquad \leq \frac{\mathbb E[Y\vert P(Z)=P(z)]-\mathbb E[Y\vert P(Z)=P(z')]}{P(z)-P(z')} \leq \\
&& \qquad \qquad \qquad \qquad \qquad \qquad \qquad \qquad \qquad \max\left\{\frac{\Delta_{YZ}(z',z)}{UB_p(z',z)},  \frac{\Delta_{YZ}(z',z)}{LB_p(z',z)}\right\}.
\end{eqnarray*}
Therefore, we can take the limit of each side when $z'$ goes to $z$. Suppose that $\lim_{z' \rightarrow z} \frac{\Delta_{YZ}(z',z)}{UB_p(z',z)}$, $\lim_{z' \rightarrow z} \frac{\Delta_{YZ}(z',z)}{LB_p(z',z)}$, and $\lim_{z' \rightarrow z} \frac{\mathbb E[Y\vert P(Z)=P(z)]-\mathbb E[Y\vert P(Z)=P(z')]}{P(z)-P(z')}$ exist.Then, using the fact that the functions $\min$ and $\max$ are continuous, and assuming that $P(z)$ is continuous in $z$, we obtain
\begin{eqnarray*}
&& \min\left\{\lim_{z' \rightarrow z} \frac{\Delta_{YZ}(z',z)}{UB_p(z',z)},  \lim_{z' \rightarrow z} \frac{\Delta_{YZ}(z',z)}{LB_p(z',z)}\right\} \nonumber\\
&& \qquad \qquad \qquad \qquad \qquad \leq MTE(P(z))\leq \\
&& \qquad \qquad \qquad \qquad \qquad \qquad \max\left\{\lim_{z' \rightarrow z} \frac{\Delta_{YZ}(z',z)}{UB_p(z',z)},  \lim_{z' \rightarrow z} \frac{\Delta_{YZ}(z',z)}{LB_p(z',z)}\right\}. \nonumber
\end{eqnarray*}

\begin{enumerate}[(i)]
\item Suppose $\bar{\alpha}=0$. Then, there is no misclassification, and the testable implications of the model are given by: for any $(z',z)$ such that $P(z') < P(z)$,
$f_{Y,D\vert Z}(y,1\vert z)-f_{Y,D\vert Z}(y,1\vert z')\geq 0$ and $f_{Y,D\vert Z}(y,0\vert z)-f_{Y,D\vert Z}(y,0\vert z')\leq 0$. Therefore, $TV_{(Y,D=1)}(z',z)=\Delta_{DZ}(z',z)=TV_{(Y,D=0)}(z',z)$, which implies $LB_p(z',z)=\Delta_{DZ}(z',z)$. In this case, we also have $UB_p(z',z)=\Delta_{DZ}(z',z)$. Thus, the result holds. 
\item Suppose $\bar{\alpha} > 0$. 
Note that $P(z) - P(z') \leq UB_p(z',z) \leq \min_{d\in\{0,1\}}\{\mathbb P(D=d \vert Z=z)+\mathbb P(D=d \vert Z=z'), 2 \alpha + \Delta_{DZ}(z',z) \} \leq \min\{\mathbb P(D=d \vert Z=z)+\mathbb P(D=d \vert Z=z'), 2 \bar{\alpha} + \Delta_{DZ}(z',z) \} \equiv UB_p^{\bar{\alpha}}(z',z)$, and $\lim_{z' \rightarrow z} UB_p^{\bar{\alpha}}(z',z) \neq 0$. 
Thus, we have $\lim_{z' \rightarrow z} \frac{\Delta_{YZ}(z',z)}{UB_p^{\bar{\alpha}}(z',z)}=0$, and $MTE(P(z)) \geq 0$ if $\Delta_{YZ}(z',z) \geq 0$ and $MTE(P(z)) \leq 0$ if $\Delta_{YZ}(z',z) < 0$.
\end{enumerate}

\section{Proof of Proposition~\ref{prop:identified_set_RS} and Corollary~\ref{cor:sign_nodewise}}\label{proof:identified_set_RS}
\begin{proof}[Proof of Proposition~\ref{prop:identified_set_RS}]
Fix a feasible pair $(P,m)$ satisfying the conditions stated in
Proposition~\ref{prop:identified_set_RS}. We prove \eqref{eq:RS_limit_prop}.

\medskip
Since $t_a(v)\le v$ for all $v\in[0,1]$ and CDFs are nondecreasing,
\[
0\le F_{P(Z)}(v)-F_{P(Z)}(t_a(v))\le 1.
\]
Hence $0\le w_P(v)\le \big(a(1-\mathbb E[P(Z)])\big)^{-1}$ by \eqref{eq:wP_def_prop}.
The denominator is positive because $0<\mathbb E[P(Z)]<1$ and $a\in(0,1)$. Since $m$ is bounded,
the function $v\mapsto m(v)w_P(v)$ is bounded and Lebesgue integrable on $[0,1]$.
Thus, for each feasible pair $(P,m)$,
\[
PRTE(P,m;a)\equiv \int_0^1 m(v)\,w_P(v)\,dv
\]
is well-defined.

\medskip
Because $P$ is of bounded
variation on $\mathcal Z$, define the signed Stieltjes measure $\mu_P$ on $\mathcal Z$ by
\[
\mu_P\big((s,t]\cap\mathcal Z\big)\equiv  P(t)-P(s),
\]
and write $dP$ for $d\mu_P$. Let $|\mu_P|$ denote its total variation measure and write $d|P|\equiv d|\mu_P|$.
The total variation of $P$ on $\mathcal Z$ is
\[
\operatorname{Var}_{\mathcal Z}(P)\equiv \sup_{\Pi}\sum_{i=1}^k |P(z_i)-P(z_{i-1})|<\infty,
\]
and a standard result for Stieltjes measures induced by BV functions yields
\[
|\mu_P|(\mathcal Z)=\operatorname{Var}_{\mathcal Z}(P)<\infty.
\]
Now define the $z$-indexed integrand
\begin{align*}
    \psi_P(z) &\equiv m(P(z))w_P(P(z)),\\
    &=m(P(z))\,\frac{F_{P(Z)}(P(z))-F_{P^a(Z)}(P(z))}{a(1-\mathbb E[P(Z)])}.
\end{align*}
As $v\mapsto m(v)w_P(v)$ is bounded, $\psi_P$ is bounded on $\mathcal Z$. Therefore,
\[
\int_{\mathcal Z}|\psi_P(z)|\,d|P|(z)
\le \|\psi_P\|_\infty\,|dP|(\mathcal Z)
= \|\psi_P\|_\infty\,\operatorname{Var}_{\mathcal Z}(P)<\infty,
\]
so $\psi_P$ is integrable with respect to the finite signed measure $dP$.

\medskip
Define
\[
g(v)\equiv m(v)\,w_P(v)
=
m(v)\,\frac{F_{P(Z)}(v)-F_{P(Z)}(t_a(v))}{a(1-\mathbb E[P(Z)])},
\qquad v\in[0,1].
\]
By the continuity of $F_{P(Z)}(\cdot)$, $m(\cdot)$, and $t_a(\cdot)$,
the function $g$ is continuous on $[0,1]$. 
Hence
\[
\Psi(u)\equiv \int_0^u g(v)\,dv
\]
belongs to $C^1([0,1])$ and satisfies $\Psi'(u)=g(u)$ for all $u\in[0,1]$.

Write $\underline z$ and $\overline z$ for the lower and upper endpoints,
possibly in the extended reals, of the support $\mathcal Z$. Because $P$ is continuous and of bounded variation,
the BV chain rule yields
\[
\int_{\mathcal Z} \Psi'(P(z))\,dP(z)
=
\Psi\!\left(\lim_{z\uparrow \overline z}P(z)\right)
-
\Psi\!\left(\lim_{z\downarrow \underline z}P(z)\right)
=
\Psi(1)-\Psi(0),
\]
where the last equality uses the endpoint normalization. Since $\Psi'(P(z))=g(P(z))=\psi_P(z)$ for all $z\in\mathcal Z$, we obtain
\begin{equation}\label{eq:LS_rep_in_proof}
PRTE(P,m;a)=\Psi(1)-\Psi(0)=\int_{\mathcal Z}\psi_P(z)\,dP(z).
\end{equation}

\medskip
Fix $M<\infty$ and work on $\mathcal Z_M=\mathcal Z\cap[-M,M]$.
By the same continuity argument above, $g$ is continuous on $[0,1]$.
Since $P$ is continuous, the composition
\[
\psi_P(z)=g(P(z)),\qquad z\in\mathcal Z_M,
\]
is continuous, hence regulated, on $\mathcal Z_M$. Because $P$ is continuous and of bounded variation on $\mathcal Z_M$,
the Riemann--Stieltjes integral $\int_{\mathcal Z_M}\psi_P\,dP$ exists, and for right-endpoint tags the canonical RS sums converge:
\begin{equation}\label{eq:proof_step4}
\int_{\mathcal Z_M}\psi_P(z)\,dP(z)
=
\lim_{\|\Pi\|\to0,\ \Pi\subset\mathcal Z_M}\sum_{i=1}^k \psi_P(z_i)\,\big(P(z_i)-P(z_{i-1})\big),
\end{equation}
which is exactly the inner limit in \eqref{eq:RS_limit_prop} with $\psi_P(z_i)$ expanded as in \eqref{eq:RS_sum_prop}.

\medskip
Since $\mathcal Z_M\uparrow\mathcal Z$ and $|\mu_P|(\mathcal Z)<\infty$, we have $|\mu_P|(\mathcal Z\setminus\mathcal Z_M)\to0$.
Because $\psi_P$ is bounded, $\int_{\mathcal Z\setminus\mathcal Z_M}\psi_P\,dP\to0$ as $M\to\infty$.
Combining \eqref{eq:LS_rep_in_proof} with \eqref{eq:proof_step4} and letting $M\to\infty$ yields \eqref{eq:RS_limit_prop}.

Finally, consider any true data-generating structure satisfying the maintained conditions and inducing a pair $(P,m)$ covered by Proposition~\ref{prop:identified_set_RS}. Its PRTE value is therefore one of the feasible values represented above. Since the envelopes
\[
\underline{PRTE}^a\equiv \inf_{P\in\mathcal P,\ m\in\mathcal M(P)} PRTE(P,m;a),
\qquad
\overline{PRTE}^a\equiv \sup_{P\in\mathcal P,\ m\in\mathcal M(P)} PRTE(P,m;a)
\]
are taken over a class that contains the feasible pairs satisfying the proposition's conditions, the resulting interval
$\left[\underline{PRTE}^a,\overline{PRTE}^a\right]$ contains every PRTE value generated by such a feasible structure. Hence it is an outer set for $\mathcal I(PRTE^a)$.

\end{proof}

\begin{proof}[Proof of Corollary~\ref{cor:sign_nodewise}]
Fix $i\in\{1,\dots,k\}$ and define
\[
C_i(P)\equiv 
\frac{F_{P(Z)}(p_i)-F_{P(Z)}(t_a(p_i))}{a(1-\mathbb E[P(Z)])}.
\]
By the identity $F_{P^a(Z)}(p)=F_{P(Z)}(t_a(p))$ used in the definition of
$w_P$, we can write
\[
B_i(P)=C_i(P)\,\Delta p_i.
\]
Since $t_a(p_i)\le p_i$, the cdf $F_{P(Z)}$ is nondecreasing, and
$a(1-\mathbb E[P(Z)])>0$ by Proposition~\ref{prop:identified_set_RS}, we have
$C_i(P)\ge 0$. Hence $B_i(P)$ has the same sign as $\Delta p_i$. Therefore,
\[
s_i B_i(P)\ge 0,
\qquad
B_i(P)=s_i\,|B_i(P)|.
\]

Next, by the local sign information for the MTE delivered by
Proposition~\ref{prop:mtebounds},
the MTE bounds satisfy
\[
LB_{MTE}(z)\le m(P(z))\le UB_{MTE}(z),\qquad
LB_{MTE}(z)\le 0\le UB_{MTE}(z),\qquad
LB_{MTE}(z)\,UB_{MTE}(z)=0
\]
for all relevant $z\in\mathcal Z$. Thus, for each $i$,
\[
m(p_i)\in[LB_{MTE}(z_i),0]
\quad\text{if } UB_{MTE}(z_i)=0,
\]
and
\[
m(p_i)\in[0,UB_{MTE}(z_i)]
\quad\text{if } LB_{MTE}(z_i)=0.
\]
If both $LB_{MTE}(z_i)$ and $UB_{MTE}(z_i)$ are zero, then both displays coincide and
$T_i(P,m)=0$, so the claimed lower and upper bounds hold trivially.

If $UB_{MTE}(z_i)=0$, then:
if $s_i=1$, so that $B_i(P)=|B_i(P)|\ge 0$, multiplying
$m(p_i)\in[LB_{MTE}(z_i),0]$ by $B_i(P)$ gives
\[
LB_{MTE}(z_i)\,|B_i(P)|\le T_i(P,m)\le 0;
\]
if $s_i=-1$, so that $B_i(P)=-|B_i(P)|\le 0$, the map
$m\mapsto mB_i(P)$ is decreasing and thus
\[
0\le T_i(P,m)\le -LB_{MTE}(z_i)\,|B_i(P)|;
\]
if $s_i=0$, then $\Delta p_i=0$, hence $B_i(P)=0$ and $T_i(P,m)=0$.

If $LB_{MTE}(z_i)=0$, then:
if $s_i=1$, we obtain
\[
0\le T_i(P,m)\le UB_{MTE}(z_i)\,|B_i(P)|;
\]
if $s_i=-1$, we obtain
\[
-UB_{MTE}(z_i)\,|B_i(P)|\le T_i(P,m)\le 0;
\]
if $s_i=0$, again $T_i(P,m)=0$.

These case-by-case bounds are exactly equivalent to
\[
\underline T_i(P)\le T_i(P,m)\le \overline T_i(P).
\]
Summing over $i=1,\dots,k$ gives
\[
\sum_{i=1}^k\underline T_i(P)
\le
\sum_{i=1}^k T_i(P,m)
\le
\sum_{i=1}^k\overline T_i(P).
\]
By definition,
\[
\sum_{i=1}^k T_i(P,m)=S(\Pi;P,m),
\]
and hence
\[
\underline S(\Pi;P)\le S(\Pi;P,m)\le \overline S(\Pi;P).
\]
Taking the infimum over feasible $P\in\mathcal P$ on the lower side and the supremum over feasible $P\in\mathcal P$ on the upper side gives the finite-partition outer bounds. Finally, applying Proposition~\ref{prop:identified_set_RS} and taking the corresponding $\liminf$ and $\limsup$ yields the stated RS-limit PRTE bounds. The final monotone-$P$ statement follows immediately by setting $s_i=1$ for all $i$.
\end{proof}

\section{Proof of Lemma \ref{symmetry}}\label{apx:prop_robust}
\begin{proof}\label{proof:symmetry}
Suppose that $\alpha_0(z)=\alpha_0$ and $\alpha_1(z)=\alpha_1$ for all $z$. Then 
\begin{eqnarray*}
\mathbb P(\varepsilon =1)&=& \mathbb P(\varepsilon=1\vert Z=z),\\
\mathbb P(\varepsilon =1)&=& \mathbb P(D^*=1 \vert Z=z) \mathbb P(\varepsilon=1\vert D^*=1, Z=z) + \mathbb P(D^*=0 \vert Z=z) \mathbb P(\varepsilon=1\vert D^*=0, Z=z),\\
\mathbb P(\varepsilon =1)&=& \alpha_1(z) \mathbb P(D^*=1 \vert Z=z)  + \alpha_0(z) \mathbb P(D^*=0 \vert Z=z),\\
\mathbb P(\varepsilon =1)&=& \alpha_1 \mathbb P(D^*=1 \vert Z=z)  + \alpha_0 \mathbb P(D^*=0 \vert Z=z),\\
\mathbb P(\varepsilon =1)&=& \alpha_1 \mathbb P(D^*=1 \vert Z=z)  + \alpha_0 (1-\mathbb P(D^*=1 \vert Z=z)),\\
\mathbb P(\varepsilon =1)&=&( \alpha_1-\alpha_0) \mathbb P(D^*=1 \vert Z=z)  + \alpha_0,
\end{eqnarray*}
where the first equality holds from Assumption \ref{RA}, the second holds from the law of total probability and Bayes' rule, the third holds from the definition of $\alpha_0(z)$ and $\alpha_1(z)$, and the fourth holds from our assumption that $\alpha_0(z)$ and $\alpha_1(z)$ are constant across $z$. 

If $\alpha_0 \neq \alpha_1$, then $\mathbb P(D^*=1 \vert Z=z)=\frac{\mathbb P(\varepsilon=1)-\alpha_0}{\alpha_1-\alpha_0},$ which is constant across $z$, which contradicts the relevance condition that $P(z)$ is a nontrivial function of $z$, since $P(z)=\mathbb P(D^*=1 \vert Z=z)$ under Assumption \ref{RA}. Therefore, $\alpha_0=\alpha_1.$ Hence, $\alpha_0(z)=\alpha_1(z)=\alpha.$

Suppose now that the misclassification is symmetric in the sense that $\alpha_0(z)=\alpha_1(z)=\alpha(z)$. Then, a similar derivation as above yields
\begin{eqnarray*}
\mathbb P(\varepsilon =1)&=& \alpha_1(z) \mathbb P(D^*=1 \vert Z=z)  + \alpha_0(z) \mathbb P(D^*=0 \vert Z=z),\\
\mathbb P(\varepsilon =1)&=& \alpha(z) \mathbb P(D^*=1 \vert Z=z)  + \alpha(z) \mathbb P(D^*=0 \vert Z=z),\\
\mathbb P(\varepsilon =1)&=& \alpha(z)\left[\mathbb P(D^*=1 \vert Z=z)  + \mathbb P(D^*=0 \vert Z=z)\right],\\
\mathbb P(\varepsilon =1)&=& \alpha(z)
\end{eqnarray*} 
From the last equality, we deduce that $\alpha(z)=\mathbb P(\varepsilon =1)=\alpha.$ Therefore, $\alpha_0(z)=\alpha_1(z)=~\alpha.$
\end{proof}

\section{Allowing for dependence between misclassification and IV}\label{apx:zeps}

Similarly to Equation \eqref{eq:fsharp2}, we have under $Z\ \indep\ (Y_d, V)$, $d=0,1$:
\begin{eqnarray*}
f_{Y \vert Z}\left(y\vert z\right)
&=&  \int^{P(z)}_0 f_{Y_1\vert V}\left(y \vert v \right)dv + \int_{P(z)}^1f_{Y_0 \vert V}\left(y \vert v \right)dv.
\end{eqnarray*}
Hence, for any $P(z') < P(z)$ we have
\begin{eqnarray*}
f_{Y \vert Z}\left(y\vert z\right) - f_{Y\vert Z}\left(y \vert z'\right)&=& \int^{P(z)}_{P(z')}f_{Y_1\vert V}(y \vert v)dv\\
&&\qquad \qquad - \int^{P(z)}_{P(z')}f_{Y_0 \vert V}(y \vert v)dv,
\end{eqnarray*}
Using the triangle inequality, we have
\begin{eqnarray*}
\left \lvert f_{Y \vert Z}\left(y \vert z\right) - f_{Y \vert Z}\left(y \vert z'\right) \right \rvert  &\leq&  \int^{P(z)}_{P(z')}f_{Y_1 \vert V}(y \vert v)dv\\
 && \qquad + \int^{P(z)}_{P(z')}f_{Y_0\vert V}(y \vert v)dv.
\end{eqnarray*}
Therefore, by integrating each side over the support $\mathcal Y$ and using the Fubini-Tonelli theorem, we have
\begin{eqnarray*}
\int_{\mathcal Y}\left \lvert f_{Y \vert Z}\left(y \vert z\right) - f_{Y \vert Z}\left(y \vert z'\right) \right \rvert d \mu_{Y}(y) &\leq&  \int^{P(z)}_{P(z')} dv+\int^{P(z)}_{P(z')}dv= 2\left(P(z)-P(z')\right).
\end{eqnarray*}
Hence, we have $TV_{Y}(z',z) \leq P(z)-P(z') \leq 1$, where 
\begin{eqnarray*}
TV_{Y}(z',z) \equiv \frac{1}{2} \int_{\mathcal Y}\left \lvert f_{Y \vert Z}\left(y \vert z\right) - f_{Y \vert Z}\left(y \vert z'\right) \right \rvert d \mu_{Y}(y).
\end{eqnarray*}
Then, we have 
\begin{eqnarray*}
&& \min\left\{\frac{\Delta_{YZ}(z',z)}{TV_{Y}(z',z)},  \Delta_{YZ}(z',z) \right\}\\
&& \qquad \qquad \leq \frac{\mathbb E[Y\vert P(Z)=P(z)]-\mathbb E[Y\vert P(Z)=P(z')]}{P(z)-P(z')} \leq \\
&& \qquad \qquad \qquad \qquad \qquad \qquad \qquad \qquad \qquad \max\left\{\frac{\Delta_{YZ}(z',z)}{TV_{Y}(z',z)},  \Delta_{YZ}(z',z)\right\}.
\end{eqnarray*}
Therefore, the following bounds hold for the MTE:
\begin{eqnarray}
&& \min\left\{\lim_{z' \rightarrow z} \frac{\Delta_{YZ}(z',z)}{TV_{Y}(z',z)}, 0 \right\} \nonumber \\
&& \qquad \qquad \leq MTE(P(z))\leq \\
&& \qquad \qquad \qquad \qquad \max\left\{\lim_{z' \rightarrow z} \frac{\Delta_{YZ}(z',z)}{TV_{Y}(z',z)},  0\right\}. \nonumber
\end{eqnarray}
These bounds are wider than those derived in Subsection~\ref{anabounds} under Assumptions~\ref{RA}--\ref{Bound:mis}.

\section{Analytical expressions of the bounds in Proposition \ref{propdiscrete}}\label{anaboundsdiscrete}
Redefining the direct pairwise upper bound as
\[
\overline{UB}_p(z',z)
\equiv
\min\bigg\{
\min_{d\in\{0,1\}}
\left[
\mathbb P(D=d\mid Z=z)+\mathbb P(D=d\mid Z=z')
\right],
2\alpha+\Delta_{DZ}(z',z),
2(1-\alpha)-\Delta_{DZ}(z',z)
\bigg\},
\]
we define
\begin{align*}
LB_p(z_{\ell-1},z_\ell)
&\equiv
\max\left\{
LB_p^1(z_{\ell-1},z_\ell),
LB_p^2(z_{\ell-1},z_\ell)
\right\},\\
UB_p(z_{\ell-1},z_\ell)
&\equiv
\min\left\{
UB_p^1(z_{\ell-1},z_\ell),
UB_p^2(z_{\ell-1},z_\ell)
\right\},\\
LB_p^1(z_{\ell-1},z_\ell)
&\equiv
\max_{d\in\{0,1\}}
\left\{
TV_{(Y,D=d)}(z_{\ell-1},z_\ell)
\right\},\\
UB_p^1(z_{\ell-1},z_\ell)
&\equiv
\overline{UB}_p(z_{\ell-1},z_\ell),\\
LB_p^2(z_{\ell-1},z_\ell)
&\equiv
\max_{d\in\{0,1\}}
\left\{
TV_{(Y,D=d)}(z_1,z_L)
\right\}
-
\sum_{k\neq \ell}
\overline{UB}_p(z_{k-1},z_k),\\
UB_p^2(z_{\ell-1},z_\ell)
&\equiv
\overline{UB}_p(z_1,z_L)
-
\sum_{k\neq \ell}
\max_{d\in\{0,1\}}
\left\{
TV_{(Y,D=d)}(z_{k-1},z_k)
\right\}.
\end{align*}

\clearpage

\renewcommand\thefigure{A.\arabic{figure}} 
\setcounter{figure}{0}

\renewcommand\thetable{A.\arabic{table}} 
\setcounter{table}{0}
\setcounter{page}{0} 
\renewcommand\theclaim{D.\arabic{claim}} 
\setcounter{claim}{0}
\pagenumbering{gobble}
\begin{center}
\vspace{2cm}
    \huge{{Online Appendix}}

\huge{Local Average and Marginal Treatment Effects with a Misclassified Treatment}\\
\bigskip
    \Large{Kyunghoon Ban \quad D\'esir\'e K\'edagni\quad Santiago Acerenza}
\end{center}
\vspace{1cm}
\setcounter{page}{0}
\pagenumbering{arabic}

\section{Empirical Relevance of the MTE Bounds}\label{EmpRel}
The identification of the MTE can help reveal the presence of treatment-effect heterogeneity. It can also be useful in estimating policy-relevant treatment effects (PRTEs) and conventional parameters such as the ATE, the average treatment effect on the treated (ATT), the average treatment effect on the untreated (ATU), and the LATE. Tables~\ref{integral} and~\ref{weights}, which we borrow from \citet{heckman2005structural}, show the link between the MTE and these parameters. Unlike the weights in \citet{heckman2005structural}, the weights for ATT, ATU, and PRTE are not point-identified in our setting. They are only partially identified, as is the true propensity score $P(Z)$. Like the MTE, these policy parameters are also partially identified. 
\begin{table}[!htbp]
\centering
\caption{Treatment effects as weighted averages of the $\MTE$}
\label{integral}
\begin{threeparttable}
\begin{tabular}{@{}l@{}}
\toprule
$ATE = \mathbb{E}\!\left[Y_{1}-Y_{0}\right]
      = \int_{0}^{1} \MTE(p)\,\omega_{ATE}(p)\,dp$ \\[0.75em]
$ATT = \mathbb{E}\!\left[Y_{1}-Y_{0}\mid D^*=1\right]
      = \int_{0}^{1} \MTE(p)\,\omega_{ATT}(p)\,dp$ \\[0.75em]
$ATU = \mathbb{E}\!\left[Y_{1}-Y_{0}\mid D^*=0\right]
      = \int_{0}^{1} \MTE(p)\,\omega_{ATU}(p)\,dp$ \\[0.75em]
$LATE(\underline p,\overline p)
      = \mathbb{E}\!\left[Y_{1}-Y_{0}\mid V\in[\underline p,\overline p]\right]
      = \int_{\underline p}^{\overline p} \MTE(p)\,\omega_{LATE}(p)\,dp$ \\[0.75em]
$PRTE
      = \dfrac{\mathbb{E}\!\left[Y_a-Y_{a'}\right]}
              {\int_0^1\{F_{P_{a'}}(p)-F_{P_a}(p)\}\,dp}
      = \int_0^1 \MTE(p)\,\omega_{PRTE}(p,a,a')\,dp$ \\
\bottomrule
\end{tabular}
\begin{tablenotes}[flushleft]
\footnotesize
\item[]\makebox[\linewidth][c]{\begin{minipage}{0.78\textwidth}
\footnotesize\textit{Notes:} The table collects standard MTE representations of average and policy-relevant treatment-effect parameters. The policies $a$ and $a'$ affect treatment only through the instrument. In the present setting, objects involving the latent true propensity score $P(Z)$ are generally partially identified because the binary treatment is misclassified.
\end{minipage}}
\end{tablenotes}
\end{threeparttable}
\end{table}

\begin{table}[!htbp]
\centering
\caption{Weights for MTE-based treatment-effect parameters}
\label{weights}
\begin{threeparttable}
\begin{tabular}{@{}l@{}}
\toprule
$\omega_{ATE}(p)=1$ \\[0.75em]
$\omega_{ATT}(p)
= \dfrac{\int_p^1 f_{P(Z)}(u)\,du}
        {\int_0^1\left\{\int_p^1 f_{P(Z)}(u)\,du\right\}dp}$ \\[1em]
$\omega_{ATU}(p)
= \dfrac{\int_0^p f_{P(Z)}(u)\,du}
        {\int_0^1\left\{\int_0^p f_{P(Z)}(u)\,du\right\}dp}$ \\[1em]
$\omega_{LATE}(p)=\dfrac{1}{\overline p-\underline p}$ \\[1em]
$\omega_{PRTE}(p,a,a')
= \dfrac{F_{P_{a'}}(p)-F_{P_a}(p)}
        {\int_0^1\{F_{P_{a'}}(p)-F_{P_a}(p)\}\,dp}$ \\
\bottomrule
\end{tabular}

\begin{tablenotes}[flushleft]
\footnotesize
\item[]\makebox[\linewidth][c]{\begin{minipage}{0.78\textwidth}
\footnotesize\textit{Notes:} Here, $f_{P(Z)}$ denotes the density of the latent true propensity score and $F_{P_a}$ denotes the distribution of the counterfactual propensity score induced by policy $a$. The $LATE$ weight is evaluated on the interval $[\underline p,\overline p]$.
\end{minipage}}
\end{tablenotes}
\end{threeparttable}
\end{table}

\section{Sharp characterization under conditional misclassification independence assumption} In this subsection, we are going to characterize the (sharp) identified set for the MTE. We add an assumption similar to the non-differential measurement error assumption to the set of our identifying assumptions, which we call \textit{conditional misclassification  independence}.
\begin{assumption}[Conditional misclassification  independence]\label{NDE}
The misclassification variable $\varepsilon$ is independent of $Y_d$ conditional on $V$, i.e., $\varepsilon\ \indep\ Y_d \vert V$, for each $d\in \left\{0,1\right\}$.
\end{assumption}
This assumption states that conditional on the unobserved heterogeneity that drives the selection into treatment, misreporting is independent of the potential outcomes. Combined with Assumption \ref{RA}, it implies that misreporting is independent of the outcome conditional on the true treatment unobserved heterogeneity. This assumption could be too restrictive, as there may exist some returns to misreporting. In our leading example, there could exist some ``returns to lying'' about college completion, as discussed in \citet{HuLewbel2012}, and \citet{DiTraglia2019}. This assumption is similar to the so-called non-differential measurement error assumption in the literature, which states that conditional on the true treatment, the measurement error is independent of the outcome.

For the sake of simplicity, we assume in this subsection that the distribution of $V$ given $\varepsilon$ is absolutely continuous. We have 
\begin{eqnarray}
\mathbb P(Y\in A, D=1 \vert P(Z)=p)  &=& \mathbb P(Y_1\in A, \varepsilon=0, V \leq p) + \mathbb P(Y_0\in A, \varepsilon=1, V > p),\nonumber \\
&=& (1-\alpha) \int^{p}_0\mathbb P\left(Y_1\in A \vert V=v, \varepsilon =0\right)f_{V\vert \varepsilon=0}(v)dv\nonumber \\
&& + \alpha \int_{p}^1\mathbb P\left(Y_0\in A \vert V=v, \varepsilon =1\right)f_{V\vert \varepsilon=1}(v)dv. \label{eq:main}
\end{eqnarray}
where the first equality follows from the results derived in the previous subsection, and the second equality holds from the law of iterated expectations.
Therefore, by taking the derivatives of both sides of this equality with respect to $p$, we obtain the following:
\begin{eqnarray}
\frac{\partial \mathbb P(Y\in A, D=1 \vert P(Z)=p) }{\partial p}
&=& (1-\alpha) f_{V\vert \varepsilon=0}(p) \mathbb P\left(Y_1\in A \vert V=p, \varepsilon =0\right) \nonumber\\
&& - \alpha f_{V\vert \varepsilon=1}(p) \mathbb P\left(Y_0\in A \vert V=p, \varepsilon =1\right), \nonumber\\
&=& (1-\alpha) f_{V\vert \varepsilon=0}(p) \mathbb P\left(Y_1\in A \vert V=p\right) \nonumber\\
&&\qquad \qquad- \alpha f_{V\vert \varepsilon=1}(p) \mathbb P\left(Y_0\in A \vert V=p\right), \label{eq:mte1}
\end{eqnarray}
where the second equality holds under Assumption \ref{NDE}.
Similarly, we can show that 
\begin{eqnarray}
\frac{\partial \mathbb P(Y\in A, D=0 \vert P(Z)=p) }{\partial p} &=&  \alpha f_{V\vert \varepsilon=1}(p) \mathbb P\left(Y_1\in A \vert V=p\right) \nonumber\\
&& \qquad \qquad  - (1-\alpha) f_{V\vert \varepsilon=0}(p) \mathbb P\left(Y_0\in A \vert V=p\right). \label{eq:mte2}
\end{eqnarray}

Applying equality (\ref{eq:mte1}) to the special case where $A=\mathcal Y$, and using the fact that $f_{V}(p)=1$ (since $V\sim \mathcal U_{[0,1]}$), we have 
\begin{eqnarray*}
(1-\alpha) f_{V\vert \varepsilon=0}(p) - \alpha f_{V\vert \varepsilon=1}(p) &=& \frac{\partial \mathbb P(D=1 \vert P(Z)=p) }{\partial p},\\
(1-\alpha) f_{V\vert \varepsilon=0}(p) +\alpha f_{V\vert \varepsilon=1}(p)&=& 1.
\end{eqnarray*}
Therefore,
\begin{eqnarray*}
f_{V\vert \varepsilon=0}(p)  &=&\frac{1+ \frac{\partial \mathbb P(D=1 \vert P(Z)=p) }{\partial p}}{2(1-\alpha)},\\
f_{V\vert \varepsilon=1}(p)  &=&\frac{1- \frac{\partial \mathbb P(D=1 \vert P(Z)=p) }{\partial p}}{2\alpha}.
\end{eqnarray*}
Hence, the function $P$ must satisfy the following conditions:
\begin{eqnarray}
 && \frac{1+ \frac{\partial \mathbb P(D=1 \vert P(Z)=p) }{\partial p}}{2(1-\alpha)} \geq 0,\label{eq:P1}\\
&&  \frac{1- \frac{\partial \mathbb P(D=1 \vert P(Z)=p) }{\partial p}}{2\alpha} \geq 0, \label{eq:P2}\\
&& \int^1_0  \frac{1+ \frac{\partial \mathbb P(D=1 \vert P(Z)=p) }{\partial p}}{2(1-\alpha)} dp = 1,\label{eq:P3}\\
&& \int^1_0 \frac{1- \frac{\partial \mathbb P(D=1 \vert P(Z)=p) }{\partial p}}{2\alpha} dp = 1 \label{eq:P4},\\
&& 0 \leq \frac{\alpha f_{V\vert \varepsilon=1}(p)\kappa_0(A;p)-\left(1-\alpha\right)f_{V\vert \varepsilon=0}(p)\kappa_1(A;p)}{\alpha f_{V\vert \varepsilon=1}(p)-\left(1-\alpha\right)f_{V\vert \varepsilon=0}(p)} \leq 1, \label{eq:P5}\\
&& 0 \leq \frac{\left(1-\alpha \right) f_{V\vert \varepsilon=0}(p)\kappa_0(A;p)-\alpha f_{V\vert \varepsilon=0}(p)\kappa_1(A;p)}{\alpha f_{V\vert \varepsilon=1}(p)-\left(1-\alpha\right)f_{V\vert \varepsilon=0}(p)} \leq 1, \label{eq:P6}\\
&&\int^1_0 \left(1-\alpha \right) f_{V\vert \varepsilon=0}(p) \frac{\alpha f_{V\vert \varepsilon=1}(p)\kappa_0(A;p)-\left(1-\alpha\right)f_{V\vert \varepsilon=0}(p)\kappa_1(A;p)}{\alpha f_{V\vert \varepsilon=1}(p)-\left(1-\alpha\right)f_{V\vert \varepsilon=0}(p)} dp\nonumber\\
&& \qquad \qquad \qquad \mathbb = P(Y\in A, D=1\vert P(Z)=1), \label{eq:P8}\end{eqnarray}
\begin{eqnarray}
&&\int^1_0 \alpha f_{V\vert \varepsilon=1}(p) \frac{\left(1-\alpha \right) f_{V\vert \varepsilon=0}(p)\kappa_0(A;p)-\alpha f_{V\vert \varepsilon=0}(p)\kappa_1(A;p)}{\alpha f_{V\vert \varepsilon=1}(p)-\left(1-\alpha\right)f_{V\vert \varepsilon=0}(p)} dp\nonumber\\
&& \qquad \qquad \qquad \mathbb = P(Y\in A, D=1\vert P(Z)=0), \label{eq:P9}\\
&&\mathbb P\left(Y\in A, D=d\vert Z=z\right) = \mathbb P\left(Y\in A, D=d\vert P(Z)=P(z)\right), \label{eq:P7}
\end{eqnarray}
for all $p$ such that $\frac{\partial \mathbb P(D=1 \vert P(Z)=p) }{\partial p}\neq 0$, all $d\in\{0,1\}$, and all Borel set $A \subset \mathcal Y$, where $\kappa_1(A;p)=\frac{\partial \mathbb P(Y\in A, D=1 \vert P(Z)=p) }{\partial p}$, and $\kappa_0(A;p)=\frac{\partial \mathbb P(Y\in A, D=0 \vert P(Z)=p) }{\partial p}$. The constraints (\ref{eq:P5}) and (\ref{eq:P6}) come from the fact that $\mathbb P\left(Y_1\in A \vert V=p\right)$ and $\mathbb P\left(Y_0\in A \vert V=p\right)$ are probabilities. In fact, using equalities (\ref{eq:mte1}) and (\ref{eq:mte2})
we can solve for $\mathbb P\left(Y_1\in A \vert V=p\right)$ and $\mathbb P\left(Y_0\in A \vert V=p\right)$ as follows:
\begin{eqnarray*}
\mathbb P\left(Y_1\in A \vert V=p\right) &=&\frac{\alpha f_{V\vert \varepsilon=1}(p)\kappa_0(A;p)-\left(1-\alpha\right)f_{V\vert \varepsilon=0}(p)\kappa_1(A;p)}{\alpha f_{V\vert \varepsilon=1}(p)-\left(1-\alpha\right)f_{V\vert \varepsilon=0}(p)} ,\\
\mathbb P\left(Y_0\in A \vert V=p\right) &=& \frac{\left(1-\alpha \right) f_{V\vert \varepsilon=0}(p)\kappa_0(A;p)-\alpha f_{V\vert \varepsilon=0}(p)\kappa_1(A;p)}{\alpha f_{V\vert \varepsilon=1}(p)-\left(1-\alpha\right)f_{V\vert \varepsilon=0}(p)}.
\end{eqnarray*}
Equations (\ref{eq:P8}) and (\ref{eq:P9}) are like terminal conditions, and come from Equation (\ref{eq:main}).
Note that conditions (\ref{eq:P3}) and (\ref{eq:P4}) are equivalent, and come from the fact that density functions integrate to 1, while Equations (\ref{eq:P1}) and (\ref{eq:P2}) are the non-negativity conditions for density functions. 
The following proposition holds.
\begin{proposition}\label{prop:sharp}
Suppose that model (\ref{seq1}) along with Assumptions \ref{RA}--\ref{NDE} hold. In addition, suppose that the distribution of $V$ given $\varepsilon$ is absolutely continuous. For a given $\alpha \in [0,\bar{\alpha}]$, the constraints (\ref{eq:P1})--(\ref{eq:P7}) yield the (sharp) identified set for the function $P: \mathcal Z \rightarrow [0,1]$, and therefore for the MTE.
\end{proposition}
The proof of Proposition \ref{prop:sharp} is given in Appendix \ref{sharpnessproof}. Although this proposition provides sharp identification region for the propensity score $P(z)$ and the MTE, this identified set is not tractable. It is important to point out that the identified set for the MTE in the proposition is \textit{uniformly sharp} in the sense that the joint distribution on $(Y_1,Y_0,V,\varepsilon,Z)$ that achieves each element in this latter is the same across $p$.

\subsection{Proof of sharpness under misclassification conditional independence} \label{sharpnessproof}
\begin{proof}
For each $\alpha \in (0,1)$, we need to find a joint distribution on the vector $\left(\tilde{Y}_1, \tilde{Y}_0, \tilde{V}, \tilde{\varepsilon}, Z\right)$, such that it satisfies model (\ref{seq1}) and Assumptions \ref{RA}, \ref{Cont}, and \ref{NDE}, and induces the joint distribution on $(Y,D,Z)$. For any function $P(z)$ satisfying the constraints in (\ref{eq:P1})-(\ref{eq:P7}), define:  
\begin{eqnarray*}
\mathbb P(\tilde{\varepsilon}=1 \vert Z=z)&=& \alpha,\\
f_{\tilde{V}\vert \tilde{\varepsilon}=0, Z=z}(p)  &=&\frac{1+ \frac{\partial \mathbb P(D=1 \vert P(Z)=p) }{\partial p}}{2(1-\alpha)},\\
f_{\tilde{V}\vert \tilde{\varepsilon}=1, Z=z}(p)  &=&\frac{1- \frac{\partial \mathbb P(D=1 \vert P(Z)=p) }{\partial p}}{2\alpha},\\
\mathbb P(\tilde{Y}_1 \leq y \vert \tilde{V}=p,\tilde{\varepsilon}=1, Z=z) &=& \frac{\alpha f_{\tilde{V}\vert \tilde{\varepsilon}=1}(p)\kappa_0(y;p)-\left(1-\alpha\right)f_{\tilde{V}\vert \tilde{\varepsilon}=0}(p)\kappa_1(y;p)}{\alpha f_{\tilde{V}\vert \tilde{\varepsilon}=1}(p)-\left(1-\alpha\right)f_{\tilde{V}\vert \tilde{\varepsilon}=0}(p)},\\
\mathbb P(\tilde{Y}_1 \leq y \vert \tilde{V}=p,\tilde{\varepsilon}=0, Z=z) &=& \frac{\alpha f_{\tilde{V}\vert \tilde{\varepsilon}=1}(p)\kappa_0(y;p)-\left(1-\alpha\right)f_{\tilde{V}\vert \tilde{\varepsilon}=0}(p)\kappa_1(y;p)}{\alpha f_{\tilde{V}\vert \tilde{\varepsilon}=1}(p)-\left(1-\alpha\right)f_{\tilde{V}\vert \tilde{\varepsilon}=0}(p)},\\
\mathbb P(\tilde{Y}_0 \leq y \vert \tilde{V}=p,\tilde{\varepsilon}=1, Z=z) &=& \frac{\left(1-\alpha\right) f_{\tilde{V}\vert \tilde{\varepsilon}=0}(p)\kappa_0(y;p)-\alpha f_{\tilde{V}\vert \tilde{\varepsilon}=1}(p)\kappa_1(y;p)}{\alpha f_{\tilde{V}\vert \tilde{\varepsilon}=1}(p)-\left(1-\alpha\right)f_{\tilde{V}\vert \tilde{\varepsilon}=0}(p)},\\
\mathbb P(\tilde{Y}_0 \leq y \vert \tilde{V}=p,\tilde{\varepsilon}=0, Z=z) &=& \frac{\left(1-\alpha\right) f_{\tilde{V}\vert \varepsilon=0}(p)\kappa_0(y;p)-\alpha f_{\tilde{V}\vert \varepsilon=1}(p)\kappa_1(y;p)}{\alpha f_{\tilde{V}\vert \tilde{\varepsilon}=1}(p)-\left(1-\alpha\right)f_{\tilde{V}\vert \varepsilon=0}(p)},
\end{eqnarray*}
where $\kappa_1(y;p)=\frac{\partial \mathbb P(Y\leq y, D=1 \vert P(Z)=p)}{\partial p}$, and $\kappa_0(y;p)=\frac{\partial \mathbb P(Y\leq y, D=0 \vert P(Z)=p) }{\partial p}$.
Define $$\mathbb P(\tilde{Y}_0 \leq y_0, \tilde{Y}_1 \leq y_1 \vert \tilde{V}=p,\varepsilon=\ell, Z=z)=\mathbb P(\tilde{Y}_0 \leq y_0 \vert \tilde{V}=p,\varepsilon=\ell, Z=z)\mathbb P(\tilde{Y}_1 \leq y_1 \vert \tilde{V}=p,\varepsilon=\ell, Z=z).$$ 
It is easy to check that the above quantities are well-defined probabilities/distributions under the constraints (\ref{eq:P1})-(\ref{eq:P9}), and the vector $\left(\tilde{Y}_1, \tilde{Y}_0, \tilde{V}, \tilde{\varepsilon}, Z\right)$ satisfies Assumptions \ref{RA}, \ref{Cont}, and \ref{NDE}.
Define
\begin{eqnarray}\label{seq2}
\left\{ \begin{array}{lcl}
     \tilde{Y}&=&\tilde{Y}_1\tilde{D}^*+\tilde{Y}_0(1-\tilde{D}^*)\\ \\
     \tilde{D}^*&=&\mathbbm{1}\left\{\tilde{V}\leq P(Z)\right\}\\ \\
     \tilde{D}&=&\tilde{D}^* (1-\tilde{\varepsilon}) +(1-\tilde{D}^*) \tilde{\varepsilon}
     \end{array} \right.
\end{eqnarray}
We will now show that the vector $\left(\tilde{Y}, \tilde{D}, Z \right)$ has the same distribution as the vector $(Y,D,Z)$. We have
\begin{eqnarray*}
\mathbb P(\tilde{Y}\leq y, \tilde{D}=1 \vert Z=z) &=&\mathbb P(\tilde{Y}\leq y, \tilde{D}=1 \vert P(Z)=P(z)),\\
&=& (1-\alpha) \int^{P(z)}_0\mathbb P\left(\tilde{Y}_1\leq y \vert \tilde{V}=v, \tilde{\varepsilon} =0\right)f_{\tilde{V}\vert \tilde{\varepsilon}=0}(v)dv\\
&& + \alpha \int_{P(z)}^1\mathbb P\left(\tilde{Y}_0\leq y \vert \tilde{V}=v, \tilde{\varepsilon} =1\right)f_{\tilde{V}\vert \tilde{\varepsilon}=1}(v)dv,\\
&=& (1-\alpha) \int^{1}_0 \mathbb P\left(\tilde{Y}_1\leq y \vert \tilde{V}=v, \tilde{\varepsilon} =0\right)f_{\tilde{V}\vert \tilde{\varepsilon}=0}(v)dv\\
&& -(1-\alpha) \int^{1}_{P(z)} \mathbb P\left(\tilde{Y}_1\leq y \vert \tilde{V}=v, \tilde{\varepsilon} =0\right)f_{\tilde{V}\vert \tilde{\varepsilon}=0}(v)dv\\
&& + \alpha \int_{P(z)}^1\mathbb P\left(\tilde{Y}_0\leq y \vert \tilde{V}=v, \tilde{\varepsilon} =1\right)f_{\tilde{V}\vert \tilde{\varepsilon}=1}(v)dv,\\
&=& (1-\alpha) \int^{1}_0 \mathbb P\left(\tilde{Y}_1\leq y \vert \tilde{V}=v, \tilde{\varepsilon} =0\right)f_{\tilde{V}\vert \tilde{\varepsilon}=0}(v)dv\\
&& + \int^{1}_{P(z)} -(1-\alpha) f_{\tilde{V}\vert \tilde{\varepsilon}=0}(v) \mathbb P\left(\tilde{Y}_1\leq y \vert \tilde{V}=v, \tilde{\varepsilon} =0\right)\\
&& + \alpha f_{\tilde{V}\vert \tilde{\varepsilon}=1}(v) \mathbb P\left(\tilde{Y}_0\leq y \vert \tilde{V}=v, \tilde{\varepsilon} =1\right)dv,
\end{eqnarray*}
where the first equality holds from Equation (\ref{eq:P7}).
Given the definition of our DGP, we have
\begin{eqnarray*}
&& -(1-\alpha) f_{\tilde{V}\vert \tilde{\varepsilon}=0}(v) \mathbb P\left(\tilde{Y}_1\leq y \vert \tilde{V}=v, \tilde{\varepsilon} =0\right)+ \alpha f_{\tilde{V}\vert \tilde{\varepsilon}=1}(v) \mathbb P\left(\tilde{Y}_0\leq y \vert \tilde{V}=v, \tilde{\varepsilon}=1\right)\\
&& \qquad = - \kappa_1(y;v)= - \frac{\partial \mathbb P(Y\leq y, D=1 \vert P(Z)=v)}{\partial v}.
\end{eqnarray*}
Therefore,
\begin{eqnarray*}
&&\int^{1}_{P(z)} -(1-\alpha) f_{\tilde{V}\vert \tilde{\varepsilon}=0}(v) \mathbb P\left(\tilde{Y}_1\leq y \vert \tilde{V}=v, \tilde{\varepsilon} =0\right) + \alpha f_{\tilde{V}\vert \tilde{\varepsilon}=1}(v) \mathbb P\left(\tilde{Y}_0\leq y \vert \tilde{V}=v, \tilde{\varepsilon} =1\right)dv\\
&& \qquad = \mathbb P(Y\leq y, D=1 \vert P(Z)=P(z))-\mathbb P(Y\leq y, D=1 \vert P(Z)=1),\\
&& \qquad = \mathbb P(Y\leq y, D=1 \vert Z=z)-\mathbb P(Y\leq y, D=1 \vert P(Z)=1)
\end{eqnarray*}
At this point, it remains to show that 
\begin{eqnarray*}
 \mathbb P(Y\leq y, D=1 \vert P(Z)=1) = (1-\alpha) \int^{1}_0 \mathbb P\left(\tilde{Y}_1\leq y \vert \tilde{V}=v, \tilde{\varepsilon} =0\right)f_{\tilde{V}\vert \tilde{\varepsilon}=0}(v)dv.
\end{eqnarray*}
This equality holds from condition (\ref{eq:P8}).

Similarly, $\mathbb P(\tilde{Y}\leq y, \tilde{D}=0 \vert Z=z)=\mathbb P(Y\leq y, D=0 \vert Z=z)$.
\end{proof}

\section{Additional results on the empirical illustration}
\subsection{More details on the implementation}\label{sec:implementation}
For each value of the misclassification
parameter $\alpha$, we first transform the fitted observed propensity score into
the corresponding latent propensity score, denoted by $\widehat P_i=\widehat
P_i(\alpha)$. We then evaluate the estimated individual-level MTE surface on a
uniform grid $\mathcal V=\{v_1,\ldots,v_G\}$ on $[0,1]$, with $G=1000$ in the
baseline specification. Let $\widehat m_i(v;\alpha)$ denote the estimated MTE for
individual $i$ at margin $v$ under candidate value $\alpha$. The simulation-based
integration averages this estimated surface over the empirical distribution of
covariates and over the uniform distribution of the latent cost variable
$V$.

For a fixed $\alpha$, the aggregate treatment-effect parameters are computed as
\[
\widehat{ATE}(\alpha)
=
\frac{1}{nG}\sum_{i=1}^n\sum_{g=1}^G
\widehat m_i(v_g;\alpha),
\]
\[
\widehat{ATT}(\alpha)
=
\frac{
\sum_{i=1}^n\sum_{g=1}^G
1\{\widehat P_i(\alpha)>v_g\}\widehat m_i(v_g;\alpha)
}{
\sum_{i=1}^n\sum_{g=1}^G
1\{\widehat P_i(\alpha)>v_g\}
},
\qquad
\widehat{ATU}(\alpha)
=
\frac{
\sum_{i=1}^n\sum_{g=1}^G
1\{\widehat P_i(\alpha)<v_g\}\widehat m_i(v_g;\alpha)
}{
\sum_{i=1}^n\sum_{g=1}^G
1\{\widehat P_i(\alpha)<v_g\}
}.
\]
The ATT and ATU weights follow from the threshold-crossing treatment rule
$D_i(v)=1\{\widehat P_i(\alpha)>v\}$: observations with
$\widehat P_i(\alpha)>v_g$ are treated at latent resistance value $v_g$, whereas
observations with $\widehat P_i(\alpha)<v_g$ are untreated.

We also report an average MTE around the individual-specific margin. For
bandwidth $\zeta$, this parameter is computed as
\[
\widehat{AMTE}_{\zeta}(\alpha)
=
\frac{
\sum_{i=1}^n\sum_{g=1}^G
1\{|\widehat P_i(\alpha)-v_g|<\zeta\}
\widehat m_i(v_g;\alpha)
}{
\sum_{i=1}^n\sum_{g=1}^G
1\{|\widehat P_i(\alpha)-v_g|<\zeta\}
}.
\]
We compute $\widehat{AMTE}_{\zeta}$ for
$\zeta\in\{0.10,0.05,0.01\}$; the main table displays the case
$\zeta=0.05$.

{
We also estimate policy relevant treatment effects for counterfactual policies
indexed by $a\in\{0.1,0.2,0.3\}$. The policy considered here shifts the latent
propensity score according to
\[
P_i^a(\alpha)
=
P_i(\alpha)+a\{1-P_i(\alpha)\}.
\]
This policy increases the treatment propensity for every individual and therefore is monotone in the sense that $\widehat P_i^a(\alpha)\geq \widehat P_i(\alpha)$.
It has the interpretation that fraction $a$ of the status-quo untreated latent-cost interval, whose length is $1-\widehat P_i(\alpha)$, is shifted
into treatment.

The PRTE is defined as the normalized policy effect, namely the change in mean outcomes induced by the counterfactual policy divided by the induced change in treatment participation:
\[
PRTE(a;\alpha)
=
\frac{
\mathbb E[Y_i^a-Y_i]
}{
\mathbb E[D_i^a-D_i]
},
\qquad
D_i^a=1\{P_i^a(\alpha)>V_i\},
\qquad
D_i=1\{P_i(\alpha)>V_i\}.
\]
Under the threshold-crossing selection model and policy invariance, this standard PRTE can be written as a weighted average of the MTE. Since the policy is monotone upward, there are no individuals shifted out of treatment. The only policy-induced switchers are those whose latent resistance values satisfy
\[
P_i(\alpha)<V_i<P_i^a(\alpha).
\]
Therefore,
\[
PRTE(a;\alpha)
=
\frac{
\mathbb E\left[
\int_{P_i(\alpha)}^{P_i^a(\alpha)}
m_i(v;\alpha)\,dv
\right]
}{
\mathbb E\left[
P_i^a(\alpha)-P_i(\alpha)
\right]
}.
\]
Thus the PRTE is an average MTE for the individuals and latent-resistance values whose treatment status changes under the counterfactual policy. 

We estimate this object by plug-in simulation-based integration. 
For each individual $i$ and grid point $v_g$, we compare the status-quo treatment rule $1\{\widehat P_i(\alpha)>v_g\}$ with the counterfactual treatment rule $1\{\widehat P_i^a(\alpha)>v_g\}$. 
The estimated PRTE is then
\[
\widehat{PRTE}(a;\alpha)
=
\frac{
\sum_{i=1}^n\sum_{g=1}^G
1\left\{
1\{\widehat P_i^a(\alpha)>v_g\}
\neq
1\{\widehat P_i(\alpha)>v_g\}
\right\}
\widehat m_i(v_g;\alpha)
}{
\sum_{i=1}^n\sum_{g=1}^G
1\left\{
1\{\widehat P_i^a(\alpha)>v_g\}
\neq
1\{\widehat P_i(\alpha)>v_g\}
\right\}
}.
\]
For the monotone policy $\widehat P_i^a(\alpha)=\widehat P_i(\alpha)+a\{1-\widehat P_i(\alpha)\}$, the switcher indicator above is equivalent, up to endpoint conventions, to $1\{\widehat P_i(\alpha)<v_g<\widehat P_i^a(\alpha)\}$. 
Hence the estimator is the grid approximation to
\[
\frac{
\mathbb E\left[
\int_{\widehat P_i(\alpha)}^{\widehat P_i^a(\alpha)}
\widehat m_i(v;\alpha)\,dv
\right]
}{
\mathbb E\left[
\widehat P_i^a(\alpha)-\widehat P_i(\alpha)
\right]
}.
\]
The same propensity-shift policy has been considered in \citet{SasakiUra2021}.\footnote{\citet{SasakiUra2021} estimate the PRTE on the same data using a double-debiased orthogonal-score estimator. 
Here the PRTE is recovered by first estimating the MTE surface and then integrating that surface over the policy-induced switcher region, with cluster-bootstrap inference.}

When $\alpha$ is treated as unknown, we compute the above aggregate parameters for each grid value of $\alpha$ in the admissible set.
For a given upper bound $\bar\alpha$, the reported identified set for any aggregate parameter $\theta\in\{ATE,ATT,ATU,AMTE_\zeta,PRTE(a)\}$ is obtained by taking the envelope over the corresponding $\alpha$ grid:
\[
\widehat\theta_L(\bar\alpha)
=
\min_{\alpha\leq\bar\alpha}\widehat\theta(\alpha),
\qquad
\widehat\theta_U(\bar\alpha)
=
\max_{\alpha\leq\bar\alpha}\widehat\theta(\alpha).
\]
Thus, for the PRTE, the final reported object is an estimated identified set for the normalized PRTE associated with the specified propensity-shift policy.
}

\subsection{Additional Empirical Results under \(V\)-Dependent Misclassification}\label{apx:sup}
This appendix reports two specifications that relax the benchmark assumption \(\varepsilon\indep V\). Both specifications use the same first-stage logit, alpha grid, Robinson residualization, Gaussian local-quadratic and degree-2 penalized B-spline branches, aggregate-parameter definitions, and cluster bootstrap procedure described in Section~\ref{App}. The only difference is the mapping from the fitted observed-treatment propensity score \(q_i=\widehat{\Pr}(D_i=1\mid Z_i,X_i)\) to the candidate latent true propensity score.

For each value of \(\bar{\alpha}\), the identified set is the envelope over all admissible alpha values in the nested grid. For MTE curves, the plotted confidence region uses the bootstrap envelope over the alpha grid. For aggregate parameters, the confidence region is the outer union of alpha-wise bootstrap intervals over the same grid. The confidence regions are therefore conservative.

\begin{figure}[h]
\centering
\includegraphics[width=\textwidth]{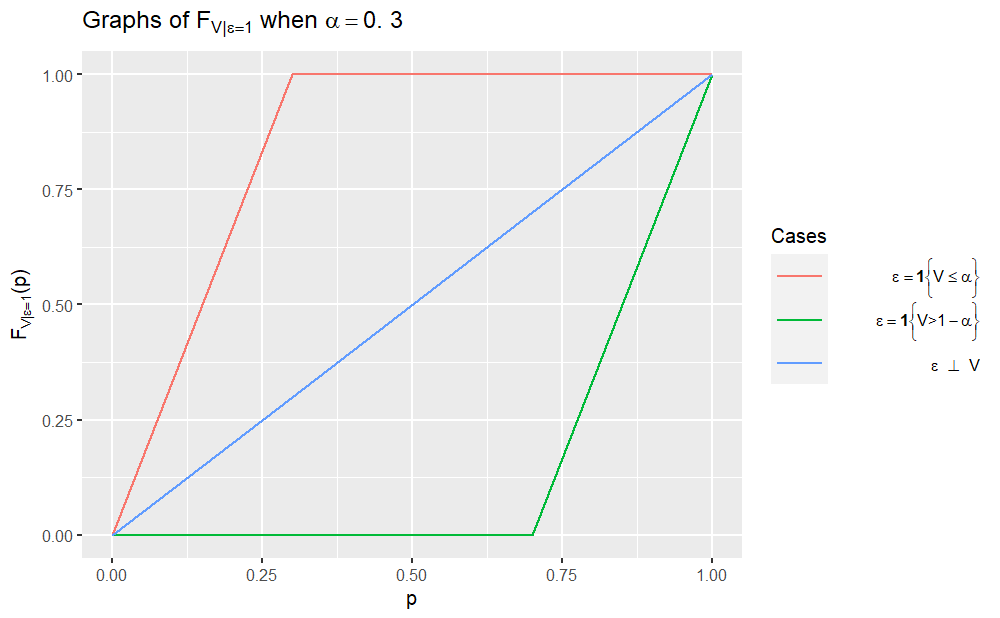}
\caption{Conditional distribution of \(V\) among misclassified observations}
\label{fig.condcdf}
\figurenote{The figure plots $F_{V\mid \varepsilon=1}(p)$ when $\alpha=0.3$ under the benchmark specification $\varepsilon\indep V$, the low-cost specification $\varepsilon=\mathbbm{1}\{V\leq\alpha\}$, and the high-cost specification $\varepsilon=\mathbbm{1}\{V>1-\alpha\}$.}
\end{figure}

The three specifications differ in how the misclassified observations are located along the latent cost distribution. Figure~\ref{fig.condcdf} illustrates this distinction by plotting \(F_{V\mid \varepsilon=1}(p)\), the distribution function of \(V\) among misclassified observations. Under \(\varepsilon\indep V\), misclassification is spread uniformly over the latent cost distribution. Under \(\varepsilon=\mathbbm{1}\{V\leq\alpha\}\), all misclassified observations come from the low-\(V\) tail. Under \(\varepsilon=\mathbbm{1}\{V>1-\alpha\}\), all misclassified observations come from the high-\(V\) tail.

This figure also clarifies the first-stage correction. Let \(q=\Pr(D=1\mid Z,X)\) denote the observed-treatment propensity score and let \(p=\Pr(D^*=1\mid Z,X)\) denote the latent true propensity score. For any specification of \(F_{V\mid\varepsilon=1}\),
\[
q
=
p+\alpha-2\alpha F_{V\mid\varepsilon=1}(p).
\]
Thus, each assumption about the location of misclassification along the \(V\)-distribution implies a different mapping from \(q\) to \(p\). When \(\varepsilon=\mathbbm{1}\{V\leq\alpha\}\), the correction treats misclassified observations as low-\(V\) false negatives and shifts \(p\) upward on the main branch. When \(\varepsilon=\mathbbm{1}\{V>1-\alpha\}\), the correction treats misclassified observations as high-\(V\) false positives and shifts \(p\) downward on the main branch.

\subsubsection{Specification \(\varepsilon=\mathbbm{1}\{V\leq\alpha\}\)}

\begin{figure}[!htbp]
\centering
\includegraphics[width=\textwidth]{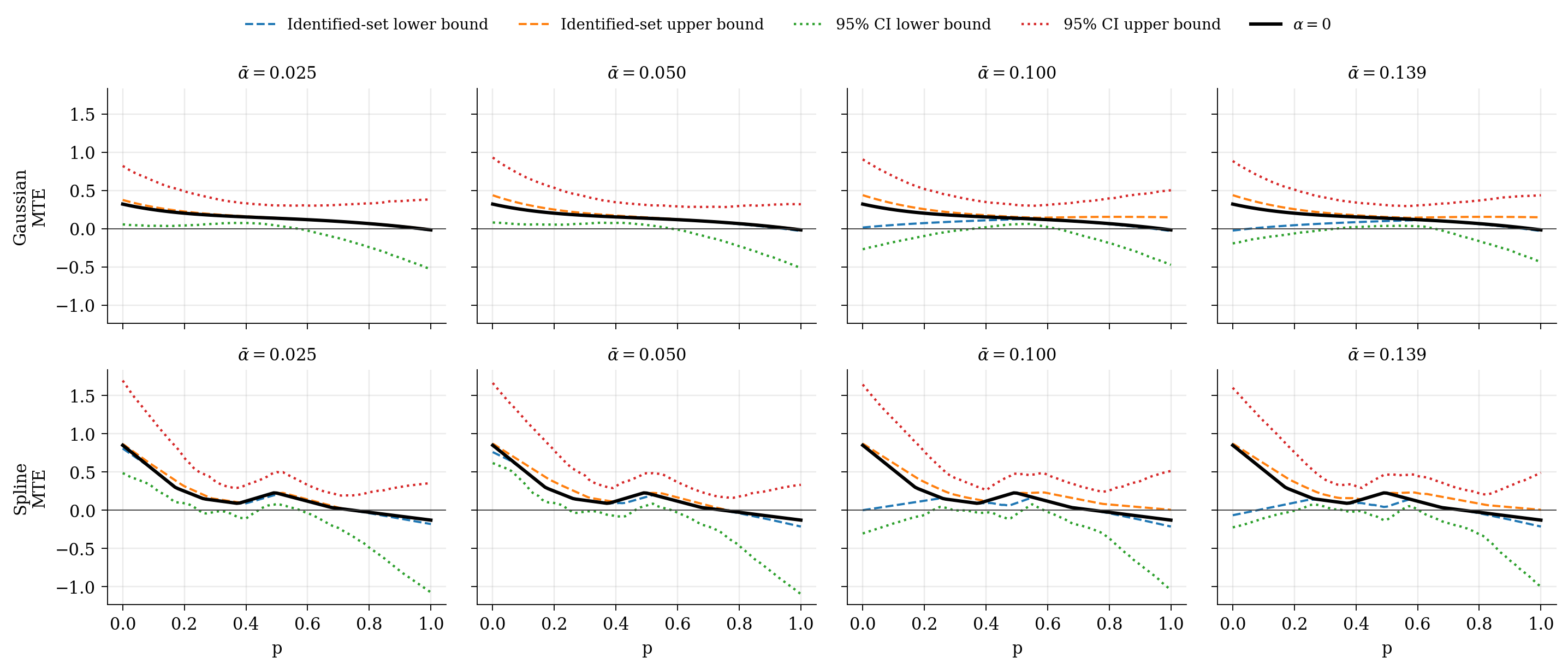}
\caption{MTE identified sets and conservative 95\% confidence regions under \(\varepsilon=\mathbbm{1}\{V\leq\alpha\}\).}
\label{fig:app-mte-vleq}
\figurenote{The upper row reports the Gaussian local-quadratic estimator and the lower row reports the degree-2 penalized B-spline estimator. The columns correspond to $\bar\alpha\in\{0.025,0.050,0.100,0.139\}$. Dashed curves are the identified-set endpoints, dotted curves are the conservative 95\% confidence-region endpoints, and the solid black curve is the estimate at $\alpha=0$. The confidence region uses 1000 community-level cluster-bootstrap replications and the alpha-grid envelope.}
\end{figure}

The first specification concentrates misclassification among individuals with lower values of the latent cost variable:
\[
P_i^{\leq}(\alpha)=
\begin{cases}
\alpha-q_i, & q_i<\alpha,\\
q_i+\alpha, & q_i\ge \alpha.
\end{cases}
\]
After this transformation, the propensity score is winsorized to \([10^{-4},1-10^{-4}]\). Figure~\ref{fig:app-mte-vleq} reports the corresponding MTE identified sets and confidence regions. The MTE remains broadly decreasing in \(p\), and the low-cost portion of the curve continues to show the largest estimated returns.

\input{tables/refined/table_app_lowv_compact.tex}

\begin{figure}[!htbp]
\centering
\includegraphics[width=0.92\textwidth]{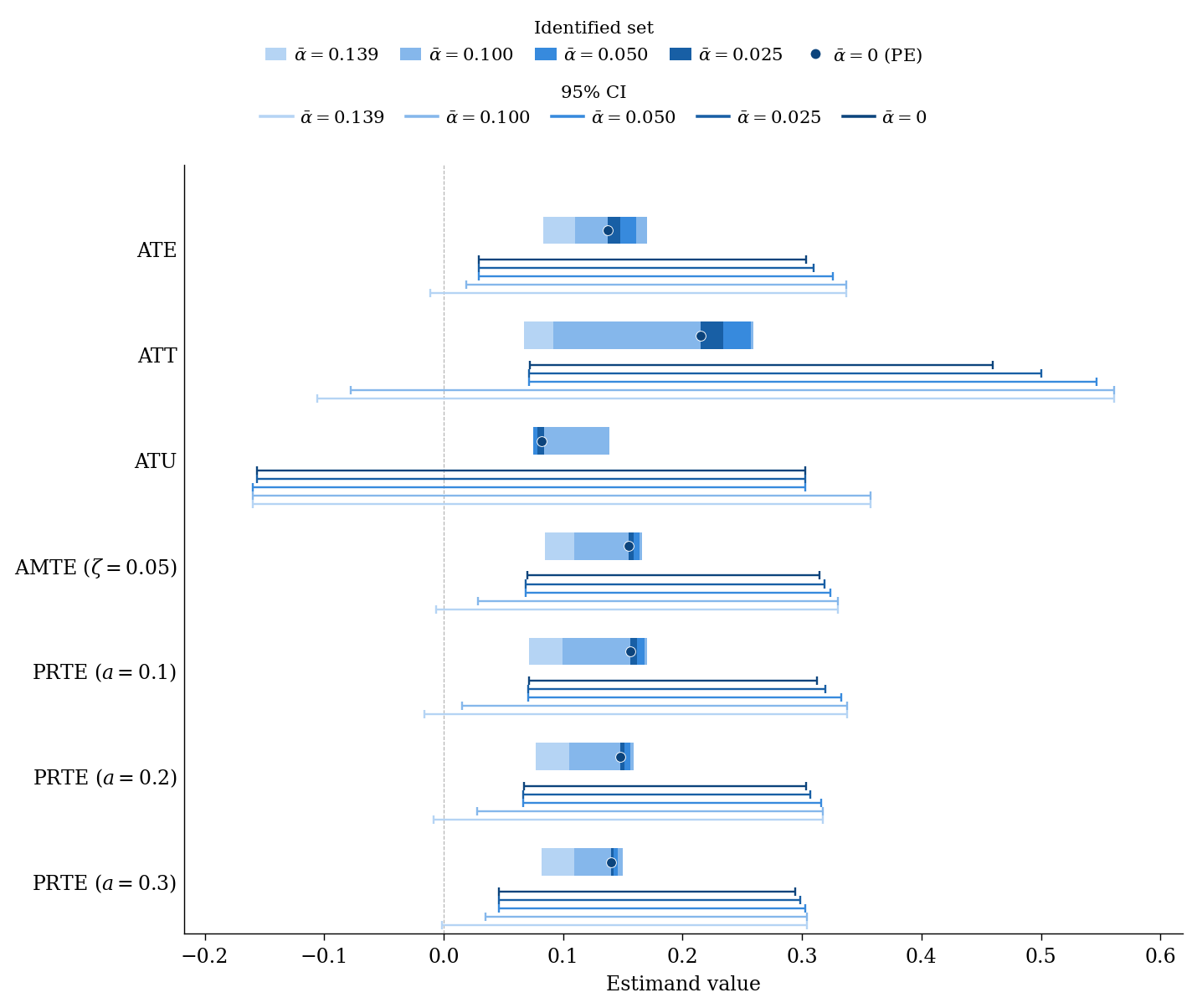}
\caption{Aggregate-return summaries under \(\varepsilon=\mathbbm{1}\{V\leq\alpha\}\) using the Gaussian local-quadratic estimator.}
\label{fig:app-aggregate-vleq-gaussian}
\figurenote{Each row is an estimand. Shaded bands report identified sets, and horizontal intervals report conservative 95\% confidence regions using 1000 community-level cluster-bootstrap replications.}
\end{figure}

\begin{figure}[!htbp]
\centering
\includegraphics[width=0.92\textwidth]{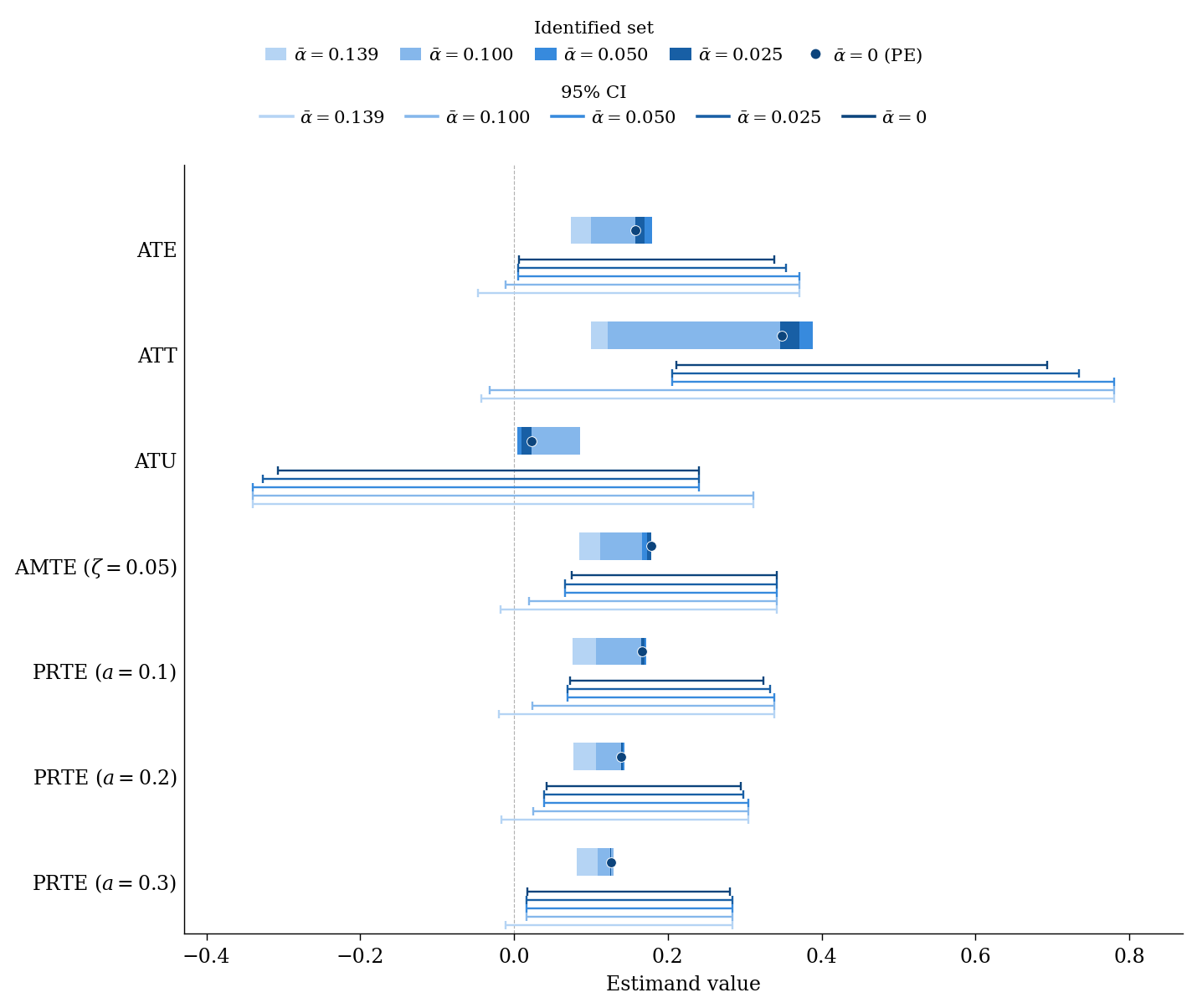}
\caption{Aggregate-return summaries under \(\varepsilon=\mathbbm{1}\{V\leq\alpha\}\) using the degree-2 penalized B-spline estimator.}
\label{fig:app-aggregate-vleq-spline}
\figurenote{Each row is an estimand. Shaded bands report identified sets, and horizontal intervals report conservative 95\% confidence regions using 1000 community-level cluster-bootstrap replications.}
\end{figure}

Under \(\bar{\alpha}=0.139\), \(\varepsilon=\mathbbm{1}\{V\leq\alpha\}\) gives wider bounds than \(\varepsilon\indep V\) but preserves the main qualitative ranking: ATT is generally larger than ATU, and the PRTE point-identified intervals remain positive for the policy shifts considered.

\subsubsection{Specification \(\varepsilon=\mathbbm{1}\{V>1-\alpha\}\)}

\begin{figure}[!htbp]
\centering
\includegraphics[width=\textwidth]{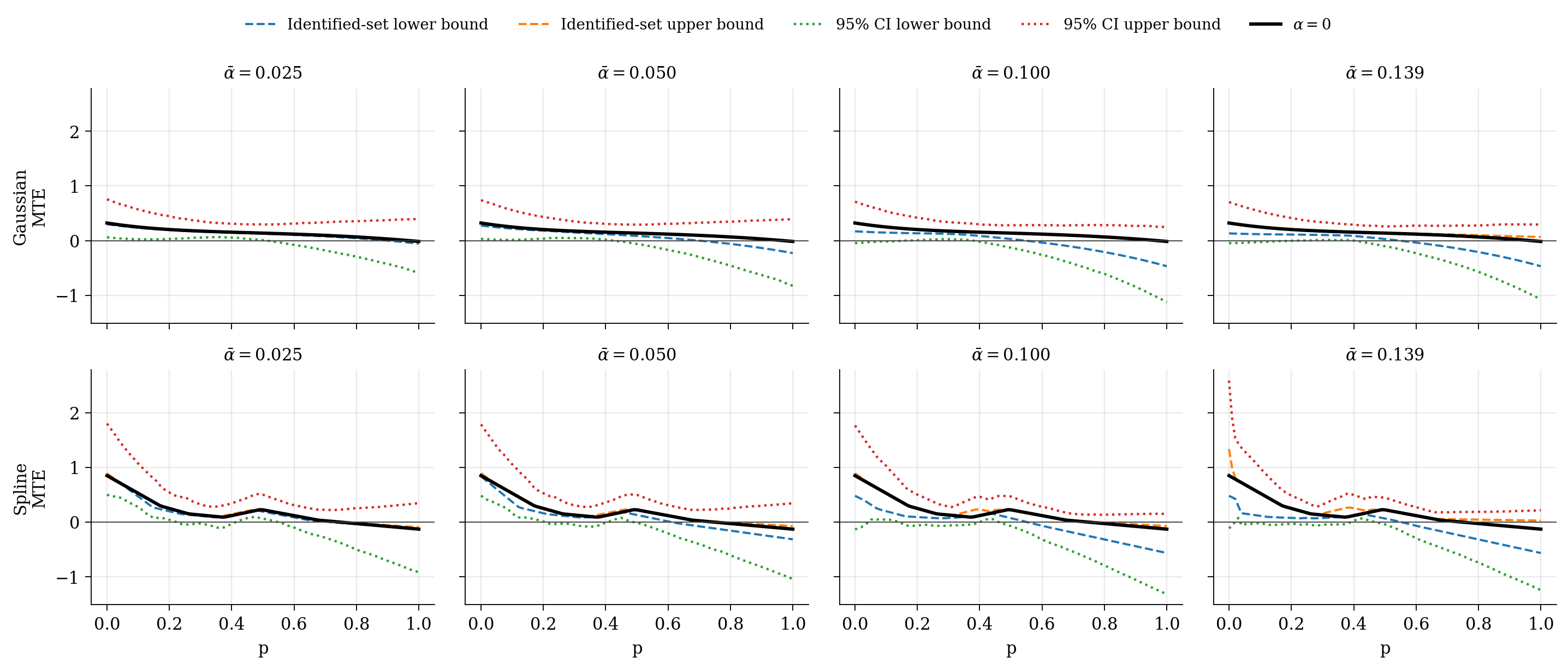}
\caption{MTE identified sets and conservative 95\% confidence regions under \(\varepsilon=\mathbbm{1}\{V>1-\alpha\}\).}
\label{fig:app-mte-vgt}
\figurenote{The upper row reports the Gaussian local-quadratic estimator and the lower row reports the degree-2 penalized B-spline estimator. The columns correspond to $\bar\alpha\in\{0.025,0.050,0.100,0.139\}$. Dashed curves are the identified-set endpoints, dotted curves are the conservative 95\% confidence-region endpoints, and the solid black curve is the estimate at $\alpha=0$. The confidence region uses 1000 community-level cluster-bootstrap replications and the alpha-grid envelope.}
\end{figure}

The second specification concentrates misclassification among individuals with higher values of the latent cost variable:
\[
P_i^{>} (\alpha)=
\begin{cases}
q_i-\alpha, & q_i<1-\alpha,\\
1-q_i+1-\alpha, & q_i\ge 1-\alpha.
\end{cases}
\]
After this transformation, the propensity score is winsorized to \([10^{-4},1-10^{-4}]\). Figure~\ref{fig:app-mte-vgt} reports the MTE results. Because this specification places measurement error near higher cost margins, it has a larger effect on portions of the MTE curve where estimated returns are lower.

\input{tables/refined/table_app_highv_compact.tex}

\begin{figure}[!htbp]
\centering
\includegraphics[width=0.92\textwidth]{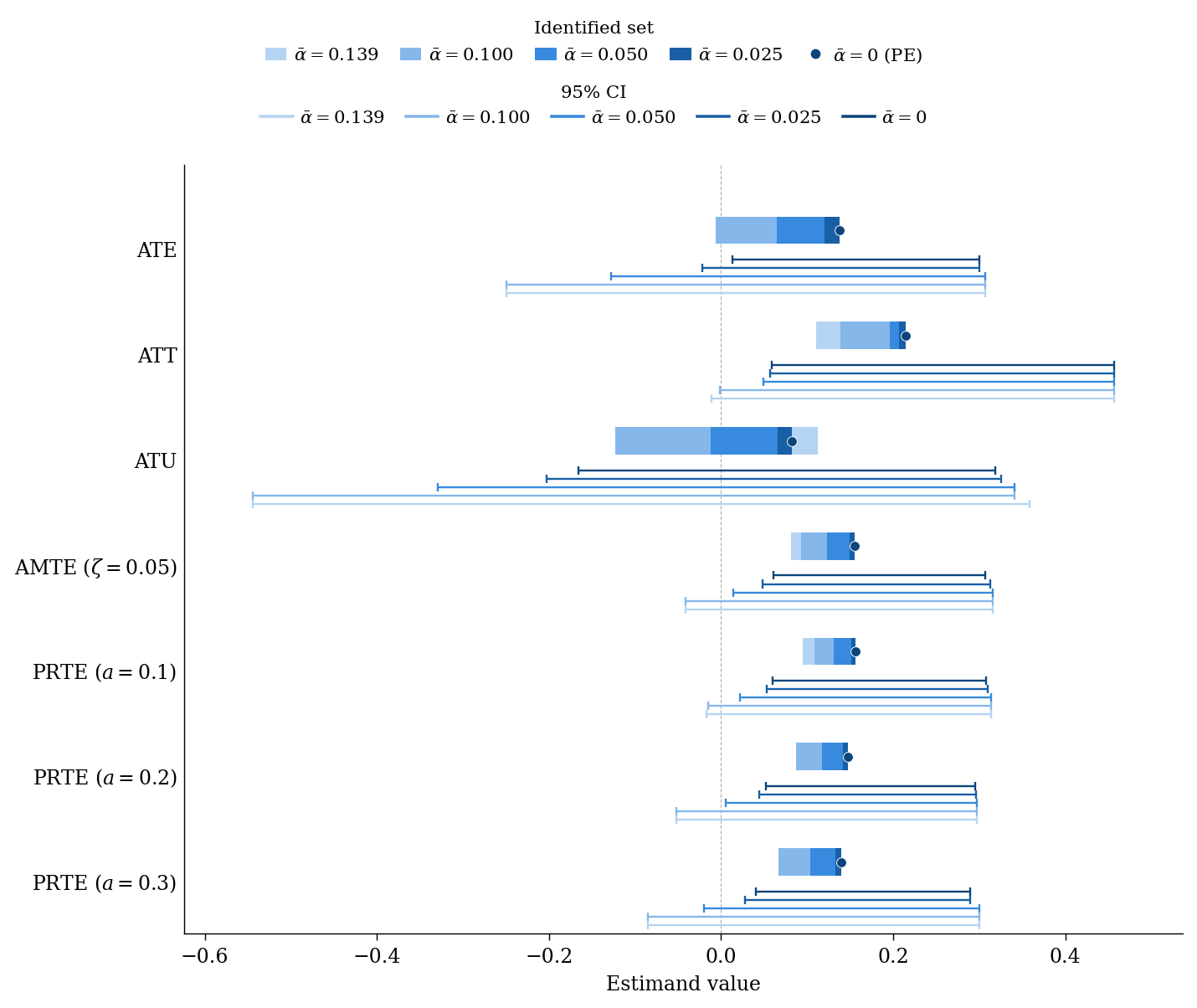}
\caption{Aggregate-return summaries under \(\varepsilon=\mathbbm{1}\{V>1-\alpha\}\) using the Gaussian local-quadratic estimator.}
\label{fig:app-aggregate-vgt-gaussian}
\figurenote{Each row is an estimand. Shaded bands report identified sets, and horizontal intervals report conservative 95\% confidence regions using 1000 community-level cluster-bootstrap replications.}
\end{figure}

\begin{figure}[!htbp]
\centering
\includegraphics[width=0.92\textwidth]{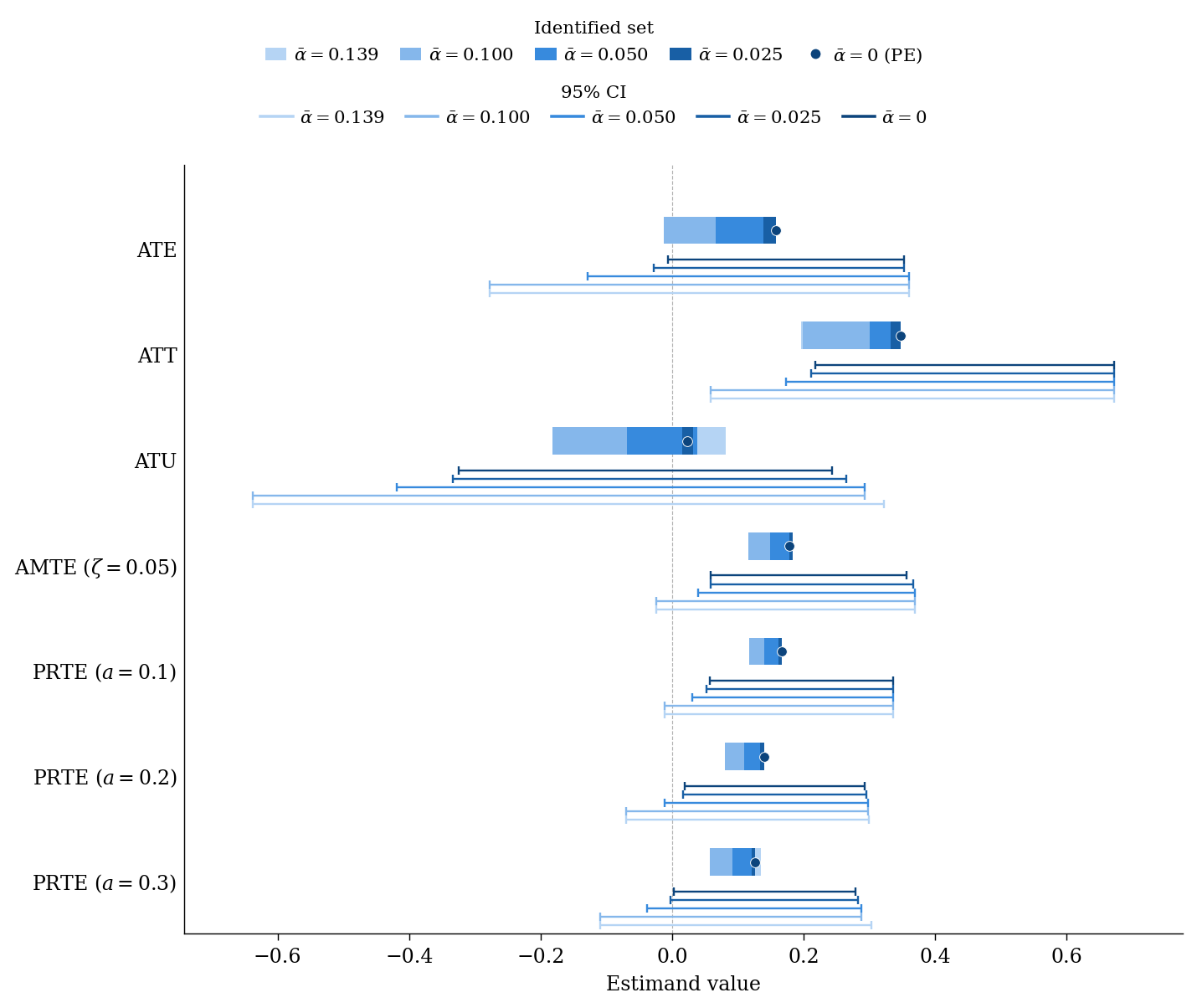}
\caption{Aggregate-return summaries under \(\varepsilon=\mathbbm{1}\{V>1-\alpha\}\) using the degree-2 penalized B-spline estimator.}
\label{fig:app-aggregate-vgt-spline}
\figurenote{Each row is an estimand. Shaded bands report identified sets, and horizontal intervals report conservative 95\% confidence regions using 1000 community-level cluster-bootstrap replications.}
\end{figure}

The specification \(\varepsilon=\mathbbm{1}\{V>1-\alpha\}\) is the least favorable of the three specifications for average returns. For large \(\bar{\alpha}\), the lower endpoints of ATE and ATU can become negative, especially in the spline branch. Nevertheless, the ATT remains positive at the point-bound level, and the PRTE upper endpoints remain close to the \(\varepsilon\indep V\) benchmark. The comparison across Figures~\ref{fig:app-mte-vleq}--\ref{fig:app-mte-vgt} reinforces the main lesson of the empirical illustration: allowing for measurement error primarily widens the identified set and the conservative confidence region, while the broad declining shape of the MTE remains visible.

Tables~\ref{tab:app-vleq-bounds}--\ref{tab:app-vgt-ci} report detailed wide identified-set and confidence-bound tables for all selected aggregate parameters under both specifications.

\input{tables/refined/table_app_lowv_bounds.tex}
\input{tables/refined/table_app_lowv_ci.tex}
\input{tables/refined/table_app_highv_bounds.tex}
\input{tables/refined/table_app_highv_ci.tex}

\clearpage
\subsection{Parametric Copula Identification: Simulation and Empirical Illustration}

\subsubsection{Monte Carlo evidence}\label{sec:mc_copula_validation}
This subsection reports a Monte Carlo exercise designed to assess the finite-sample performance of the Gaussian-copula identification approach. The data-generating process follows the structural model described above. Specifically,
\[
Z\sim \mathrm{Unif}[0,1],\qquad
P(z)=\Phi(\delta_0+\delta_1 z),\qquad
D^*=\mathbf 1\{V\le P(Z)\},
\]
with $(\delta_0,\delta_1)=(-2.5,5.0)$. Misclassification is generated by
\[
D=D^*(1-\varepsilon)+(1-D^*)\varepsilon,
\qquad
\varepsilon=\mathbf 1\{\xi\le \alpha\},
\]
where $(V,\xi)$ have uniform margins and are coupled through a Gaussian copula with correlation parameter $\rho$. The calibration uses $(\alpha,\rho)=(0.2,0.5)$. The outcome equation is
\[
Y=\beta D^*+U,\qquad
\beta=\beta_0+\beta_1(1-\sqrt V)+\eta,
\]
with $(\beta_0,\beta_1)=(1.5,1.0)$, $U\sim N(2,1)$, and $\eta\sim N(0,1)$.

For each Monte Carlo replication, we estimate the first-stage structural parameters $(\alpha,\delta_0,\delta_1,\rho)$ using both nonlinear least squares (NLS) and Bernoulli maximum likelihood (MLE) applied to the observed treatment equation
\[
\mathbb E[D\mid Z=z]
=
P(z)+\alpha-2C_G(P(z),\alpha;\rho).
\]
Given the estimated first-stage parameters, the outcome parameters are estimated from the implied conditional mean equation
\[
\mathbb E[Y\mid Z=z]
=
\mu_U+\beta_0P(z)+\beta_1\left\{P(z)-\frac{2}{3}P(z)^{3/2}\right\}.
\]
The selected MTE and PRTE objects are then computed by plug-in. Table~\ref{tab:mc_full_sequential} reports results based on $R=500$ replications with sample size $n=25000$ in each replication.

\begin{table}[!htbp]
\centering
\scriptsize
\caption{Monte Carlo performance of the full sequential estimator}
\label{tab:mc_full_sequential}
\begin{tabular}{llrrrrrrr}
\toprule
Method & Object & True & Mean & Bias & RMSE & SD & Mean SE & Coverage \\
\midrule
MLE & $\alpha$ & 0.200 & 0.200 & 0.000 & 0.006 & 0.006 & 0.006 & 0.956 \\
MLE & $\delta_0$ & -2.500 & -2.535 & -0.035 & 0.193 & 0.190 & 0.199 & 0.962 \\
MLE & $\delta_1$ & 5.000 & 5.040 & 0.040 & 0.248 & 0.245 & 0.263 & 0.970 \\
MLE & $\rho$ & 0.500 & 0.487 & -0.013 & 0.085 & 0.084 & 0.077 & 0.956 \\
MLE & $\beta_0$ & 1.500 & 1.468 & -0.032 & 0.236 & 0.234 & 0.235 & 0.980 \\
MLE & $\beta_1$ & 1.000 & 1.082 & 0.082 & 0.628 & 0.624 & 0.620 & 0.982 \\
MLE & $MTE(0.50)$ & 1.793 & 1.785 & -0.008 & 0.061 & 0.060 & 0.063 & 0.972 \\
MLE & $PRTE(0.10)$ & 2.036 & 2.058 & 0.022 & 0.134 & 0.132 & 0.125 & 0.972 \\
\addlinespace
NLS & $\alpha$ & 0.200 & 0.200 & 0.000 & 0.006 & 0.006 & 0.006 & 0.956 \\
NLS & $\delta_0$ & -2.500 & -2.539 & -0.039 & 0.204 & 0.201 & 0.209 & 0.958 \\
NLS & $\delta_1$ & 5.000 & 5.045 & 0.045 & 0.258 & 0.254 & 0.270 & 0.972 \\
NLS & $\rho$ & 0.500 & 0.485 & -0.015 & 0.092 & 0.091 & 0.084 & 0.964 \\
NLS & $\beta_0$ & 1.500 & 1.463 & -0.037 & 0.250 & 0.247 & 0.250 & 0.976 \\
NLS & $\beta_1$ & 1.000 & 1.095 & 0.095 & 0.663 & 0.657 & 0.658 & 0.978 \\
NLS & $MTE(0.50)$ & 1.793 & 1.783 & -0.009 & 0.064 & 0.063 & 0.066 & 0.966 \\
NLS & $PRTE(0.10)$ & 2.036 & 2.061 & 0.025 & 0.143 & 0.141 & 0.135 & 0.970 \\
\bottomrule
\end{tabular}

\vspace{0.5em}
\begin{minipage}{0.98\textwidth}
\footnotesize
\textit{Notes}: The table reports Monte Carlo summaries for the full sequential estimator under the Gaussian-copula DGP. The first stage estimates $(\alpha,\delta_0,\delta_1,\rho)$ from the observed treatment equation, and the second stage estimates $(\beta_0,\beta_1)$ from the outcome mean equation using the generated propensity score. $MTE(0.50)$ denotes the marginal treatment effect evaluated at $p=0.50$. $PRTE(0.10)$ denotes the policy-relevant treatment effect for the policy shift $P^a(z)=a+(1-a)P(z)$ with $a=0.10$. ``SD'' is the Monte Carlo standard deviation of the estimates, ``Mean SE'' is the average estimated standard error, and ``Coverage'' is the empirical coverage rate of nominal 95\% confidence intervals based on the reported standard errors. Results are based on 500 replications with sample size 25000 per replication.
\end{minipage}
\end{table}

The results indicate that the Gaussian-copula first stage is numerically informative at this sample size. Both MLE and NLS recover the structural parameters with small bias, and the empirical coverage rates for $(\alpha,\delta_0,\delta_1,\rho)$ are close to the nominal 95\% level. The estimates of the downstream outcome parameters and the selected treatment-effect functionals are also centered close to their true values. Overall, the exercise supports the interpretation that the proposed parametric identification strategy is feasible in sufficiently large samples. 

\clearpage

\subsection{Empirical illustration using a parametric copula first stage}\label{sec:empirical_copula_illustration}
This subsection applies the parametric-copula first-stage approach to an empirical illustration based on a covariate-adjusted earnings outcome. The goal is not to replace the baseline partial-identification analysis, but to examine what the proposed first-stage parametrization implies when combined with a flexible second stage. The analysis starts from the raw data. The working sample restricts to the estimation sample used in the original empirical exercise and contains 2608 observations from 304 clusters. The observed treatment mean is 0.416; the mean and standard deviation of the residualized outcome are 6.835 and 0.903, respectively; and the transformed instrument $Z_{\mathrm{cdf}}$ has 41 distinct support points in $[0.006,1.000]$.

\textbf{Pre-first-stage residualization.} To control for predetermined covariates while keeping the illustration stable, we construct a residualized outcome once before estimating the copula first stage. Let $Y_i=\log(\texttt{earnhr00}_i)$ and let $p_i^0$ denote the legacy observed-treatment propensity score saved in the raw data. The preprocessing step estimates the partially linear projection
\[
Y_i = W_i'\beta + K(p_i^0)+u_i,
\qquad
W_i=\begin{bmatrix}X_i' & p_i^0X_i'\end{bmatrix}',
\]
where $X_i$ contains predetermined demographic, family-background, religion/rural, school-type, and province controls, and $K(\cdot)$ is approximated by a cubic B-spline in $p_i^0$. Operationally, we implement a Robinson-style partialling-out step: residualize $Y_i$ and $W_i$ on the spline basis in $p_i^0$, regress the residualized outcome on the residualized $W_i$, and then form
\[
Y_i^\perp = Y_i-W_i'\widehat\beta.
\]
This one-time residualization is intended as a transparent covariate-adjustment step. It is deliberately performed before the copula first stage and outside the bootstrap loop. Therefore, the bootstrap intervals below should be interpreted as conditional on this preprocessing step. This is less ambitious than a fully generated-propensity Robinson procedure that would residualize $Y$, $X$, and $\widehat P X$ after every first-stage estimate, but it avoids a high-variance residualization step inside each bootstrap replication. The preprocessing uses 50 design columns after interacting covariates with $p_i^0$ and the spline basis and explains about 6.2\% of the variation in log earnings.

\textbf{Copula first stage and flexible second stage.} For each copula family $c\in\{\mathrm{Gaussian},\mathrm{AMH},\mathrm{Gumbel}\}$, the latent propensity is parameterized by a low-dimensional spline index:
\[
P_c(z)=\Phi\{B_K(z)'\gamma_c\},\qquad
Q_c(z)=\Pr(D=1\mid Z=z)=P_c(z)+\alpha-2C_c(P_c(z),\alpha;\theta_c),
\]

where the Gaussian case uses $\theta_c=\rho$. The estimates below use Bernoulli maximum likelihood, with the upper bound $\alpha\leq 0.139$ and the copula-parameter bounds used in the numerical implementation. Given the recovered propensity score $\widehat P_i=\widehat P_c(Z_i)$, the second stage is deliberately nonparametric. We estimate
\[
\mu_c(p)\equiv E[Y_i^\perp\mid \widehat P_c(Z_i)=p]
\]
by a penalized cubic B-spline and compute
\[
\widehat{MTE}_c(p)=\frac{d\widehat\mu_c(p)}{dp}.
\]
For the policy shift $T_a(p)=a+(1-a)p$, $a\in\{0.1,0.2,0.3\}$, we report the observed-support PRTE,
\[
\widehat{PRTE}_{obs,c}(a)
=\int_{[\min_i\widehat P_i,\max_i\widehat P_i]}
\widehat{MTE}_c(p)\widehat\omega_a(p)\,dp.
\]
The policy weight outside the recovered support is not renormalized. Confidence intervals are cluster-bootstrap percentile intervals. In each bootstrap replication, clusters are resampled and the copula first stage, recovered propensity score, spline outcome regression, MTE curve, and PRTE are re-estimated, while the pre-first-stage residualization is held fixed.

\input{output_app3_sympara/tab_emp_copula_first_stage.tex}

Table~\ref{tab:emp_copula_first_stage} shows that the Gaussian and Gumbel point estimates place both $\alpha$ and the dependence parameter at their imposed upper bounds. AMH produces a smaller point estimate of $\alpha$ and a dependence-parameter estimate at the imposed upper bound. The bootstrap boundary-hit rates are non-negligible, especially for $\alpha$ in the Gaussian and Gumbel specifications and for the AMH copula parameter. The Gumbel condition number is large, and the Gaussian and AMH condition numbers also point to limited numerical separation of the first-stage parameters.

\input{output_app3_sympara/tab_emp_copula_spline.tex}

\begin{figure}[H]
\centering
\includegraphics[width=0.98\textwidth]{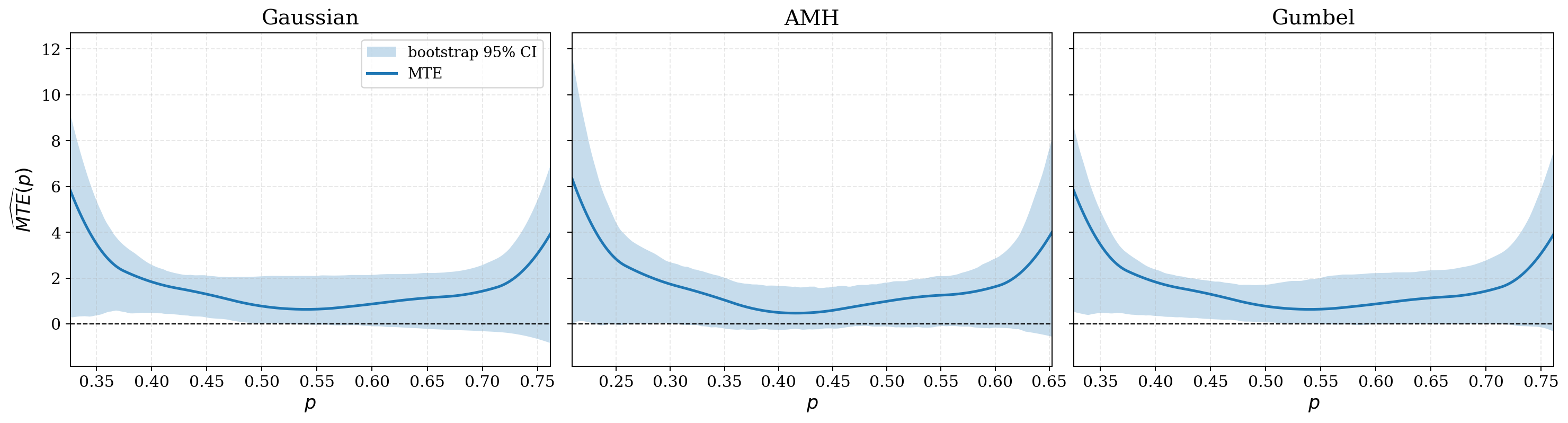}
\caption{Nonparametric MTE curves under alternative copula first stages}
\label{fig:emp_copula_mte}
\figurenote{Each panel plots the derivative of the fitted spline outcome curve with respect to the recovered latent propensity score. The shaded bands are 95\% cluster-bootstrap percentile intervals based on 499 successful bootstrap replications. The horizontal dashed line marks zero. The plotted support is the support of $\widehat P_c(Z)$ recovered under each copula, so the horizontal scale differs across panels.}
\end{figure}

Figure~\ref{fig:emp_copula_mte} shows similar U-shaped point estimates across copulas. The point estimates are positive over the plotted support, largest near the lower edge, smaller in the middle, and rising toward the upper edge. Bootstrap uncertainty is concentrated near the boundaries of the recovered support.

\input{output_app3_sympara/tab_emp_copula_mte_point.tex}

\begin{figure}[H]
\centering
\includegraphics[width=0.92\textwidth]{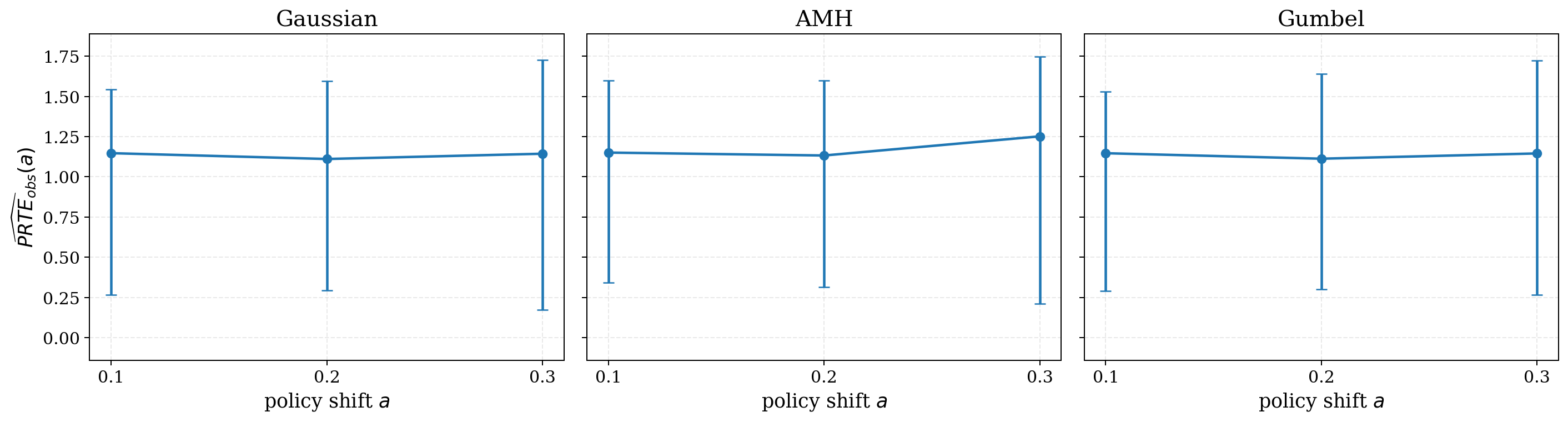}
\caption{Observed-support PRTE estimates under alternative copula first stages}
\label{fig:emp_copula_prte}
\figurenote{Each panel reports observed-support PRTE estimates for $a\in\{0.1,0.2,0.3\}$. Error bars are 95\% cluster-bootstrap percentile intervals. The PRTE is computed by integrating the estimated MTE only over the recovered support of $\widehat P_c(Z)$; omitted policy-relevant weight outside that support is not renormalized.}
\end{figure}

\input{output_app3_sympara/tab_emp_copula_prte.tex}

Overall, the empirical exercise illustrates both the usefulness and the fragility of the parametric-copula first stage. The spline-index first stage yields recovered propensity supports of $[0.326,0.762]$ for Gaussian and Gumbel and $[0.209,0.652]$ for AMH, with positive observed-support PRTE estimates for all displayed policy shifts. The MTE point estimates are qualitatively similar across copulas and positive over the plotted support. Boundary-hit rates and the large Gumbel condition number indicate weak numerical discipline in the first stage. The results should be interpreted as a structured empirical illustration of the copula-based first stage combined with a flexible second stage and pre-specified covariate adjustment; they do not substitute for the more robust bounds analysis.

\subsection{Empirical TV-Based Bounds under General Misclassification}\label{subsec:empirical-tv-mte-prte}

This subsection provides an empirical illustration of the TV-based marginal treatment effect (MTE) and policy-relevant treatment effect (PRTE) bounds. The exercise uses 
the working outcome $Y_i^\perp$, where the outcome adjustment follows the Robinson-style residualization used in Subsection~\ref{sec:empirical_copula_illustration}, and the observed treatment $D_i$.
Also, we take
\[
    Z_i = \widehat F_{\mathrm{kmsmp}}(\mathrm{kmsmp}_i)
\]
as the scalar instrument, so that the empirical instrument is normalized to the unit interval.  For notational simplicity, the equations below write $Y$ for the working outcome $Y^\perp$.  The outcome regression
$\mu(z)=E[Y\mid Z=z]$, the observed propensity score
$Q(z)=E[D\mid Z=z]$, and the joint subdensities
$f_{Y,D=d\mid Z=z}(y)$, $d\in\{0,1\}$, are estimated using Gaussian-kernel smoothers.  The implementation uses a 15-point grid for $z$, a 200-point grid for the numerical integration over $y$, and kernel-density TV bounds without additional relaxation.  The PRTE confidence intervals reported in the figures are based on a full cluster bootstrap with 1,000 replications; conditional on the one-time residualization above, each bootstrap replication recomputes the first-stage smoothing estimates, TV bounds, MTE bounds, feasible paths, and PRTE endpoints.

For adjacent grid points $z_{i-1}<z_i$, the empirical TV lower bound on the latent propensity-score increment is
\begin{equation}
\label{eq:emp-lbp-tv}
    \widehat{LB}_{p,i}
    =
    \max_{d\in\{0,1\}}
    \int
    \left|
        \widehat f_{Y,D=d\mid Z=z_i}(y)
        -
        \widehat f_{Y,D=d\mid Z=z_{i-1}}(y)
    \right|dy .
\end{equation}
The corresponding nodewise MTE bounds are obtained from the local ratio
\begin{equation}
\label{eq:emp-local-ratio}
    \widehat R_i
    =
    \frac{\widehat\mu(z_i)-\widehat\mu(z_{i-1})}
         {\max\{\widehat{LB}_{p,i},\varepsilon\}},
    \qquad
    \widehat{\underline{m}}(z_i)=\min\{0,\widehat R_i\},
    \qquad
    \widehat{\overline{m}}(z_i)=\max\{0,\widehat R_i\},
\end{equation}
where $\varepsilon$ is a small numerical floor.  For a fixed misclassification ceiling $\bar\alpha$, feasible latent propensity-score paths $P$ are required to satisfy the pointwise restrictions
\begin{equation}
\label{eq:emp-pointwise-p}
    \max\{0,\widehat Q(z_i)-\bar\alpha\}
    \le P(z_i) \le
    \min\{1,\widehat Q(z_i)+\bar\alpha\},
\end{equation}
together with the signed increment restrictions implied by $\widehat Q$ and $\bar\alpha$, the TV magnitude restriction $|\Delta P_i|\ge \widehat{LB}_{p,i}$, and the SignTrack restriction on $\Delta P_i$.  In the reported run, the SignTrack direction is taken from the empirical sign of each nonzero $\Delta\widehat Q_i$.  The PRTE bound for policy parameter $a$ is then computed by applying the Riemann--Stieltjes representation to each feasible path.  Specifically, for
\[
    t_a(p)=\max\left\{0,\frac{p-a}{1-a}\right\},
\]
the pathwise coefficient for node $i$ is
\begin{equation}
\label{eq:emp-prte-coef}
    \widehat B_i(P)
    =
    \frac{
        \widehat F_{P(Z)}(P(z_i))
        -
        \widehat F_{P(Z)}(t_a(P(z_i)))
    }
    {a\{1-\widehat E[P(Z)]\}}
    \Delta P_i,
\end{equation}
where $\widehat F_{P(Z)}$ is the empirical distribution of $P(Z_i)$.  Combining $\widehat B_i(P)$ with the one-sided MTE bounds in \eqref{eq:emp-local-ratio} gives a pathwise lower and upper PRTE value; the reported empirical PRTE bounds are the minimum and maximum over the searched feasible paths.

\begin{figure}[tbp]
    \centering
    \includegraphics[width=0.92\linewidth]{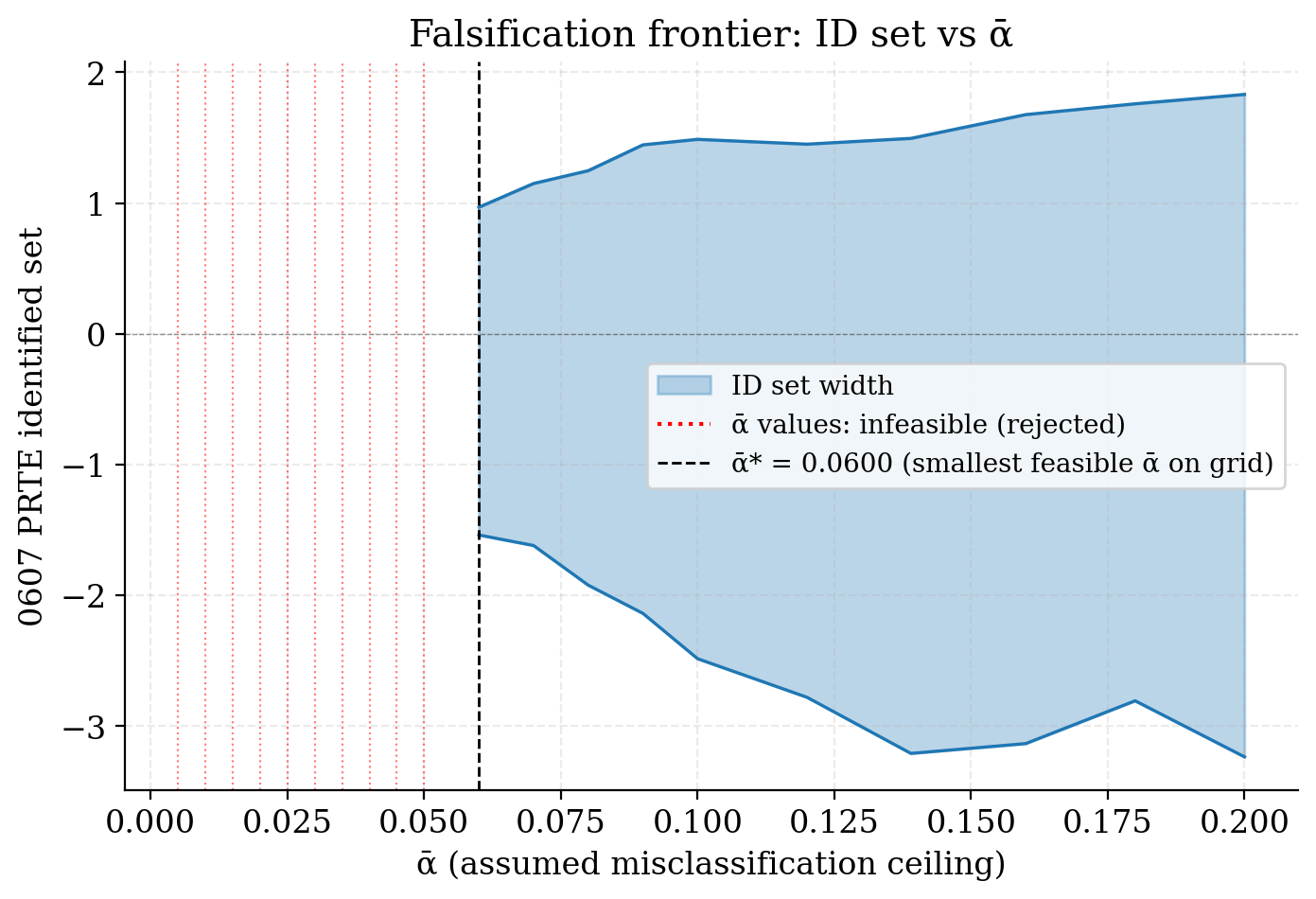}
    \caption{Falsification frontier for the empirical TV-based PRTE bounds.}
    \label{fig:emp-alpha-frontier}
    \figurenote{The dashed vertical line marks the first feasible value of the misclassification ceiling on the grid, $\bar\alpha^*=0.060$. Red vertical lines indicate infeasible values of $\bar\alpha$. The shaded region shows the PRTE identified set for the reference policy parameter $a=0.2$ used in the frontier calculation.}
\end{figure}

Figure~\ref{fig:emp-alpha-frontier} reports the falsification-frontier calculation for the reference policy parameter $a=0.2$.  Values of $\bar\alpha$ below 0.060 are infeasible on the search grid, while $\bar\alpha=0.060$ is the first feasible value.  At this first feasible value, the estimated PRTE identified set is approximately $[-1.539,0.968]$.  Thus, in this empirical illustration, the TV-based restrictions imply that small misclassification ceilings are rejected by the data under the maintained assumptions.

\begin{figure}[tbp]
    \centering
    \includegraphics[width=0.92\linewidth]{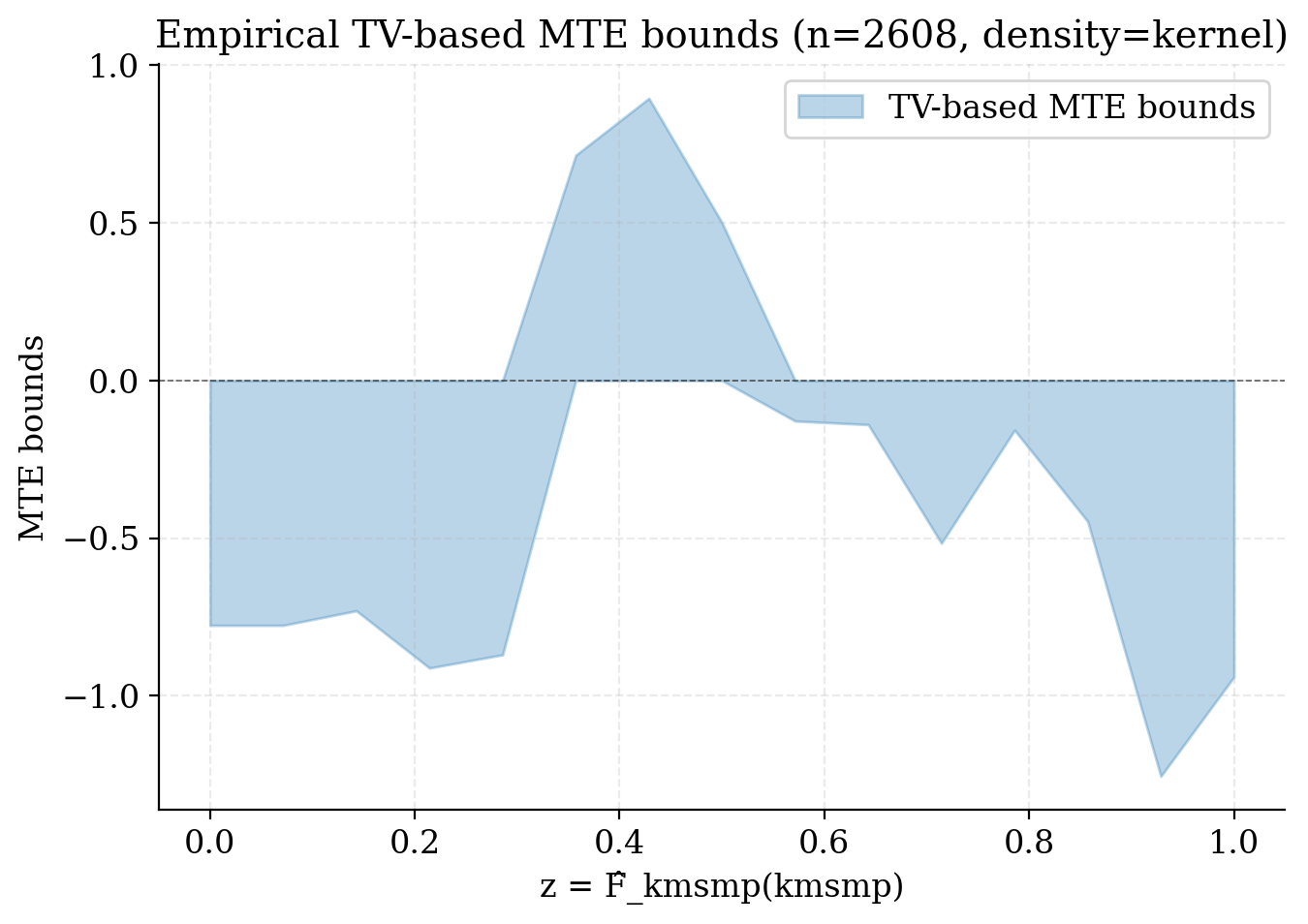}
    \caption{Empirical TV-based MTE bounds.}
    \label{fig:emp-mte-bounds}
    \figurenote{The horizontal axis is the empirical-CDF-transformed instrument $Z=\widehat F_{\mathrm{kmsmp}}(\mathrm{kmsmp})$. The shaded region reports the nodewise bounds $[\widehat{\underline m}(z),\widehat{\overline m}(z)]$ constructed from the local ratio in \eqref{eq:emp-local-ratio}.}
\end{figure}

Figure~\ref{fig:emp-mte-bounds} shows the resulting TV-based MTE bounds over the normalized instrument.  In the latest run, the estimated nodewise bounds are one-sided nonpositive over the lower part of the instrument support, roughly through $z=0.29$, and again over the upper support, roughly from $z=0.57$ onward.  They allow positive MTE values over the middle of the support, with the upper bound peaking at about 0.90 near $z=0.43$.  The most negative lower bound is about $-1.25$ near the upper tail of the instrument support.  This pattern reflects variation in the residualized outcome regression relative to the TV-implied lower bound on latent propensity-score movement.

\begin{figure}[tbp]
    \centering
    \includegraphics[width=0.92\linewidth]{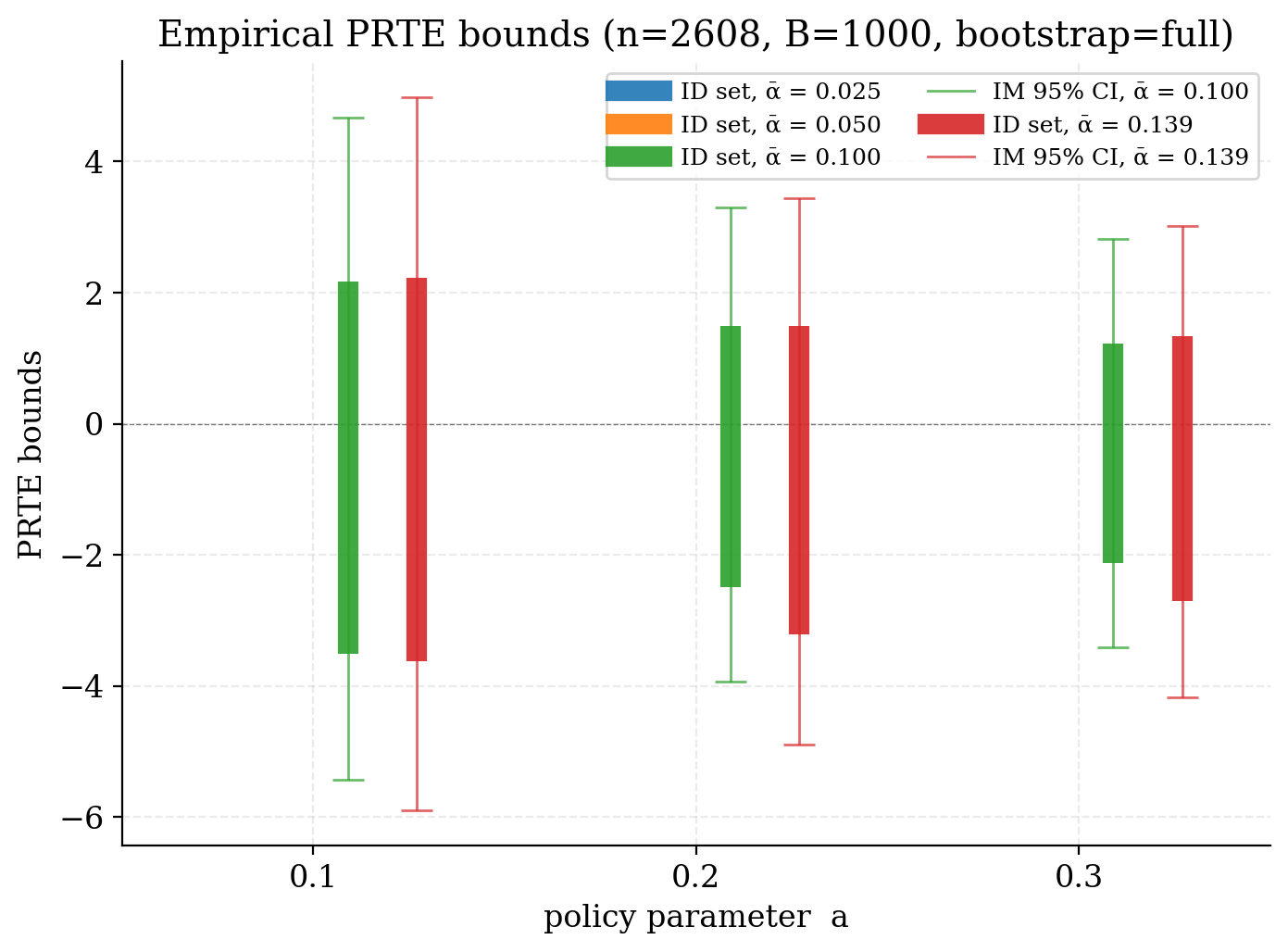}
    \caption{Empirical PRTE bounds by policy parameter and misclassification ceiling.}
    \label{fig:emp-prte-bounds}
    \figurenote{Thick vertical bars denote the estimated identified sets. Thin error bars denote full-cluster-bootstrap Imbens--Manski-style intervals for the bound endpoints. Infeasible ceilings in the original sample are omitted from the plotted identified-set bars and endpoint intervals.}
\end{figure}

Figure~\ref{fig:emp-prte-bounds} reports the PRTE bounds for policy parameters $a\in\{0.1,0.2,0.3\}$.  The ceilings $\bar\alpha=0.025$ and $\bar\alpha=0.050$ are infeasible in the original sample and therefore do not produce identified-set bars.  For $\bar\alpha=0.100$, the estimated PRTE bounds are approximately $[-3.503,2.161]$, $[-2.485,1.487]$, and $[-2.124,1.216]$ for $a=0.1,0.2,0.3$, respectively.  For $\bar\alpha=0.139$, the corresponding bounds are approximately $[-3.625,2.229]$, $[-3.209,1.494]$, and $[-2.705,1.333]$.  All displayed identified sets include zero.  The thin error bars report full-cluster-bootstrap Imbens--Manski-style intervals for the bound endpoints where the original-sample identified set is nonempty.  These intervals should be interpreted subject to the regularity conditions required for the first-stage smoothing estimators, the cluster bootstrap, and the numerical approximation to the finite-grid identified set.

\section{Details on the Numerical Illustrations}

\subsection{Details on the numerical illustrations in Section \ref{numeric1}}\label{numeric:apx1}
\subsubsection{DGP}

Let
\[
Z\sim \mathrm{Unif}[0,1],\qquad
P(z)\equiv \Phi(\delta_0+\delta_1 z),\qquad
D^*\equiv \mathbbm 1\{V\le P(Z)\},
\]
with $\delta_1>0$. In the numerical implementation,
\[
\delta_0=-2.5,\qquad \delta_1=5.0,
\]
so that $P(0)=\Phi(\delta_0)$ and $P(1)=\Phi(\delta_0+\delta_1)$ are close to, but not exactly, 0 and 1.
Let the observed treatment be
\[
D\equiv D^*(1-\varepsilon)+(1-D^*)\varepsilon,\qquad
\varepsilon\equiv \mathbbm 1\{\xi\le \alpha\}.
\]
The latent pair $(V,\xi)$ has uniform margins and a Gaussian copula:
\[
V\equiv \Phi(V^*),\qquad \xi\equiv \Phi(\xi^*),\qquad
\begin{pmatrix}V^*\\ \xi^*\end{pmatrix}
\sim N\!\left(
\begin{pmatrix}0\\ 0\end{pmatrix},
\begin{pmatrix}1 & \rho\\ \rho & 1\end{pmatrix}
\right).
\]

The outcome equation is
\[
Y\equiv \beta D^*+U,\qquad
\beta \equiv \beta_0+\beta_1(1-\sqrt V)+\eta,
\]
where $U$ and $\eta$ are independent of $(V,\xi)$, with
\[
U\sim N(\mu_U,\sigma_U^2),\qquad \eta\sim N(0,\sigma_\eta^2).
\]

Hence the potential outcomes are
\[
Y_0\equiv U,\qquad
Y_1\equiv \beta+U.
\]

The implementation uses the same calibration as in the code:
\[
\beta_0=1.5,\qquad \beta_1=1.0,\qquad \mu_U=2,\qquad \sigma_U^2=1,\qquad \sigma_\eta^2=1.
\]

\subsubsection{Structural objects}

The true propensity score is
\[
P(z)\equiv \Pr(D^*=1\mid Z=z)=\Phi(\delta_0+\delta_1z),
\]
so $z$ denotes the raw instrument index, while $p=P(z)$ denotes the true propensity score. The true MTE is a function of the scalar resistance index:

\[
MTE(p)\equiv \mathbb E[\beta\mid V=p]=\beta_0+\beta_1(1-\sqrt p),
\qquad p\in[0,1].
\]
Therefore
\[
MTE(P(z))=\beta_0+\beta_1\{1-\sqrt{P(z)}\}
=\beta_0+\beta_1\left\{1-\sqrt{\Phi(\delta_0+\delta_1z)}\right\},
\qquad z\in[0,1],
\]

which is strictly positive and nonconstant.

The conditional mean of $Y$ is
\[
\mathbb E[Y\mid Z=z]
=
\mu_U+\int_0^{P(z)} MTE(v)\,dv
=
\mu_U+\beta_0 P(z)+\beta_1\left(P(z)-\frac{2}{3}P(z)^{3/2}\right).
\]

The observed propensity score is
\[
Q(z)\equiv \mathbb E[D\mid Z=z]=P(z)+\alpha-2C_G(P(z),\alpha;\rho),
\]
where
\[
C_G(u,v;\rho)\equiv \Phi_2(\Phi^{-1}(u),\Phi^{-1}(v);\rho)
\]
is the Gaussian copula CDF. Its derivative is
\[
Q'(z)=P'(z)\left[1-2\Phi\!\left(
\frac{\Phi^{-1}(\alpha)-\rho\,\Phi^{-1}(P(z))}{\sqrt{1-\rho^2}}
\right)\right],
\qquad
P'(z)=\delta_1\phi(\delta_0+\delta_1z).
\]

The true PRTE under the policy shift $P^a(z)\equiv a+(1-a)P(z)$ is
\[
PRTE(a)
=
\int_0^1 MTE(p)\,\omega_a(p)\,dp,
\]
with
\[
\omega_a(p)\equiv
\frac{F_{P(Z)}(p)-F_{P(Z)}(t_a(p))}
{a\{1-\mathbb E[P(Z)]\}},
\qquad
 t_a(p)\equiv \frac{p-a}{1-a},
\]
where $F_{P(Z)}$ is the distribution function of the generated propensity score $P(Z)$. Since $Z\sim\mathrm{Unif}[0,1]$ and $P(z)=\Phi(\delta_0+\delta_1z)$,
\[
F_{P(Z)}(p)=
\left[\frac{\Phi^{-1}(p)-\delta_0}{\delta_1}\right]_{0}^{1},
\]
where $[x]_0^1\equiv\min\{1,\max\{0,x\}\}$. The true PRTE is evaluated numerically from the score distribution induced by $P(Z)$.

\subsubsection{Naive LIV, naive MTE, and naive PRTE}

The local naive LIV ratio based on the observed treatment is
\[
R^{naive}(z)\equiv
\frac{\frac{d}{dz}\mathbb E[Y\mid Z=z]}{\frac{d}{dz}\mathbb E[D\mid Z=z]}
=
\frac{\beta_0+\beta_1\{1-\sqrt{P(z)}\}}
{1-2\Phi\!\left(
\frac{\Phi^{-1}(\alpha)-\rho\,\Phi^{-1}(P(z))}{\sqrt{1-\rho^2}}
\right)}.
\]
Here the factor $P'(z)$ cancels from the numerator and denominator, but the numerator and the conditional reporting probability are both evaluated at $P(z)$.

When $Q(z)$ is monotone, the same object can be plotted as a score-indexed naive LIV curve $q\mapsto LIV(q)$. 
In this DGP, this occurs for
\[
(\alpha,\rho)=(0.1,0),\ (0.1,0.3),\ (0.1,0.5),\ (0.3,0).
\]

When $Q$ is non-monotone, that score-indexed object is not single-valued. In that case the natural practitioner-side object is the path
\[
z\mapsto \bigl(Q(z),R^{naive}(z)\bigr).
\]
It shows the horizontal backtracking caused by non-monotone $Q$, and therefore makes clear why the score-indexed naive LIV is not single-valued.

\begin{figure}[htbp]
\centering
{\color{red}\includegraphics[width=0.82\textwidth]{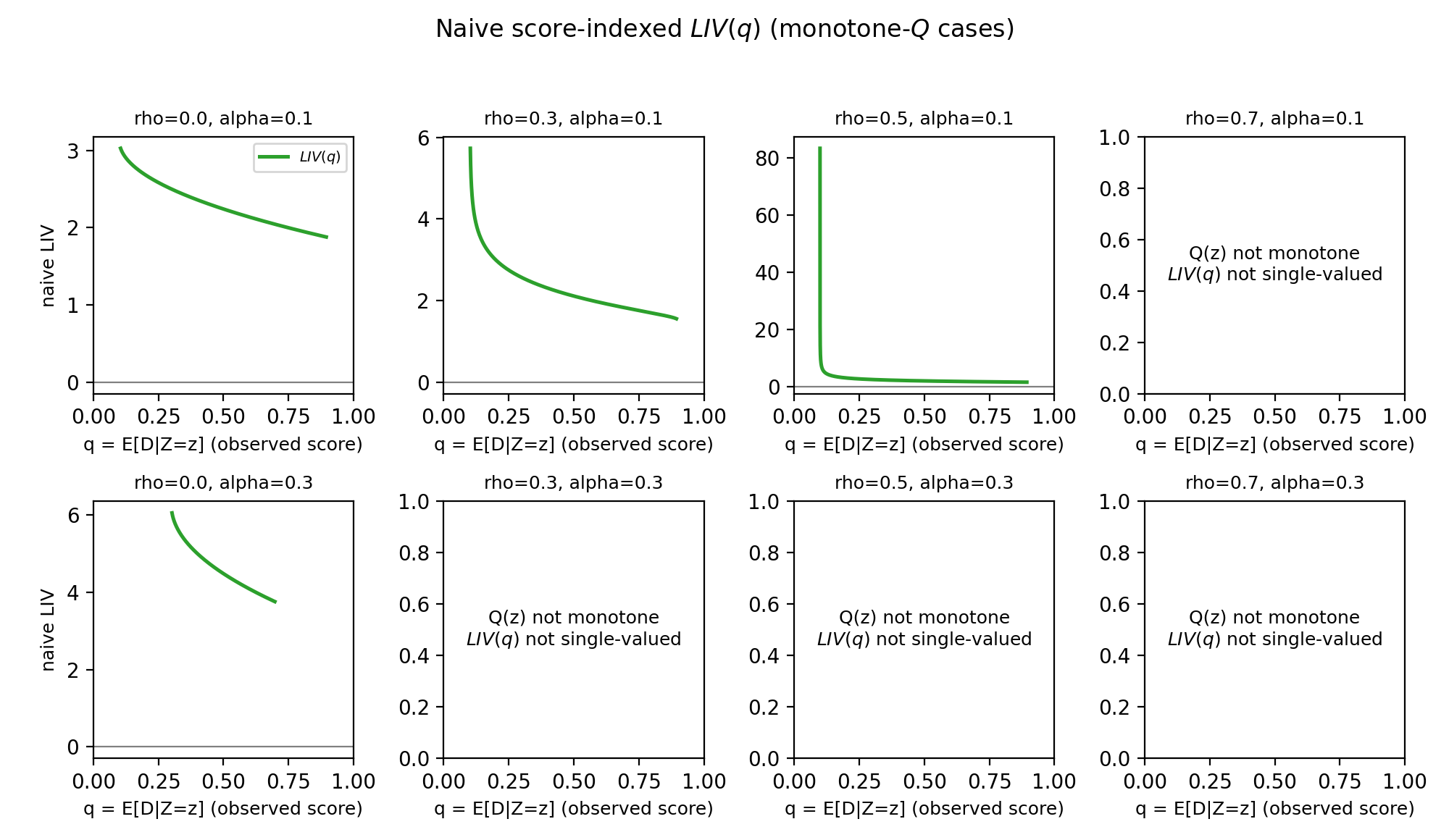}}
\caption{Score-indexed naive LIV in the monotone-$Q$ cases.}\label{fig.ey1.exam0921_qmono}
\end{figure}

\begin{figure}[htbp]
\centering
{\color{red}\includegraphics[width=0.82\textwidth]{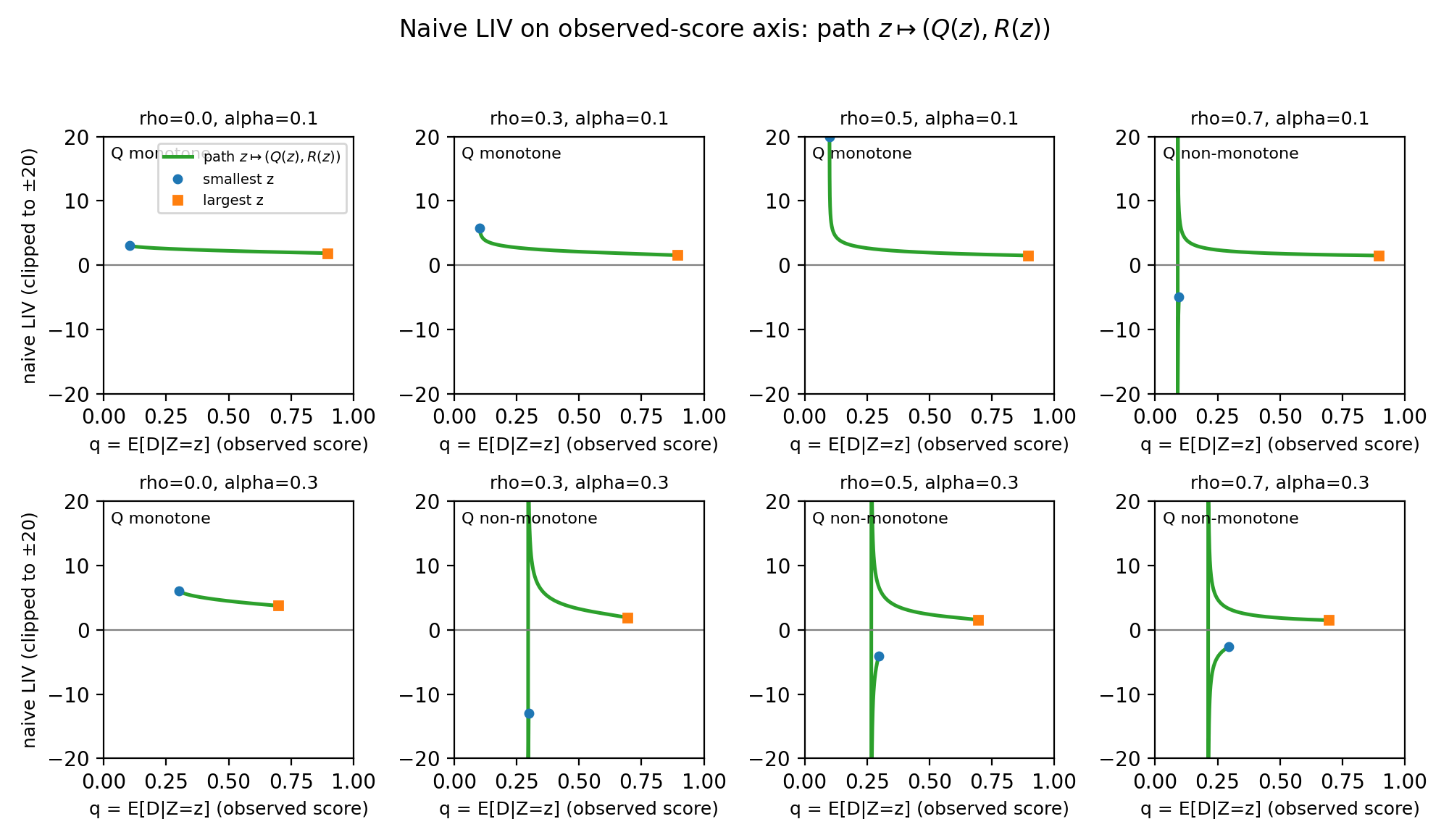}}
\caption{Pathwise naive LIV: $z\mapsto(Q(z),R^{naive}(z))$.}\label{fig.ey1.exam0921_qpath}
\end{figure}

The naive PRTE used in the code is the pathwise score-weighted object
\[
PRTE^{naive,path}(a)
\equiv
\int_0^1
R^{naive}(z)\,
\frac{F_{Q(Z)}(Q(z))-F_{Q(Z)}(t_a(Q(z)))}
{a(1-\mathbb E[Q(Z)])}\,Q'(z)\,dz,
\]
where
\[
t_a(q)\equiv \max\left\{0,\frac{q-a}{1-a}\right\}.
\]
This object is well-defined even when $Q$ is non-monotone because it is evaluated in $z$-space rather than by inverting $Q$.

\subsubsection{TV-based MTE bounds}

The pointwise TV-based bound takes the form
\[
\min\{0,R^{TV}(z)\}\le MTE(P(z))\le \max\{0,R^{TV}(z)\},
\]
where
\[
R^{TV}(z)\equiv
\frac{\Delta_{YZ}(z',z)}{LB_p^{TV}(z',z)}
\]
for a small local interval and
\[
LB_p^{TV}(z',z)\equiv
\max\{TV_{(Y,D=1)}(z',z),\,TV_{(Y,D=0)}(z',z)\}.
\]

Under this DGP,
\[
f_1(y\mid v)\equiv f_{Y_1\mid V=v}(y)
=
\phi\!\left(y;\mu_U+\beta_0+\beta_1(1-\sqrt v),\,\sigma_U^2+\sigma_\eta^2\right),
\]
\[
f_0(y)\equiv f_{Y_0}(y)=\phi(y;\mu_U,\sigma_U^2).
\]
The conditional misclassification probability is
\[
F_{\xi\mid V}(u\mid v)\equiv
\Phi\!\left(
\frac{\Phi^{-1}(u)-\rho\,\Phi^{-1}(v)}{\sqrt{1-\rho^2}}
\right).
\]
Hence
\[
f_{Y,D=1\mid Z=z}(y)
=
\int_0^{P(z)} f_1(y\mid v)\bigl[1-F_{\xi\mid V}(\alpha\mid v)\bigr]\,dv
+
f_0(y)\bigl[\alpha-C_G(P(z),\alpha;\rho)\bigr],
\]
\[
f_{Y,D=0\mid Z=z}(y)
=
\int_0^{P(z)} f_1(y\mid v)F_{\xi\mid V}(\alpha\mid v)\,dv
+
f_0(y)\bigl[1-P(z)-\alpha+C_G(P(z),\alpha;\rho)\bigr].
\]
The TV denominator is obtained by numerically integrating the absolute differences of these subdensities over $y$.

\subsubsection{TV-based PRTE bounds}

The PRTE bounds are computed from a monotone path search over $p_i=P(z_i)$ on a coarse grid:
\[
0=z_1<\cdots<z_K=1,\qquad p_i\equiv P(z_i),\qquad p_1=P(0),\quad p_K=P(1).
\]
The implementation has two layers.

First, the coarse observable layer constructs:
\begin{itemize}
\item pointwise score bounds $LB(z_i)\le p_i\le UB(z_i)$,
\item coarse increment bounds $LB_p(z_{i-1},z_i)\le p_i-p_{i-1}\le UB_p(z_{i-1},z_i)$.
\end{itemize}

Second, the local MTE layer builds nodewise one-sided MTE bounds from semi-analytic local TV denominators on a fine step $h_{\mathrm{MTE}}$.

Feasible monotone paths are then explored in four steps.
\begin{enumerate}
\item \textit{Midpoint feasible path.} The algorithm first converts the primitive node and increment restrictions into effective increment bounds
\[
LBd_i \equiv \max\{0,LB_p(z_{i-1},z_i),LB(z_i)-UB(z_{i-1})\},
\qquad
UBd_i \equiv \min\{UB_p(z_{i-1},z_i),UB(z_i)-LB(z_{i-1})\},
\]
and then computes the remaining-sum envelopes
\[
rem\_min(i) \equiv \sum_{j=i}^K LBd_j,
\qquad
rem\_max(i) \equiv \sum_{j=i}^K UBd_j.
\]
Given the preceding value $p_{i-1}$, the admissible interval for $p_i$ is

\[
\ell_i \equiv \max\{LB(z_i),\ p_{i-1}+LBd_i,\ P(1)-rem\_max(i+1)\},
\]
\[
u_i \equiv \min\{UB(z_i),\ p_{i-1}+UBd_i,\ P(1)-rem\_min(i+1)\}.
\]
The midpoint path sets $p_i=(\ell_i+u_i)/2$ whenever this interval is nonempty. This gives one deterministic feasible interior path.

\item \textit{Random interior feasible paths.} Starting from $p_1=P(0)$, the sampler proceeds sequentially over the grid. At each interior node it recomputes the admissible interval $[\ell_i,u_i]$ conditional on the already chosen predecessor and the requirement that the remaining nodes can still reach $p_K=P(1)$. It then draws $p_i$ uniformly from $[\ell_i,u_i]$. Every such draw is feasible by construction, but the resulting paths tend to lie in the interior of the feasible polytope rather than on its boundary.

\item \textit{Linear-programming-guided (LP-guided) boundary/extreme paths.} To target boundary points more directly, the code solves a family of linear programs of the form
\[
\min_p c^\top p
\]
subject to the endpoint equalities
\[
p_1=P(0),\qquad p_K=P(1),
\]
variable bounds
\[
LB(z_i)\le p_i\le UB(z_i),
\]
and increment inequalities
\[
LB_p(z_{i-1},z_i)\le p_i-p_{i-1}\le UB_p(z_{i-1},z_i).
\]
Different choices of the linear objective $c$ push the optimizer toward different faces of the feasible polytope. The implementation uses a zero objective (to get any feasible point), several deterministic shape objectives (for example constant, left-tilting, right-tilting, and curved directions), and randomized Gaussian directions. Since linear objectives attain their optima on extreme faces, these LP solutions are typically much more boundary-oriented than the random interior draws.

\item \textit{Coordinate-wise boundary refinement.} Starting from the best candidate paths found so far, the algorithm updates one interior coordinate at a time while keeping its neighbors fixed. For node $i$, this yields a local feasible interval $[\ell_i,u_i]$ implied by the pointwise and adjacent increment constraints. The code then evaluates the PRTE objective at the finite candidate set
\[
\{p_i,\ \ell_i,\ u_i,\ (\ell_i+u_i)/2\},
\]
where $p_i$ is the incumbent value. It replaces $p_i$ by whichever candidate improves the lower- or upper-bound objective the most. Repeating this sweep over coordinates pushes a candidate path toward nearby boundary points or vertices that are more favorable for extremizing PRTE.
\end{enumerate}

Taken together, these four steps do not deliver an exhaustive search over all feasible paths, but they combine interior exploration, boundary exploration, and local refinement in a way that is much more informative than using random feasible paths alone.

\subsection{Numerical verification of the implication of Lemma \ref{morett_samedirection}}

$\,$

\begin{figure}[h]
    \includegraphics[width=\textwidth]{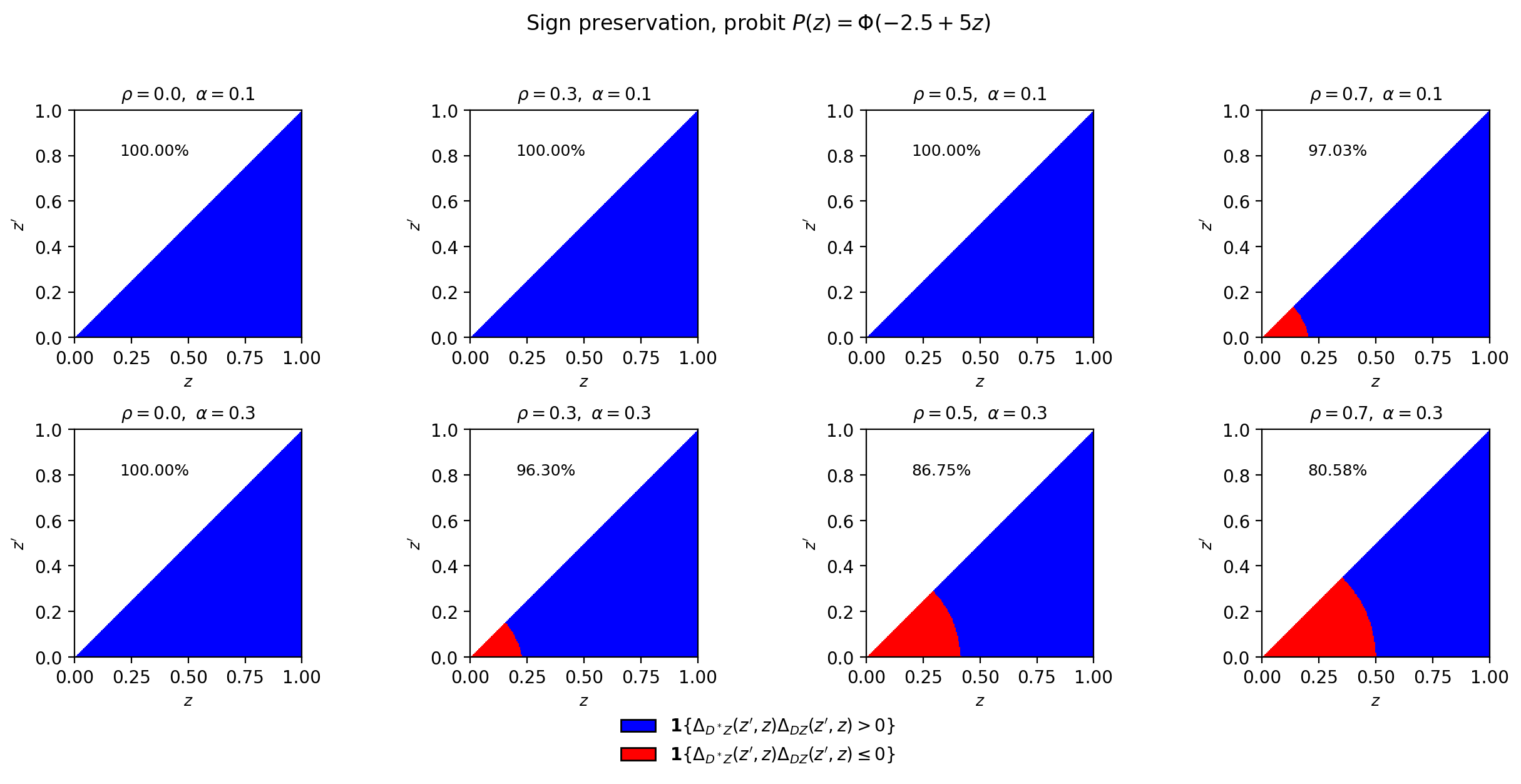}\\
    \figurenote{The blue region represents combinations for which $\Delta_{D^*Z}(z',z)\cdot\Delta_{DZ}(z',z)>0$ is satisfied, whereas the red region represents combinations that violate this condition. The percentage in each panel reports the share of the grid covered by the blue region.}
\caption{Numerical verification of $\Delta_{D^*Z}(z',z)\cdot\Delta_{DZ}(z',z)>0$}
	\label{fig.samedirection}
\end{figure}

\clearpage

\subsection{Details on the numerical illustration of the special case} \label{numeric2:apx}

We assume in this illustration that the researcher knows that the rate of misclassification $\alpha$ is less than 1/2.
Consider the same example from the previous section (\ref{eq:ex}) where $\rho=0$ (i.e., $\varepsilon$ is independent of $V$). 

Note that have $P(z)=z$ and therefore
\[
Q(z) \equiv \Pr(D=1\mid Z=z) = z + \alpha - 2 z\alpha = (1-2\alpha)z + \alpha.
\]
Hence, the following can be verified for the identification region for $\alpha$:
\[
\inf_z Q(z)=\alpha,
\qquad
\sup_z Q(z)=1-\alpha.
\]
Also,
\[
E[Y\mid Z=z] = \mu_U + \int_0^z \bigl[\beta_0+\beta_1(1-\sqrt v)\bigr]dv
= \mu_U + \beta_0 z + \beta_1\left(z-\frac{2}{3}z^{3/2}\right).
\]
Since $Q(z)$ is invertible when $\alpha<1/2$,
\[
z = \frac{q-\alpha}{1-2\alpha},
\qquad q\in[\alpha,1-\alpha],
\]
so
\[
E[Y\mid Q(Z)=q]
= \mu_U + \beta_0\frac{q-\alpha}{1-2\alpha}
+ \beta_1\left(\frac{q-\alpha}{1-2\alpha} - \frac{2}{3}\left(\frac{q-\alpha}{1-2\alpha}\right)^{3/2}\right).
\]
Differentiating with respect to $q$ gives the naive LIV:
\[
LIV(q)
= \frac{1}{1-2\alpha}\left[\beta_0 + \beta_1\left(1-\sqrt{\frac{q-\alpha}{1-2\alpha}}\right)\right].
\]
Thus for any candidate $\tilde\alpha<1/2$,
\[
MTE(p;\tilde\alpha)
\equiv
(1-2\tilde\alpha)\,LIV\bigl((1-2\tilde\alpha)p+\tilde\alpha\bigr)
= \frac{1-2\tilde\alpha}{1-2\alpha}
\left[\beta_0 + \beta_1\left(1-\sqrt{\frac{(1-2\tilde\alpha)p+\tilde\alpha-\alpha}{1-2\alpha}}\right)\right],
\]
whenever the square-root argument lies in $[0,1]$. If $\tilde\alpha=\alpha$, this collapses to the true MTE:
\[
MTE(p;\alpha)=\beta_0+\beta_1(1-\sqrt p).
\]

Let
\[
P(z)\equiv \Phi(\delta_0+\delta_1 z),
\qquad
p_L\equiv P(0)=\Phi(\delta_0),
\qquad
p_U\equiv P(1)=\Phi(\delta_0+\delta_1),
\]
with $\delta_1>0$. Since $\rho=0$, $V$ and $\xi$ are independent, and therefore
\[
Q(z)\equiv \Pr(D=1\mid Z=z)
=
P(z)+\alpha-2\alpha P(z)
=
\alpha+(1-2\alpha)P(z).
\]
Hence, because $\alpha<1/2$ and $P(z)$ is increasing in $z$,
\[
\inf_z Q(z)=\alpha+(1-2\alpha)p_L,
\qquad
\sup_z Q(z)=\alpha+(1-2\alpha)p_U.
\]
Equivalently, if the support endpoints of $Q(Z)$ are denoted by
\[
q_L\equiv \inf_z Q(z),
\qquad
q_U\equiv \sup_z Q(z),
\]
then the support restrictions imply
\[
q_L=\alpha+(1-2\alpha)p_L,
\qquad
q_U=\alpha+(1-2\alpha)p_U.
\]
Thus, for known $(p_L,p_U)$, the support endpoints provide the restrictions
\[
\alpha=\frac{q_L-p_L}{1-2p_L}
=
\frac{q_U-p_U}{1-2p_U},
\]
whenever the denominators are nonzero. In the numerical calibration used here,
$p_L$ is close to zero and $p_U$ is close to one, so these restrictions are close to the full-support identities
$q_L=\alpha$ and $q_U=1-\alpha$.

Also,
\[
E[Y\mid Z=z]
=
\mu_U+\int_0^{P(z)}
\bigl[\beta_0+\beta_1(1-\sqrt v)\bigr]\,dv
=
\mu_U+\beta_0 P(z)
+\beta_1\left(P(z)-\frac{2}{3}P(z)^{3/2}\right).
\]
Since $Q(z)$ is invertible when $\alpha<1/2$, the corresponding propensity-score value is
\[
p(q)
=
\frac{q-\alpha}{1-2\alpha},
\qquad
q\in
\left[
\alpha+(1-2\alpha)p_L,\,
\alpha+(1-2\alpha)p_U
\right].
\]
Therefore
\[
E[Y\mid Q(Z)=q]
=
\mu_U+\beta_0\frac{q-\alpha}{1-2\alpha}
+
\beta_1\left[
\frac{q-\alpha}{1-2\alpha}
-\frac{2}{3}
\left(\frac{q-\alpha}{1-2\alpha}\right)^{3/2}
\right].
\]
Differentiating with respect to $q$ gives the naive LIV:
\[
LIV(q)
=
\frac{1}{1-2\alpha}
\left[
\beta_0+\beta_1\left(
1-\sqrt{\frac{q-\alpha}{1-2\alpha}}
\right)
\right],
\]
for $q$ in the observed support of $Q(Z)$.

Thus, for any candidate $\tilde\alpha<1/2$,
\[
MTE(p;\tilde\alpha)
\equiv
(1-2\tilde\alpha)\,
LIV\bigl((1-2\tilde\alpha)p+\tilde\alpha\bigr),
\]
or equivalently
\[
MTE(p;\tilde\alpha)
=
\frac{1-2\tilde\alpha}{1-2\alpha}
\left[
\beta_0+\beta_1\left(
1-
\sqrt{
\frac{(1-2\tilde\alpha)p+\tilde\alpha-\alpha}{1-2\alpha}
}
\right)
\right],
\]
whenever
\[
(1-2\tilde\alpha)p+\tilde\alpha
\in
\left[
\alpha+(1-2\alpha)p_L,\,
\alpha+(1-2\alpha)p_U
\right].
\]
Equivalently, the square-root argument must lie in $[p_L,p_U]$. If $\tilde\alpha=\alpha$, this collapses to the true MTE on the support of $P(Z)$:
\[
MTE(p;\alpha)=\beta_0+\beta_1(1-\sqrt p),
\qquad p\in[p_L,p_U].
\]

\end{document}

%% file: tables/refined/table_main_aggregate_exogenous.tex
\begin{table}[!htbp]\centering
\caption{Aggregate-return summaries under $\varepsilon\indep V$.}\label{tab:main-aggregate-exogenous}
\scriptsize
\begin{tabular}{llcccc}
\toprule
Method & Estimand & $\alpha=0$ PE [95\% CI] & $\bar\alpha=0.025$ & $\bar\alpha=0.139$ & Conservative 95\% CI at $\bar\alpha=0.139$\\
\midrule
Gaussian & ATE & 0.137 [0.016, 0.295] & [0.129, 0.137] & [0.080, 0.137] & [-0.006, 0.295]\\
Gaussian & ATT & 0.215 [0.084, 0.460] & [0.200, 0.215] & [0.099, 0.215] & [0.007, 0.460]\\
Gaussian & ATU & 0.082 [-0.154, 0.314] & [0.078, 0.082] & [0.069, 0.099] & [-0.154, 0.314]\\
Gaussian & AMTE ($\zeta=0.05$) & 0.155 [0.061, 0.306] & [0.146, 0.155] & [0.079, 0.155] & [0.007, 0.306]\\
Gaussian & PRTE ($a=0.1$) & 0.157 [0.067, 0.310] & [0.149, 0.157] & [0.082, 0.157] & [0.004, 0.310]\\
Gaussian & PRTE ($a=0.2$) & 0.148 [0.056, 0.294] & [0.140, 0.148] & [0.080, 0.148] & [0.011, 0.294]\\
Gaussian & PRTE ($a=0.3$) & 0.140 [0.048, 0.281] & [0.133, 0.140] & [0.079, 0.140] & [0.008, 0.281]\\
Spline & ATE & 0.158 [0.004, 0.344] & [0.136, 0.158] & [0.091, 0.158] & [-0.044, 0.344]\\
Spline & ATT & 0.348 [0.213, 0.663] & [0.325, 0.348] & [0.165, 0.348] & [0.028, 0.663]\\
Spline & ATU & 0.023 [-0.301, 0.255] & [0.004, 0.023] & [-0.003, 0.067] & [-0.323, 0.274]\\
Spline & AMTE ($\zeta=0.05$) & 0.178 [0.065, 0.344] & [0.167, 0.178] & [0.118, 0.178] & [0.017, 0.344]\\
Spline & PRTE ($a=0.1$) & 0.166 [0.059, 0.325] & [0.158, 0.166] & [0.119, 0.166] & [0.025, 0.325]\\
Spline & PRTE ($a=0.2$) & 0.140 [0.038, 0.286] & [0.130, 0.140] & [0.094, 0.140] & [0.024, 0.286]\\
Spline & PRTE ($a=0.3$) & 0.126 [0.018, 0.275] & [0.116, 0.126] & [0.086, 0.126] & [0.004, 0.275]\\
\bottomrule
\end{tabular}
\par\smallskip
\begin{minipage}{0.78\textwidth}
{\footnotesize * Note: The point-estimate column reports the value at $\alpha=0$ together with its 95\% confidence interval from 1000 community-level cluster bootstrap replications. The columns indexed by $\bar\alpha$ report identified-set bounds. The final column reports the conservative 95\% confidence region at $\bar\alpha=0.139$, obtained by taking the outer union of alpha-wise bootstrap intervals over the nested grid. ATE, ATT, ATU, and AMTE are computed using the simulation-based integration approach in \cite{Carneiroal2017}. The PRTE is the policy-relevant treatment effect of \citet{heckman2005structural} under the counterfactual propensity shift $P_i^{a}=P_i+a(1-P_i)$ \citep{checkman2010}, averaged over the margins at which latent treatment status changes.}
\end{minipage}
\end{table}

%% file: tables/refined/table_app_lowv_compact.tex
\begin{table}[!htbp]\centering
\caption{Aggregate-return summaries under $\varepsilon=\mathbbm{1}\{V\leq\alpha\}$.}\label{tab:app-vleq-compact}
\scriptsize
\begin{tabular}{llcccc}
\toprule
Method & Estimand & $\alpha=0$ PE [95\% CI] & $\bar\alpha=0.025$ & $\bar\alpha=0.139$ & Conservative 95\% CI at $\bar\alpha=0.139$\\
\midrule
Gaussian & ATE & 0.137 [0.030, 0.304] & [0.137, 0.148] & [0.083, 0.170] & [-0.011, 0.337]\\
Gaussian & ATT & 0.215 [0.072, 0.460] & [0.215, 0.234] & [0.067, 0.259] & [-0.106, 0.561]\\
Gaussian & ATU & 0.082 [-0.156, 0.303] & [0.078, 0.084] & [0.075, 0.138] & [-0.160, 0.358]\\
Gaussian & AMTE ($\zeta=0.05$) & 0.155 [0.070, 0.315] & [0.155, 0.159] & [0.085, 0.166] & [-0.006, 0.330]\\
Gaussian & PRTE ($a=0.1$) & 0.157 [0.071, 0.312] & [0.157, 0.162] & [0.071, 0.170] & [-0.016, 0.337]\\
Gaussian & PRTE ($a=0.2$) & 0.148 [0.067, 0.303] & [0.148, 0.151] & [0.077, 0.159] & [-0.009, 0.317]\\
Gaussian & PRTE ($a=0.3$) & 0.140 [0.046, 0.294] & [0.140, 0.142] & [0.082, 0.150] & [-0.001, 0.304]\\
Spline & ATE & 0.158 [0.006, 0.338] & [0.158, 0.169] & [0.074, 0.179] & [-0.047, 0.372]\\
Spline & ATT & 0.348 [0.211, 0.693] & [0.347, 0.371] & [0.100, 0.388] & [-0.043, 0.780]\\
Spline & ATU & 0.023 [-0.307, 0.241] & [0.010, 0.023] & [0.004, 0.086] & [-0.340, 0.312]\\
Spline & AMTE ($\zeta=0.05$) & 0.178 [0.075, 0.342] & [0.173, 0.178] & [0.085, 0.178] & [-0.017, 0.342]\\
Spline & PRTE ($a=0.1$) & 0.166 [0.073, 0.325] & [0.165, 0.169] & [0.076, 0.172] & [-0.020, 0.338]\\
Spline & PRTE ($a=0.2$) & 0.140 [0.042, 0.294] & [0.139, 0.141] & [0.078, 0.143] & [-0.016, 0.305]\\
Spline & PRTE ($a=0.3$) & 0.126 [0.017, 0.280] & [0.125, 0.126] & [0.082, 0.129] & [-0.011, 0.284]\\
\bottomrule
\end{tabular}
\par\smallskip
\begin{minipage}{0.78\textwidth}
{\footnotesize * Note: The point-estimate column reports the value at $\alpha=0$ together with its 95\% confidence interval from 1000 community-level cluster bootstrap replications. The columns indexed by $\bar\alpha$ report identified-set bounds. The final column reports the conservative 95\% confidence region at $\bar\alpha=0.139$, obtained by taking the outer union of alpha-wise bootstrap intervals over the nested grid. ATE, ATT, ATU, and AMTE are computed using the simulation-based integration approach in \cite{Carneiroal2017}. The PRTE is the policy-relevant treatment effect of \citet{heckman2005structural} under the counterfactual propensity shift $P_i^{a}=P_i+a(1-P_i)$ \citep{checkman2010}, averaged over the margins at which latent treatment status changes.}
\end{minipage}
\end{table}

%% file: tables/refined/table_app_highv_compact.tex
\begin{table}[!htbp]\centering
\caption{Aggregate-return summaries under $\varepsilon=\mathbbm{1}\{V>1-\alpha\}$.}\label{tab:app-vgt-compact}
\scriptsize
\begin{tabular}{llcccc}
\toprule
Method & Estimand & $\alpha=0$ PE [95\% CI] & $\bar\alpha=0.025$ & $\bar\alpha=0.139$ & Conservative 95\% CI at $\bar\alpha=0.139$\\
\midrule
Gaussian & ATE & 0.137 [0.013, 0.300] & [0.121, 0.137] & [-0.006, 0.137] & [-0.250, 0.307]\\
Gaussian & ATT & 0.215 [0.058, 0.457] & [0.206, 0.215] & [0.111, 0.215] & [-0.011, 0.457]\\
Gaussian & ATU & 0.082 [-0.166, 0.319] & [0.066, 0.082] & [-0.123, 0.112] & [-0.544, 0.358]\\
Gaussian & AMTE ($\zeta=0.05$) & 0.155 [0.060, 0.307] & [0.149, 0.155] & [0.082, 0.155] & [-0.042, 0.316]\\
Gaussian & PRTE ($a=0.1$) & 0.157 [0.060, 0.307] & [0.151, 0.157] & [0.095, 0.157] & [-0.017, 0.313]\\
Gaussian & PRTE ($a=0.2$) & 0.148 [0.052, 0.295] & [0.142, 0.148] & [0.087, 0.148] & [-0.051, 0.298]\\
Gaussian & PRTE ($a=0.3$) & 0.140 [0.041, 0.289] & [0.133, 0.140] & [0.066, 0.140] & [-0.085, 0.300]\\
Spline & ATE & 0.158 [-0.006, 0.353] & [0.139, 0.158] & [-0.013, 0.158] & [-0.277, 0.361]\\
Spline & ATT & 0.348 [0.217, 0.672] & [0.332, 0.348] & [0.195, 0.348] & [0.059, 0.672]\\
Spline & ATU & 0.023 [-0.325, 0.243] & [0.015, 0.031] & [-0.183, 0.081] & [-0.637, 0.321]\\
Spline & AMTE ($\zeta=0.05$) & 0.178 [0.059, 0.356] & [0.178, 0.183] & [0.115, 0.183] & [-0.025, 0.369]\\
Spline & PRTE ($a=0.1$) & 0.166 [0.057, 0.335] & [0.162, 0.166] & [0.117, 0.166] & [-0.011, 0.336]\\
Spline & PRTE ($a=0.2$) & 0.140 [0.019, 0.293] & [0.134, 0.140] & [0.080, 0.140] & [-0.070, 0.299]\\
Spline & PRTE ($a=0.3$) & 0.126 [0.003, 0.279] & [0.121, 0.126] & [0.057, 0.134] & [-0.110, 0.303]\\
\bottomrule
\end{tabular}
\par\smallskip
\begin{minipage}{0.78\textwidth}
{\footnotesize * Note: The point-estimate column reports the value at $\alpha=0$ together with its 95\% confidence interval from 1000 community-level cluster bootstrap replications. The columns indexed by $\bar\alpha$ report identified-set bounds. The final column reports the conservative 95\% confidence region at $\bar\alpha=0.139$, obtained by taking the outer union of alpha-wise bootstrap intervals over the nested grid. ATE, ATT, ATU, and AMTE are computed using the simulation-based integration approach in \cite{Carneiroal2017}. The PRTE is the policy-relevant treatment effect of \citet{heckman2005structural} under the counterfactual propensity shift $P_i^{a}=P_i+a(1-P_i)$ \citep{checkman2010}, averaged over the margins at which latent treatment status changes.}
\end{minipage}
\end{table}

%% file: tables/refined/table_app_lowv_bounds.tex
\begin{landscape}
\begin{table}[!htbp]\centering
\caption{Aggregate-return summaries under $\varepsilon=\mathbbm{1}\{V\leq\alpha\}$. Identified-set bounds}\label{tab:app-vleq-bounds}
\scriptsize
\begin{tabular}{llrrrrrrrrr}
\toprule
Method & Estimand & $LB_{0.139}$ & $LB_{0.100}$ & $LB_{0.050}$ & $LB_{0.025}$ & $PE_0$ & $UB_{0.025}$ & $UB_{0.050}$ & $UB_{0.100}$ & $UB_{0.139}$\\
\midrule
Gaussian & ATE & 0.083 & 0.110 & 0.137 & 0.137 & 0.137 & 0.148 & 0.161 & 0.170 & 0.170\\
Gaussian & ATT & 0.067 & 0.092 & 0.215 & 0.215 & 0.215 & 0.234 & 0.257 & 0.259 & 0.259\\
Gaussian & ATU & 0.075 & 0.075 & 0.075 & 0.078 & 0.082 & 0.084 & 0.084 & 0.138 & 0.138\\
Gaussian & AMTE ($\zeta=0.05$) & 0.085 & 0.109 & 0.155 & 0.155 & 0.155 & 0.159 & 0.164 & 0.166 & 0.166\\
Gaussian & PRTE ($a=0.1$) & 0.071 & 0.100 & 0.157 & 0.157 & 0.157 & 0.162 & 0.168 & 0.170 & 0.170\\
Gaussian & PRTE ($a=0.2$) & 0.077 & 0.105 & 0.148 & 0.148 & 0.148 & 0.151 & 0.156 & 0.159 & 0.159\\
Gaussian & PRTE ($a=0.3$) & 0.082 & 0.109 & 0.140 & 0.140 & 0.140 & 0.142 & 0.146 & 0.150 & 0.150\\
Spline & ATE & 0.074 & 0.101 & 0.158 & 0.158 & 0.158 & 0.169 & 0.179 & 0.179 & 0.179\\
Spline & ATT & 0.100 & 0.122 & 0.347 & 0.347 & 0.348 & 0.371 & 0.388 & 0.388 & 0.388\\
Spline & ATU & 0.004 & 0.004 & 0.004 & 0.010 & 0.023 & 0.023 & 0.023 & 0.086 & 0.086\\
Spline & AMTE ($\zeta=0.05$) & 0.085 & 0.112 & 0.166 & 0.173 & 0.178 & 0.178 & 0.178 & 0.178 & 0.178\\
Spline & PRTE ($a=0.1$) & 0.076 & 0.107 & 0.165 & 0.165 & 0.166 & 0.169 & 0.172 & 0.172 & 0.172\\
Spline & PRTE ($a=0.2$) & 0.078 & 0.107 & 0.139 & 0.139 & 0.140 & 0.141 & 0.143 & 0.143 & 0.143\\
Spline & PRTE ($a=0.3$) & 0.082 & 0.108 & 0.125 & 0.125 & 0.126 & 0.126 & 0.127 & 0.129 & 0.129\\
\bottomrule
\end{tabular}
\par\smallskip
\begin{minipage}{0.78\textwidth}
{\footnotesize * Note: $LB_b$ and $UB_b$ denote the lower and upper endpoints of the identified set for $\bar\alpha=b$. $PE_0$ is the point estimate at $\alpha=0$. The reported sets use the nested alpha grid and the same MTE-to-aggregate integration rule as the compact tables.}
\end{minipage}
\end{table}
\end{landscape}

%% file: tables/refined/table_app_lowv_ci.tex
\begin{landscape}
\begin{table}[!htbp]\centering
\caption{Aggregate-return summaries under $\varepsilon=\mathbbm{1}\{V\leq\alpha\}$. Conservative confidence bounds}\label{tab:app-vleq-ci}
\scriptsize
\begin{tabular}{llrrrrrrrrrr}
\toprule
Method & Estimand & $LCB_{0.139}$ & $LCB_{0.100}$ & $LCB_{0.050}$ & $LCB_{0.025}$ & $LCB_0$ & $UCB_0$ & $UCB_{0.025}$ & $UCB_{0.050}$ & $UCB_{0.100}$ & $UCB_{0.139}$\\
\midrule
Gaussian & ATE & -0.011 & 0.019 & 0.030 & 0.030 & 0.030 & 0.304 & 0.309 & 0.326 & 0.337 & 0.337\\
Gaussian & ATT & -0.106 & -0.078 & 0.072 & 0.072 & 0.072 & 0.460 & 0.500 & 0.546 & 0.561 & 0.561\\
Gaussian & ATU & -0.160 & -0.160 & -0.160 & -0.157 & -0.156 & 0.303 & 0.303 & 0.303 & 0.358 & 0.358\\
Gaussian & AMTE ($\zeta=0.05$) & -0.006 & 0.029 & 0.069 & 0.069 & 0.070 & 0.315 & 0.319 & 0.324 & 0.330 & 0.330\\
Gaussian & PRTE ($a=0.1$) & -0.016 & 0.015 & 0.071 & 0.071 & 0.071 & 0.312 & 0.320 & 0.333 & 0.337 & 0.337\\
Gaussian & PRTE ($a=0.2$) & -0.009 & 0.028 & 0.067 & 0.067 & 0.067 & 0.303 & 0.307 & 0.316 & 0.317 & 0.317\\
Gaussian & PRTE ($a=0.3$) & -0.001 & 0.035 & 0.046 & 0.046 & 0.046 & 0.294 & 0.298 & 0.303 & 0.304 & 0.304\\
Spline & ATE & -0.047 & -0.011 & 0.006 & 0.006 & 0.006 & 0.338 & 0.353 & 0.372 & 0.372 & 0.372\\
Spline & ATT & -0.043 & -0.032 & 0.205 & 0.205 & 0.211 & 0.693 & 0.735 & 0.780 & 0.780 & 0.780\\
Spline & ATU & -0.340 & -0.340 & -0.340 & -0.327 & -0.307 & 0.241 & 0.241 & 0.241 & 0.312 & 0.312\\
Spline & AMTE ($\zeta=0.05$) & -0.017 & 0.019 & 0.066 & 0.066 & 0.075 & 0.342 & 0.342 & 0.342 & 0.342 & 0.342\\
Spline & PRTE ($a=0.1$) & -0.020 & 0.024 & 0.070 & 0.070 & 0.073 & 0.325 & 0.333 & 0.338 & 0.338 & 0.338\\
Spline & PRTE ($a=0.2$) & -0.016 & 0.025 & 0.040 & 0.040 & 0.042 & 0.294 & 0.298 & 0.305 & 0.305 & 0.305\\
Spline & PRTE ($a=0.3$) & -0.011 & 0.016 & 0.016 & 0.016 & 0.017 & 0.280 & 0.284 & 0.284 & 0.284 & 0.284\\
\bottomrule
\end{tabular}
\par\smallskip
\begin{minipage}{0.78\textwidth}
{\footnotesize * Note: $LCB_b$ and $UCB_b$ denote the lower and upper endpoints of the conservative 95\% confidence region for $\bar\alpha=b$. For $b>0$, the confidence region is the outer union of alpha-wise bootstrap intervals over the nested grid. $LCB_0$ and $UCB_0$ report the 95\% confidence interval at $\alpha=0$. All confidence regions use 1000 community-level cluster bootstrap replications.}
\end{minipage}
\end{table}
\end{landscape}

%% file: tables/refined/table_app_highv_bounds.tex
\begin{landscape}
\begin{table}[!htbp]\centering
\caption{Aggregate-return summaries under $\varepsilon=\mathbbm{1}\{V>1-\alpha\}$. Identified-set bounds}\label{tab:app-vgt-bounds}
\scriptsize
\begin{tabular}{llrrrrrrrrr}
\toprule
Method & Estimand & $LB_{0.139}$ & $LB_{0.100}$ & $LB_{0.050}$ & $LB_{0.025}$ & $PE_0$ & $UB_{0.025}$ & $UB_{0.050}$ & $UB_{0.100}$ & $UB_{0.139}$\\
\midrule
Gaussian & ATE & -0.006 & -0.006 & 0.065 & 0.121 & 0.137 & 0.137 & 0.137 & 0.137 & 0.137\\
Gaussian & ATT & 0.111 & 0.139 & 0.196 & 0.206 & 0.215 & 0.215 & 0.215 & 0.215 & 0.215\\
Gaussian & ATU & -0.123 & -0.123 & -0.012 & 0.066 & 0.082 & 0.082 & 0.082 & 0.082 & 0.112\\
Gaussian & AMTE ($\zeta=0.05$) & 0.082 & 0.093 & 0.123 & 0.149 & 0.155 & 0.155 & 0.155 & 0.155 & 0.155\\
Gaussian & PRTE ($a=0.1$) & 0.095 & 0.109 & 0.131 & 0.151 & 0.157 & 0.157 & 0.157 & 0.157 & 0.157\\
Gaussian & PRTE ($a=0.2$) & 0.087 & 0.087 & 0.117 & 0.142 & 0.148 & 0.148 & 0.148 & 0.148 & 0.148\\
Gaussian & PRTE ($a=0.3$) & 0.066 & 0.066 & 0.104 & 0.133 & 0.140 & 0.140 & 0.140 & 0.140 & 0.140\\
Spline & ATE & -0.013 & -0.013 & 0.067 & 0.139 & 0.158 & 0.158 & 0.158 & 0.158 & 0.158\\
Spline & ATT & 0.195 & 0.199 & 0.300 & 0.332 & 0.348 & 0.348 & 0.348 & 0.348 & 0.348\\
Spline & ATU & -0.183 & -0.183 & -0.068 & 0.015 & 0.023 & 0.031 & 0.039 & 0.039 & 0.081\\
Spline & AMTE ($\zeta=0.05$) & 0.115 & 0.115 & 0.149 & 0.178 & 0.178 & 0.183 & 0.183 & 0.183 & 0.183\\
Spline & PRTE ($a=0.1$) & 0.117 & 0.117 & 0.141 & 0.162 & 0.166 & 0.166 & 0.166 & 0.166 & 0.166\\
Spline & PRTE ($a=0.2$) & 0.080 & 0.080 & 0.109 & 0.134 & 0.140 & 0.140 & 0.140 & 0.140 & 0.140\\
Spline & PRTE ($a=0.3$) & 0.057 & 0.057 & 0.092 & 0.121 & 0.126 & 0.126 & 0.127 & 0.127 & 0.134\\
\bottomrule
\end{tabular}
\par\smallskip
\begin{minipage}{0.78\textwidth}
{\footnotesize * Note: $LB_b$ and $UB_b$ denote the lower and upper endpoints of the identified set for $\bar\alpha=b$. $PE_0$ is the point estimate at $\alpha=0$. The reported sets use the nested alpha grid and the same MTE-to-aggregate integration rule as the compact tables.}
\end{minipage}
\end{table}
\end{landscape}

%% file: tables/refined/table_app_highv_ci.tex
\begin{landscape}
\begin{table}[!htbp]\centering
\caption{Aggregate-return summaries under $\varepsilon=\mathbbm{1}\{V>1-\alpha\}$. Conservative confidence bounds}\label{tab:app-vgt-ci}
\scriptsize
\begin{tabular}{llrrrrrrrrrr}
\toprule
Method & Estimand & $LCB_{0.139}$ & $LCB_{0.100}$ & $LCB_{0.050}$ & $LCB_{0.025}$ & $LCB_0$ & $UCB_0$ & $UCB_{0.025}$ & $UCB_{0.050}$ & $UCB_{0.100}$ & $UCB_{0.139}$\\
\midrule
Gaussian & ATE & -0.250 & -0.250 & -0.128 & -0.022 & 0.013 & 0.300 & 0.300 & 0.307 & 0.307 & 0.307\\
Gaussian & ATT & -0.011 & -0.002 & 0.050 & 0.057 & 0.058 & 0.457 & 0.457 & 0.457 & 0.457 & 0.457\\
Gaussian & ATU & -0.544 & -0.544 & -0.329 & -0.203 & -0.166 & 0.319 & 0.325 & 0.341 & 0.341 & 0.358\\
Gaussian & AMTE ($\zeta=0.05$) & -0.042 & -0.042 & 0.014 & 0.049 & 0.060 & 0.307 & 0.313 & 0.316 & 0.316 & 0.316\\
Gaussian & PRTE ($a=0.1$) & -0.017 & -0.015 & 0.022 & 0.053 & 0.060 & 0.307 & 0.310 & 0.313 & 0.313 & 0.313\\
Gaussian & PRTE ($a=0.2$) & -0.051 & -0.051 & 0.006 & 0.044 & 0.052 & 0.295 & 0.297 & 0.298 & 0.298 & 0.298\\
Gaussian & PRTE ($a=0.3$) & -0.085 & -0.085 & -0.019 & 0.028 & 0.041 & 0.289 & 0.290 & 0.300 & 0.300 & 0.300\\
Spline & ATE & -0.277 & -0.277 & -0.128 & -0.028 & -0.006 & 0.353 & 0.353 & 0.361 & 0.361 & 0.361\\
Spline & ATT & 0.059 & 0.059 & 0.173 & 0.211 & 0.217 & 0.672 & 0.672 & 0.672 & 0.672 & 0.672\\
Spline & ATU & -0.637 & -0.637 & -0.418 & -0.334 & -0.325 & 0.243 & 0.264 & 0.293 & 0.293 & 0.321\\
Spline & AMTE ($\zeta=0.05$) & -0.025 & -0.025 & 0.040 & 0.059 & 0.059 & 0.356 & 0.367 & 0.369 & 0.369 & 0.369\\
Spline & PRTE ($a=0.1$) & -0.011 & -0.011 & 0.031 & 0.053 & 0.057 & 0.335 & 0.336 & 0.336 & 0.336 & 0.336\\
Spline & PRTE ($a=0.2$) & -0.070 & -0.070 & -0.012 & 0.017 & 0.019 & 0.293 & 0.295 & 0.298 & 0.298 & 0.299\\
Spline & PRTE ($a=0.3$) & -0.110 & -0.110 & -0.038 & -0.003 & 0.003 & 0.279 & 0.282 & 0.288 & 0.288 & 0.303\\
\bottomrule
\end{tabular}
\par\smallskip
\begin{minipage}{0.78\textwidth}
{\footnotesize * Note: $LCB_b$ and $UCB_b$ denote the lower and upper endpoints of the conservative 95\% confidence region for $\bar\alpha=b$. For $b>0$, the confidence region is the outer union of alpha-wise bootstrap intervals over the nested grid. $LCB_0$ and $UCB_0$ report the 95\% confidence interval at $\alpha=0$. All confidence regions use 1000 community-level cluster bootstrap replications.}
\end{minipage}
\end{table}
\end{landscape}

%% file: output_app3_sympara/tab_emp_copula_first_stage.tex
\begin{table}[H]
\centering
\scriptsize
\caption{First-stage estimates and identification diagnostics}
\label{tab:emp_copula_first_stage}
\begin{threeparttable}
\resizebox{\textwidth}{!}{%
\begin{tabular}{lrrrrrrrrrrr}
\toprule
Copula & $\hat\alpha$ & $\hat\gamma_{0}$ & $\hat\gamma_{1}$ & $\hat\gamma_{2}$ & $\hat\gamma_{3}$ & $\hat\gamma_{4}$ & $\hat\theta_c$ & $\widehat P$ support & Cond. no. & $\alpha$ upper hit (\%) & Copula upper hit (\%) \\
\midrule
Gaussian & 0.139 & 0.746 & -0.285 & 0.859 & -0.054 & -0.450 & 0.970 & [0.326, 0.762] & 8.29e+02 & 76.8 & 36.9 \\
AMH & 0.024 & 0.425 & -0.558 & 0.539 & -0.339 & -0.809 & 0.970 & [0.209, 0.652] & 9.61e+02 & 28.1 & 45.5 \\
Gumbel & 0.139 & 0.746 & -0.284 & 0.858 & -0.053 & -0.451 & 8.000 & [0.326, 0.762] & 3.96e+04 & 52.7 & 26.5 \\
\bottomrule
\end{tabular}}
\begin{tablenotes}[flushleft]
\scriptsize
\item[] 
\begin{minipage}[t]{\textwidth}
\textit{Notes:} The table reports preferred MLE first-stage estimates. 
The recovered latent propensity is $P_c(z)=\Phi(B_K(z)'\gamma_c)$; 
the $\hat\gamma_j$ columns report the spline-index coefficients. 
For the Gaussian copula, $\theta_c=\rho$; for AMH and Gumbel, $\theta_c$ 
denotes the corresponding copula parameter. The support column reports the 
minimum and maximum of the recovered latent propensity score $\widehat P_c(Z)$. 
The condition number is computed from the scaled first-stage Jacobian. The last 
two columns report the fraction of cluster-bootstrap replications in which the 
estimate is at the imposed upper bound for $\alpha$ or for the copula parameter.
\end{minipage}
\end{tablenotes}
\end{threeparttable}
\end{table}

%% file: output_app3_sympara/tab_emp_copula_spline.tex
\begin{table}[H]
\centering
\scriptsize
\caption{Spline outcome regression and recovered propensity support}
\label{tab:emp_copula_spline}
\begin{threeparttable}
\begin{tabular}{lrrrrr}
\toprule
Copula & Basis dim. & $\lambda$ & Effective df & GCV & $\widehat P$ support \\
\midrule
Gaussian & 12 & 562.341 & 3.575 & 0.805 & [0.326, 0.762] \\
AMH & 12 & 316.228 & 3.955 & 0.805 & [0.209, 0.652] \\
Gumbel & 12 & 562.341 & 3.575 & 0.805 & [0.326, 0.762] \\
\bottomrule
\end{tabular}
\begin{tablenotes}[flushleft]
\footnotesize
\item[]\makebox[\linewidth][c]{%
\begin{minipage}{0.88\textwidth}
\footnotesize\textit{Notes:} The second stage fits a penalized cubic B-spline regression of the residualized outcome $Y_i^\perp$ on the recovered propensity score $\widehat P_c(Z_i)$. The MTE is the derivative of this fitted spline. The penalty parameter $\lambda$ is selected by the implementation, and effective degrees of freedom summarize the amount of smoothing. The recovered support is bounded away from the full unit interval, so the MTE figures should be read as local over the estimated support of $\widehat P_c(Z)$.
\end{minipage}}
\end{tablenotes}
\end{threeparttable}
\end{table}

%% file: output_app3_sympara/tab_emp_copula_mte_point.tex
\begin{table}[H]
\centering
\scriptsize
\caption{Selected MTE estimates at $p=0.50$}
\label{tab:emp_copula_mte_point}
\begin{threeparttable}
\begin{tabular}{lrrrr}
\toprule
Copula & $p$ & $\widehat{MTE}(p)$ & 95\% bootstrap CI & Bootstrap reps. \\
\midrule
Gaussian & 0.50 & 0.775 & [0.028, 2.078] & 499 \\
AMH & 0.50 & 0.988 & [-0.114, 1.809] & 499 \\
Gumbel & 0.50 & 0.774 & [0.063, 1.712] & 499 \\
\bottomrule
\end{tabular}
\begin{tablenotes}[flushleft]
\footnotesize
\item[]\makebox[\linewidth][c]{%
\begin{minipage}{0.78\textwidth}
\footnotesize\textit{Notes:} The table reports the MTE at $p=0.50$, which lies inside the recovered support for all three copula specifications. Requested points such as $p=0.25$ or $p=0.75$ are not common interior points across all three specifications, so the table reports only the common interior point.
\end{minipage}}
\end{tablenotes}
\end{threeparttable}
\end{table}

%% file: output_app3_sympara/tab_emp_copula_prte.tex
\begin{table}[H]
\centering
\scriptsize
\caption{Observed-support PRTE estimates}
\label{tab:emp_copula_prte}
\begin{threeparttable}
\begin{tabular}{llrrrr}
\toprule
Copula & $a$ & $\widehat{PRTE}_{obs}(a)$ & 95\% bootstrap CI & Observed weight (\%) & Outside support (\%) \\
\midrule
Gaussian & 0.1 & 1.147 & [0.267, 1.544] & 99.6 & 0.3 \\
Gaussian & 0.2 & 1.111 & [0.293, 1.596] & 99.0 & 1.2 \\
Gaussian & 0.3 & 1.144 & [0.173, 1.726] & 98.5 & 1.6 \\
AMH & 0.1 & 1.151 & [0.343, 1.600] & 99.3 & 0.7 \\
AMH & 0.2 & 1.133 & [0.313, 1.600] & 98.6 & 1.3 \\
AMH & 0.3 & 1.252 & [0.211, 1.749] & 97.5 & 2.6 \\
Gumbel & 0.1 & 1.147 & [0.288, 1.529] & 99.8 & 0.0 \\
Gumbel & 0.2 & 1.113 & [0.300, 1.639] & 99.2 & 0.6 \\
Gumbel & 0.3 & 1.146 & [0.264, 1.724] & 98.7 & 1.2 \\
\bottomrule
\end{tabular}
\begin{tablenotes}[flushleft]
\footnotesize
\item[]\textit{Notes:} The table reports observed-support PRTE estimates for the policy shift $T_a(p)=a+(1-a)p$. ``Observed weight'' is the integrated policy weight falling inside the recovered support of $\widehat P_c(Z)$; ``Outside support'' is the omitted policy-relevant weight. The confidence intervals are cluster-bootstrap percentile intervals from 499 successful replications. Across copulas, the PRTE estimates are positive for all three policy shifts; for the displayed shifts, most policy weight falls inside the recovered support.
\end{tablenotes}
\end{threeparttable}
\end{table}